\documentclass[journal]{IEEEtran}
\usepackage[T1]{fontenc}
\usepackage{color}
\usepackage{array,authblk,amssymb}
\usepackage{multirow}
\usepackage{fixltx2e}
\usepackage{graphicx}
\usepackage{cite}
\usepackage{amsfonts,xcolor,colortbl} 
\usepackage{amsmath}

\usepackage{xcolor}
\usepackage{hyperref}

\usepackage{xurl}

\usepackage[T1]{fontenc}       
\usepackage[utf8]{inputenc}    

\usepackage{amsthm,epstopdf }
\theoremstyle{definition}

\makeatletter
\def\endthebibliography{%
	\def\@noitemerr{\@latex@warning{Empty `thebibliography' environment}}%
	\endlist
}
\makeatother

\newcommand{\plus}{\texttt{\char43}}

\makeatother

\begin{document}

\title{Network Digital Twin for 6G and Beyond: An End-to-End View Across Multi-Domain Network Ecosystems}

\author{ Dinh-Hieu Tran\IEEEauthorrefmark{1}, Nazar Waheed \IEEEauthorrefmark{9}, Yuris Mulya Saputra\IEEEauthorrefmark{7}\IEEEmembership{(Senior, IEEE)},  Xingqin Lin\IEEEauthorrefmark{12}\IEEEmembership{(Senior, IEEE)}, Cong T. Nguyen\IEEEauthorrefmark{5,6}, Tedros Salih Abdu\IEEEauthorrefmark{1},  Van Nhan Vo\IEEEauthorrefmark{4}, Van-Quan Pham\IEEEauthorrefmark{8},  Madyan Alsenwi\IEEEauthorrefmark{1}, Abuzar BABIKIR MOHAMMAD ADAM \IEEEauthorrefmark{1}, Symeon Chatzinotas\IEEEauthorrefmark{1}\IEEEmembership{(Fellow, IEEE)}, Eva Lagaunas \IEEEauthorrefmark{1}\IEEEmembership{(Senior, IEEE)}, Hung Tran\IEEEauthorrefmark{10}, Tu Ho Dac\IEEEauthorrefmark{11}\IEEEmembership{(Senior, IEEE)}, Nguyen Van Huynh\IEEEauthorrefmark{2} 
\thanks{Dinh-Hieu Tran, Tedros Salih Abdu, Madyan Alsenwi, Abuzar BABIKIR MOHAMMAD ADAM, Eva Lagaunas, and Symeon Chatzinotas are with the Interdisciplinary Centre for Security, Reliability and Trust (SnT), the University of Luxembourg, Luxembourg. (e-mail: \{hieu.tran-dinh, tedros-salih.abdu,
madyan.alsenwi, abuzar.babikir, eva.lagunas, symeon.chatzinotas\} @uni.lu). }
\thanks{Corresponding author: Dinh-Hieu~Tran (e-mail: hieu.tran-dinh@uni.lu).}
\thanks{Nazar Waheed is with the nazar.waheed@gmail.com Higher colleges of Technology, Abu dhabi (email: nwaheed@hct.ac.ae ). }
\thanks{Yuris Mulya Saputra is with the Internet Engineering Technology, Department of Electrical Engineering and Informatics, Vocational College, Universitas Gadjah Mada,
Yogyakarta, 55281, Indonesia (email: ym.saputra@ugm.ac.id).}
\thanks{Cong T. Nguyen is with the Ho Chi Minh City University of Technology (email: congnguyen@hcmut.edu.vn) }
\thanks{Xingqin Lin is with NVIDIA, USA (email: xingqinl@nvidia.com).}
\thanks{Van Nhan Vo is with the Institute of Fundamental and Applied Sciences, Duy Tan University, Ho Chi Minh City 70000, Vietnam
and  Faculty of Information Technology, Duy Tan University, Da Nang 50000, Vietnam (email: vonhanvan@dtu.edu.vn).}
\thanks{Van Quan Pham is with the Nokia Bell Labs, Murray Hill, NJ 07974, USA (email: quan.pham$\_$van@nokia-bell-labs.com).}
\thanks{Hung Tran is with the college of Technology, National Economics University, Vietnam (email: hung.tran@neu.edu.vn).}
\thanks{Tu Ho Dac is with the the Department of Electrical Engineering of the Arctic University of Norway (UiT) (email: tu.d.ho@ntnu.no).}
\thanks{Nguyen Van Huynh is with the Department of Electrical Engineering and Electronics, University of Liverpool, Liverpool, L69 3GJ, United Kingdom (email: huynh.Nguyen@liverpool.ac.uk).}
}
\maketitle
\begin{abstract}
With the rapid development of technology, the number of smart mobile users is increasing, accompanied by growing demands from applications such as virtual/augmented reality (VR/XR), remote surgery, autonomous vehicles, and real-time holographic communications, all of which require high transmission rates and ultra-low latency in 6G and beyond networks (6G\plus{}). This poses enormous challenges in efficiently deploying large-scale networks, including network design, planning, troubleshooting, optimization, and maintenance, without affecting the user experience. Network Digital Twin (NDT) has emerged as a potential solution, enabling the creation of a virtual model that reflects the actual network, supporting the simulation of various network designs, applying diverse operating policies, and reproducing complex fault scenarios under real-world conditions. This motivate us for this study, where we provide a comprehensive survey of NDT in the context of 6G\plus{}, covering areas such as radio access networks (RAN), transport networks, 5G core networks and beyond (5GCORE\plus{}), cloud/edge computing, applications (blockchain, health system, manufacturing, security, and vehicular networks), non-terrestrial networks (NTNs), and quantum networks, from both academic and industrial perspectives. In particular, we are the first to provide an in-depth guide and usage of RAN and 5GCORE+ for NDT. Then, we provide an extensive review of foundation technologies such as transport networks, cloud/edge computing, applications, NTNs, and quantum networks in NDT. Finally, we discuss the key challenges, open issues, and future research directions for NDT in the context of 6G\plus{}.
\end{abstract}


\section{Introduction}
The rapid evolution of telecommunications has paved the way for the sixth generation (6G) wireless system, envisioned to provide terabit-per-second data rates, ultra-low latency, and massive device connectivity~\cite{jiang2021road}. To enable this vision, 6G networks are expected to operate across an expanded spectrum, including sub-THz and THz bands. Moreover, advanced communication technologies such as massive multiple-input multiple output (MIMO), intelligent reflecting surface, and open radio access network (ORAN) will be integrated into 6G networks for better communication performance~\cite{nguyen20216g}. Furthermore, non-terrestrial networks (NTNs), including satellites, high-altitude platforms, and unmanned aerial vehicles, are expected to be integrated into 6G networks to achieve ubiquitous global coverage. With its unique features and advantages, 6G can bring us to a new era of digital transformation by enabling transformative applications such as autonomous vehicles, smart cities, and immersive augmented reality experiences. However, realizing these ambitious goals comes with significant technical challenges that must be addressed to ensure the robustness and efficiency of 6G networks.

In particular, 6G networks are envisioned to be highly autonomous with abilities of self-optimization, self-healing, and self-configuration~\cite{khan2022digital}. Advanced artificial intelligence (AI) techniques can be used for dynamically adapting to new network conditions and user demands. Unfortunately, AI is notorious for its long training time and massive data requirement, which may not feasible for real-time and mission-critical applications in 6G. Moreover, to provide high-rate and ultra reliable connections to a massive number of diverse users and services, 6G is expected to be proactive with online learning capability~\cite{khan2022digital}. This is to make sure that 6G networks can quickly capture and respond to new system conditions and users' behaviors, ensuring their key performance indicators for different types of services, especially under heterogeneous architectures of future communication systems.

To effectively address these challenges that conventional approaches are likely to be infeasible, digital twins (DTs) are recently emerging as a key enabler for 6G networks. The concept of DTs was originally introduced by Grieves in 2003~\cite{tao2024wireless, grieves2017digital} with the focus on the manufacturing industry~\cite{mihai2022digital}. The official term, ``digital twin'', was then introduced by John Vickers at NASA in 2010. Since then, DTs have received great attention from both industry and academia with applications in various domains from healthcare, transportation to smart cities and robotics~\cite{mihai2022digital}, especially since the creation of deep neural networks and advanced AI technologies. In general, a DT is a sophisticated virtual replica of a physical system that is continuously updated with real-time data from its physical counterpart to enable real-time monitoring, simulation, and predictive analytics. In addition, DTs are designed to have a two-way flow of information where they not only collect real-time data from physical systems but also can share any insights/actions with these physical systems corresponding to specific tasks. In addition, a DT can run any number of simulations to study/monitor multiple processes, while a traditional simulation typically studies one particular process~\cite{IBM_2025}. This underscores a fundamental distinction between simulators and DTs. Traditional simulators, long utilized across various layers of telecommunications networks such as the physical, MAC, network, and application layers are generally designed for static or scenario-specific analyses. They provide valuable insights with manageable complexity by modeling specific network conditions or behaviors, often operating offline without real-time feedback. Conversely, a DT extends these capabilities by integrating real-time data from the physical system into the simulation loop, enabling continuous synchronization, dynamic model updates, and proactive decision-making. While simulators offer snapshots under predefined assumptions, DTs evolve alongside their physical counterparts, facilitating predictive analysis, anomaly detection, and automated control at scale. Beyond the scope of conventional simulators, DTs possess the ability to simultaneously monitor and influence multiple interconnected components or layers, reflecting the holistic and adaptive nature of modern telecommunications systems. 

In the context of 6G, DTs can serve as a real-time and data-driven mirror of the physical network. With advanced AI techniques, DTs can simulate, predict, and make decisions in a risk-free virtual environment before deploying them in the physical network. This is particularly useful in 6G to significantly mitigate network problems that may affect a massive number of users and services, thus further enhancing its ultra-reliable capacity. In addition, by simulating future network conditions and reconfiguring resources in advance, DTs can help to prevent potential network failures and congestion, resulting in reduced response time and high robustness of the system. As mentioned, 6G is envisioned to integrate multiple technologies and also unifies terrestrial, aerial, and space-based networks. Coordinating across this complex and heterogeneous architecture is difficult without a centralized and synchronized view. DTs can cope with these challenges by modeling and predicting the behavior of each domain/entity to enable seamless orchestration and joint optimization. Finally, DTs can be used as a framework to test and evaluate new protocols, applications, and architectures in a virtual space.

Given the above advantages, DTs have been widely adopted to address various problems in 6G and beyond networks \cite{IBM_2025,tao2024wireless, grieves2017digital, 10742580}. This survey aims to provide a comprehensive review of the applications of DTs in the key enablers of 6G, including RAN, transport networks, 5G+, cloud/ edge computing, applications (blockchain, health system, manufacturing, security, vehicular networks), NTN, and quantum networking, . By systematically examining the implementations and applications of DTs in these domains, we then provide a holistic understanding of how DTs will shape the future of wireless communications, along with their existing design challenges and future research directions. 

\begin{figure}[!]
	\centering
	\includegraphics[scale=0.5]{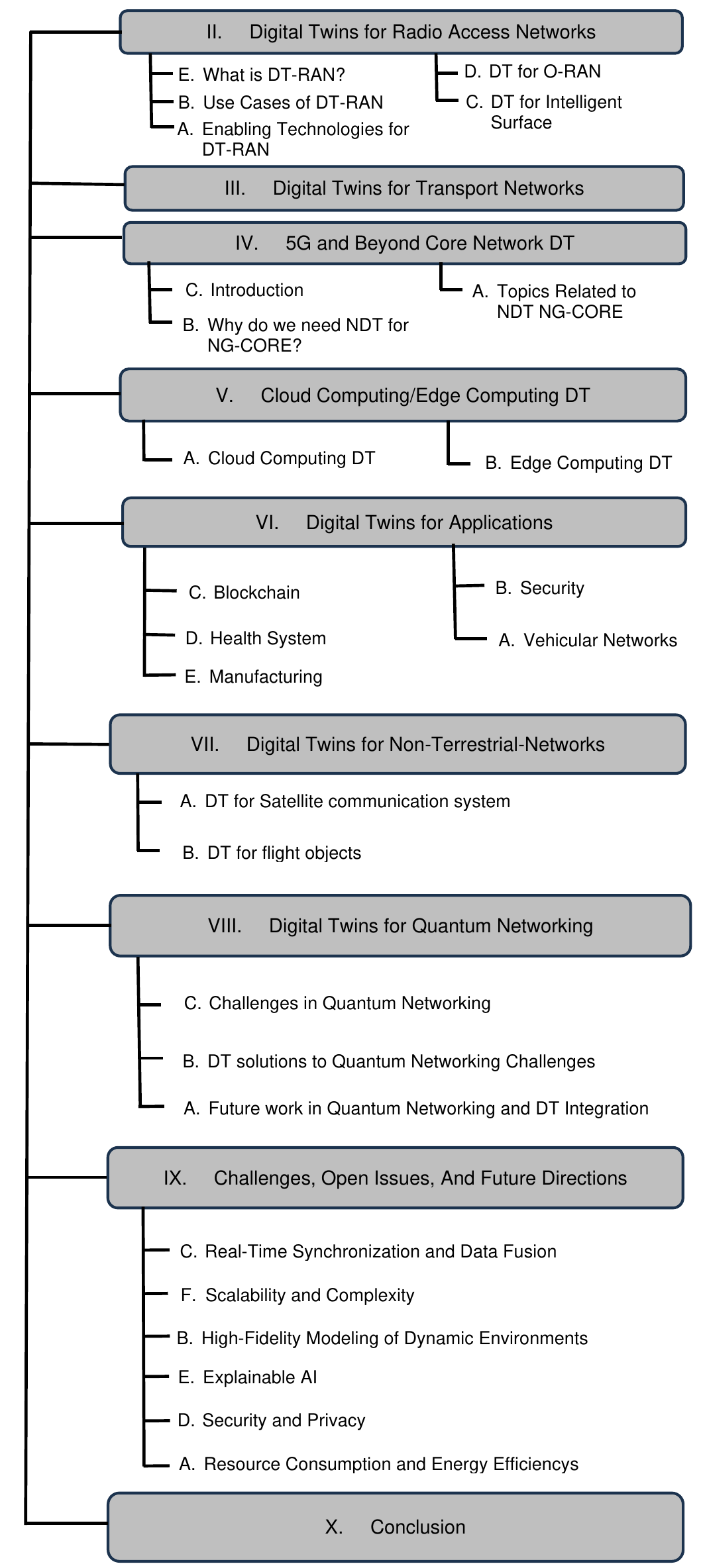}
	\caption{The overall structure of this paper.}
	\label{fig:paper_structure}
\end{figure} 

\begin{figure*}[!]
	\centering
	\includegraphics[scale=0.5]{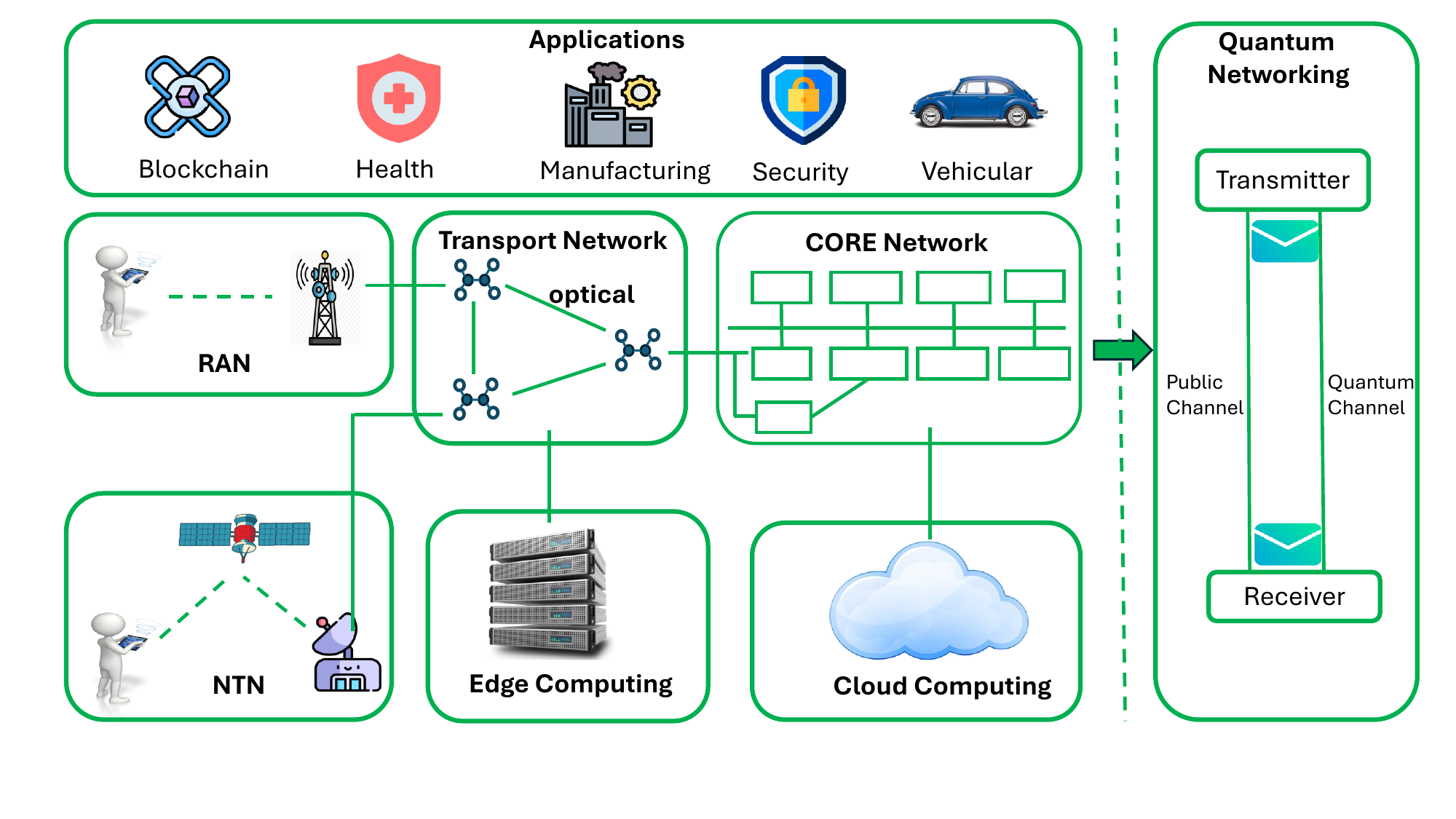}
	\caption{End-to-End Perspective of Network Digital Twin Architecture.}
	\label{fig:Illustration_structure}
\end{figure*} 

It is worth noting that there are a few surveys in the literature focusing on applications of DTs. For example, the authors in~\cite{wang2023survey} provide a review on the Internet of DTs with a focus on distributed architectures, enabling technologies, and security/privacy issues. Similarly, the authors in~\cite{mihai2022digital} present a comprehensive survey on key enabling technologies of DTs with three use cases in smart factory and healthcare. Instead of focusing on DTs for multiple domains, the authors in~\cite{tang2022survey} solely focus on the applications of DTs in 6G edge networks. The authors in~\cite{kuruvatti2022empowering} also survey key issues in the DT deployment for 6G and highlight some key applications of DTs such as AI model training, radio resource management, and security. Different from these surveys and others in the literature, our paper will provide a systematic and comprehensive review of the applications of DTs in key enablers of 6G, including RAN, ORAN, 5GCORE+, transport networks, edge/cloud computing, applications (blockchain, health system, manufacturing, security, and vehicular networks), NTN, and quantum networks. Moreover, this paper will provide insights and up-to-date implementations of DTs in industry with various use cases and standards. To the best of our knowledge, this is the first survey that comprehensively review a wide range of DT applications in various domains of 6G, both from academia and industry viewpoints. For comparison, we provide table \ref{tab:dt-surveys} to provide more details. 

\begin{table*}[ht]
\centering
\caption{Comparison of Survey Papers on Digital Twin Applications}
\begin{tabular}{|p{1.5cm}|p{4cm}|p{5cm}|p{5cm}|}
\hline
\textbf{Reference} & \textbf{Focus Area} & \textbf{Key Topics} & \textbf{Use Cases / Highlights} \\
\hline
\cite{wang2023survey} & Internet of Digital Twins (IoDT) & Distributed architectures, enabling technologies, security and privacy & General IoDT applications across domains \\
\hline
\cite{mihai2022digital} & Key enabling technologies of DTs & Enabling technologies, challenges, trends, future prospects & Use cases in smart factory and healthcare \\
\hline
\cite{tang2022survey} & DTs in 6G edge networks & Edge computing, DT-based communication systems & Sole focus on DTs in 6G edge networks \\
\hline
\cite{kuruvatti2022empowering} & DTs for 6G systems & DT deployment, key challenges, integration in 6G & AI model training, radio resource management, security \\
\hline
\textbf{Our Work} & DTs in 6G and beyond key enablers & RAN, ORAN, 5GCORE+, transport networks, edge/cloud computing, applications (blockchain, health system, manufacturing, security, and vehicular networks), NTN, and quantum networks& Comprehensive review from academia and industry, recent implementations, use cases \\
\hline
\end{tabular}
\label{tab:dt-surveys}
\end{table*}

To provide a comprehensive view of the scope and connectivity of NDTs, Figure \ref{fig:Illustration_structure} presents an end-to-end architecture encompassing all the major domains of 6G and beyond communication systems. The foundation begins with Digital Twins for Radio Access Networks (RAN), which virtualize the behavior, structure, and performance of terrestrial base stations and user equipment. The concept of DT-RAN includes an overview of what it is and explores key use cases such as predictive maintenance, fault management, and performance optimization. The architecture then evolves into more advanced paradigms such as Open RAN (O-RAN), where digital twins enhance modularity and interoperability among disaggregated components. Additionally, Digital Twins for Intelligent Surfaces represent a futuristic extension of RAN, allowing dynamic modeling and control of reconfigurable intelligent surfaces (RIS) to improve wireless coverage and energy efficiency.

Next, DTs for Transport Networks address the backbone of the communications system, providing ultra-high throughput and low-latency links between edge nodes, the core network, and data centers. DTs in this domain facilitate real-time traffic monitoring, fault localization, and resource optimization, ensuring stable and efficient transport performance under diverse traffic conditions. Moving upward, DTs for the 5G and Beyond Core Network (5GCORE+) form the core system of the network, responsible for tasks such as routing, mobility management, and network slicing. An introduction to this section outlines the relevance of DTs in modeling both logical and physical elements of the core, enabling fine-grained control and orchestration. The need for NDT in 5GCORE+ arises from the increasing complexity of service delivery, where predictive analytics and dynamic reconfiguration can significantly enhance reliability and performance. Topics in this area range from control-plane optimization to intelligent slice management and policy enforcement. 

Supporting both RAN and core layers are Cloud and Edge Computing DTs, which reflect the distributed computing infrastructure underpinning modern networks. DTs for cloud computing model centralized resources for long-term data storage and compute-heavy tasks, while edge computing DTs handle low-latency and proximity-based services near the end-user. These DTs enable dynamic resource provisioning, workload orchestration, and energy-aware operation, contributing to a flexible and scalable network architecture. 

At the application layer, DTs for Applications span a diverse set of verticals including blockchain, healthcare systems, manufacturing processes, security frameworks, and vehicular networks. These application-specific DTs play a pivotal role in aligning network behavior with service-level expectations and operational constraints. 

Beyond terrestrial networks, DTs for Non-Terrestrial Networks (NTN) extend coverage through satellite communication systems and airborne platforms. DTs for satellite communication systems support global monitoring, trajectory planning, and link adaptation, while DTs for flight objects such as UAVs enable efficient path planning and coordination within integrated air-ground networks. NTN acts as a crucial complement to terrestrial systems, ensuring continuous coverage in remote or disaster-struck areas.


Finally, DTs for Quantum Networking represent a forward-looking domain where both classical and quantum communication channels co-exist. Challenges in quantum networking, such as entanglement degradation and channel noise, are addressed through simulation and analysis using DTs. These DTs assist in designing robust solutions for quantum key distribution (QKD) and evaluating the performance of quantum channels under different physical conditions. Future directions in this area explore the integration of quantum networking with DT-based control for ultra-secure and intelligent communications. The survey concludes with a discussion on Challenges, Open Issues, and Future Directions, emphasizing the need for standardized frameworks, scalable architectures, and privacy-preserving mechanisms across all domains. Together, this layered and interconnected digital twin architecture forms the foundation of a unified, intelligent, and resilient NDT ecosystem for 6G and beyond.

As illustrated in Fig.~\ref{fig:paper_structure}, the rest of this paper is organized as follows. Section~\ref{sec:RAN}, we present an overview of RAN then O-RAN and how DTs can help to improve the performance of O-RAN as well as ensure its successful deployment in 6G. Then, in Section~\ref{sec:optical} and Section~\ref{sec:5gcore} we highlight the applications of DT in transport networks and applications DT in 5G and beyond core networks. In Section~\ref{sec:cloud} summaries the applications of DTs in edge and cloud computing.  After that, section \ref{sec:aiml} presents various DT for radio access networks in a number of domains such as blockchain, healthcare, manufacturing, security and vehicular networks. In Section~\ref{sec:ntn}, we present our insights in applications of DT in non-terrestrial networks. In addition, the applications of DTs for quantum networking are presented in Section~\ref{sec:quantum}. Then, we highlight the existing challenges of DTs in 6G and present various research directions in Section~\ref{sec:challenge}. Finally, Section~\ref{sec:Summary} concludes our paper.

\section{Digital Twins for Radio Access Networks}\label{sec:RAN}
A DT for a radio access network (RAN), i.e., DT-RAN, is a specialized network digital twin focused on the RAN domain, i.e., the base stations (BSs) and user equipments (UEs) that deliver wireless connectivity in a certain environment \cite{lin20236g}. In this section, we introduce DT-RAN concept, discuss the use cases of DT-RAN, and describe the enabling technologies for DT-RAN.

\subsection{What Is DT-RAN?}

A DT-RAN is a detailed, virtual replica of the corresponding physical RAN and its environment, integrating ray tracing, channel emulation, antenna modeling, mobility patterns, algorithms, and protocols, among others \cite{itur2022future}. It is usually built as a layered, software-defined platform that mirrors both the physical RAN infrastructure and its operational context in as close to real-time as possible \cite{itut2022digital}. At its foundation lies the physical twin and data-acquisition layer, where network nodes (e.g., central units (CUs), distributed units (DUs), radio units (RUs), or monolithic BSs), UEs, and communication links (e.g., access, fronthaul, midhaul, or backhaul link) are instrumented to continuously stream metrics as appropriate. The instrumentation may be carried out either via embedded internet of things (IoT) sensors or native telemetry interfaces. Example metrics include signal strengths, interference levels, hardware health indicators (e.g., temperature and power), and traffic counters. 

A high-throughput data-ingestion pipeline then funnels the heterogeneous streams into the twin with minimal latency, ensuring that the virtual model maintains tight synchronization with the live network state \cite{itut2022digital}. Once ingested, the raw telemetry is funneled into a consolidated data platform and repository. Time-series databases or data lakes cleanse, align, and persist both real-time and historical data. Schema registries and metadata services enforce consistency and auditability across geographically distributed sites \cite{oran2024digital}. The centralized data store not only underpins real-time key performance indicator (KPI) monitoring but also enables retrospective analysis, which is critical for model training, anomaly investigations, and ``what-if'' scenario replay.

Above the data layer sits the modeling and simulation engine, which integrates detailed three-dimensional (3D) environment models (e.g., geographic information system (GIS) maps, building outlines, and material properties) with physics-based channel models (e.g., ray tracing for multipath emulation) and virtualized RAN functions \cite{hoydis2024learning, ruah2024calibrating}. By simulating user mobility, antenna beamforming patterns, scheduling algorithms, interference dynamics, etc., this modeling and simulation engine can accurately predict coverage, capacity, and quality-of-experience (QoE) under hypothetical deployment scenarios. The high-fidelity simulations form the sandbox in which network planners and artificial intelligence (AI) agents can evaluate parameter changes risk-free before any real-world rollout.

Running on top of the simulation layer, an analytics tier can leverage machine learning (ML) models, which may be trained on both live and synthetic datasets, to perform tasks such as anomaly detection, traffic forecasting, and closed-loop optimization of RAN controls (e.g., adaptive antenna tilt or dynamic radio resource management). Specifically, AI agents can propose configuration adjustments within the DT-RAN, where outcomes are assessed against defined KPIs and service level agreements (SLAs) \cite{ren2023end}. Only those changes that demonstrate safety and performance gains are automatically validated for deployment back to the physical RAN, completing a fully automated feedback loop that minimizes manual intervention and risk. In addition, orchestration and control can be handled by a service management layer. This layer may be implemented via Kubernetes or network functions virtualization management and orchestration (NFV MANO) that governs the lifecycle, scaling, and multi-tenant isolation of the DT-RAN's microservices \cite{3gpp2025study, 3gpp2025management}. As the bridge between the DT-RAN and its corresponding physical RAN, standardized network-bound interfaces enable seamless telemetry ingestion and the safe application of validated configurations back into the physical RAN. This integration ensures that the DT-RAN not only reflects the network but also acts as a policy-compliant control plane for the live physical system.

Finally, users can interact with the DT-RAN through rich visualization dashboards that render two-dimensional (2D) or 3D coverage maps, heat maps of performance metrics, and predictive alerts \cite{oran2024digital}. Interactive tools let the users drill down to site-level details, replay historical scenarios, or launch automated optimization workflows conveniently.

\subsection{Use Cases of DT-RAN}

One of the key use cases for a DT-RAN is in enhanced network planning and deployment \cite{akgun2024advancing}. By creating a virtual replica of individual cell sites, including equipment layouts, cabling, and site geography, operators can simulate rollout and upgrade processes without physically modifying hardware. Extending this to multiple sites within a geographic region, a DT-RAN can merge real-world measurements and physics-based propagation models to predict coverage, capacity, and interference patterns under various ``what-if'' scenarios. This enables more informed site selection, antenna configuration, and spectrum planning, ultimately reducing capital expenditure (CapEx) and accelerating time-to-market for new deployments. Capacity planning and congestion management can also be improved through DT-RANs \cite{thomas2024digital}. Operators can simulate traffic surges (e.g., from large events, emergency scenarios, or seasonal peaks) and evaluate strategies like traffic steering, dynamic spectrum sharing, and priority queuing within the DT-RANs. These preemptive trials help to identify potential bottlenecks and validate mitigation measures, ensuring that the live network maintains satisfactory QoE even under stress.

Beyond network planning, DT-RANs can be used for closed-loop network optimization powered by AI/ML. In practice, real-time telemetry streams feed a high-fidelity simulation engine where algorithms for beamforming, mobility management, and dynamic load balancing can be tested at scale \cite{gao2023digital}. Research prototypes have demonstrated that neural receivers and learned beamforming models tuned within a DT-RAN outperform traditional approaches when applied in the corresponding physical RAN \cite{wiesmayr2024design, salehi2024multiverse}. With detailed modeling of cell-sleep modes, dynamic power scaling, and antenna pattern adjustments, DT-RANs can enhance network energy efficiency \cite{oran2024research}. For instance, massive multiple-input multiple-output (MIMO) arrays can be partially deactivated or radio frequency (RF) circuits can be shut off when traffic is low, guided by historical and simulated load profiles in DT-RANs. Furthermore, operators can leverage DT-RANs to generate synthetic training data for AI/ML models in the early phases of rollout and then iterate continuously as the network evolves \cite{akgun2024advancing}. This is motivated by the fact that creating robust training datasets from live-network telemetry is often hampered by data sparsity in rare or edge-condition scenarios. To fill these gaps, a DT-RAN can generate synthetic yet realistic channel and mobility traces which may be further augmented by generative AI models. The synthetic datasets allow developers to train, test, and fine-tune AI/ML models before any code reaches production. This sandboxed approach reduces risk and accelerates model maturity.

Predictive maintenance is another critical DT-RAN use case \cite{lin20236g}. By ingesting sensor data (e.g., temperature, vibration, and power draw) from RAN elements into the DT-RANs, operators can apply anomaly detection analytics to forecast component failures days or weeks in advance. This allows maintenance teams to schedule proactive interventions during low-impact windows, minimizing unplanned downtime and reducing repair costs. Such predictive insights can extend equipment lifespan and drive operating expense (OpEx) savings. Resilience and security testing is likewise enhanced by DT-RANs \cite{gao2023digital}. Operators can simulate equipment failures, backhaul disruptions, or even cyber-attack scenarios in a controlled environment, evaluating failover strategies and hardening measures without affecting subscribers. These simulations refine contingency plans and improve incident response playbooks, ensuring that the real network can sustain service continuity under adverse conditions.

Service assurance and network slicing are other domains where DT-RANs can add value. By emulating individual slices in DT-RANs, operators can validate SLAs through testing isolation, bandwidth guarantees, and latency targets under varying load patterns. The ability to emulate slice behavior ensures that new slice configurations meet performance criteria before they reach live infrastructure, reducing the risk of SLA violations and costly penalties \cite{ren2023end}. In addition, DT-RANs facilitate deeper integration with industrial and enterprise networks. In Industry 4.0 environments, DTs of both the factory floor and the RAN can interoperate, enabling use cases such as real-time robotics control, predictive asset maintenance, and automated logistics \cite{groshev2021toward}.

\subsection{Enabling Technologies for DT-RAN}

DT-RANs depend on a wide ecosystem of data, modeling, and computation technologies that together enable a live, high-fidelity ``mirror'' of the physical RAN.

\subsubsection{Data acquisition and management}

Data acquisition for a DT-RAN is the foundational enabler that bridges the physical RAN with its virtual counterpart \cite{oran2024digital}. It encompasses not only the breadth of telemetry sources but also the precision of instrumentation, the fidelity of timestamping and geolocation, and the robustness of data pipelines that deliver raw measurements to the twin engine in a timely manner \cite{itut2022digital}. The DT-RAN can ingest both control plane and user plane telemetry, as well as environmental context. Control plane metrics include signaling counters (e.g., attach/detach, handover requests, paging interactions) and protocol-stack KPIs (e.g., radio resource control (RRC) reconfigurations, radio link failures). They are typically exposed by the baseband unit or by minimization of drive test (MDT) reports from UEs, which can periodically upload location-tagged measurement logs. User plane metrics includes throughput, latency, packet-loss statistics, and jitter measurements, which may come from deep packet inspection probes or accessible counters in the CU/DU/RU stack. For environmental context, temperature, humidity, and vibration sensors on towers and at sites feed into the DT-RAN to help correlate thermal throttling or hardware degradation with performance dips. By aggregating this heterogeneous data, the DT-RAN can mirror not just the logical state of the network but also its environmental context.

Data management turns streams of raw telemetry and logs into a reliable, searchable, and secure foundation for DT-RAN functions. Data management  must address five intertwined aspects: ingestion, storage, cataloging, accessibility, and governance \cite{itut2022digital}. Raw measurements can be ingested via a distributed streaming backbone. Edge preprocessing entities can perform lightweight aggregation and anomaly filtering before forwarding, reducing bandwidth while preserving fidelity for downstream consumers. For real-time dashboarding and control loops, high-performance time-series databases can index metrics by timestamp and cell identity, enabling fast queries and aggregations. Meanwhile, a central data lake can archive raw and preprocessed telemetry, protocol traces, and batch simulation outputs. As the RAN evolves, data schemas shift. A federated metadata catalog can track schema versions, source provenance, and data lineage so twin-engine components can adapt to changes. Futhermore, There are numerous RAN-NDT applications, each with distinct feature requirements that may vary significantly in terms of KPIs, data types, time granularity, and other characteristics. To fully leverage available data and system resources, the data collection framework must be more intelligent and programmable. For instance, one such platform \cite{10437430} proposed a programmable telemetry system tailored for next-generation open RANs. Together, these components ensure data is properly captured, stored and served for DT-RAN applications.

\subsubsection{Physics-based simulation and data-driven AI/ML models}

Physics-based simulators form the ``first principles'' core of DT-RANs, providing an accurate, physics-grounded view of how electromagnetic waves propagate, interact with the environment, and ultimately deliver connectivity to UEs \cite{li2024digitalpart1, li2024digitalpart2}. By explicitly modeling the physical laws that govern electromagnetic wave behavior, these simulators enable DT-RAN usage with high confidence. At the heart of physics-based simulation is ray tracing \cite{hoydis2024learning, ruah2024calibrating}. Deterministic ray tracing engines ingest detailed 3D environment models and launch virtual rays from each antenna element. Each ray is tracked as it reflects, refracts, diffracts, or penetrates materials, accumulating path loss, delay spread, and angular spread data. This level of geometric fidelity is especially critical for higher-frequency bands, where small obstructions can result in deep shadows and multipath clusters become highly site-specific. The physical propagation models feed into system-level simulators, which overlay protocol stacks, scheduler logic, and mobility modeling \cite{pegurri2024toward}. The system-level simulators can then emulate many cells and hundreds or thousands of UEs to evaluate system-wide KPIs, such as throughput distributions, handover success rates, and interference heat maps.

Data-driven AI/ML models bring the DT-RAN beyond first-principles physics, allowing it to learn and adapt to real-world operational nuances, detect subtle anomalies, and suggest optimal control actions \cite{nguyen2021digital}. High-fidelity physics simulators may be slow for large ``what-if'' sweeps. Surrogate models can compress these simulators into lightweight approximators by using, for example, neural networks \cite{chen20215g}. A RAN naturally forms a graph: sites and cells are nodes, and edges represent interference coupling or handover adjacency. Graph neural networks (GNNs) can exploit this structure to infer per-cell performance or to recommend parameter adjustments that optimize network-wide objectives \cite{zhang2023gnn}. Model-based or model-free reinforcement learning agents trained in the twin environment can discover dynamic resource management policies that adapt to time-varying traffic and channel conditions \cite{liu2024coverage}. Beyond prediction and control, AI/ML models enable continuous monitoring and fault detection. Generative adversarial networks (GANs) can synthesize rare failure modes, helping twin operators to stress test recovery procedures under corner case conditions \cite{huang2024digital}.

\subsubsection{Accelerated computing}

As high-fidelity virtual replicas of physical RANs and their environments, DT-RANs are compute-intensive. Accelerated computing, e.g., the use of graphics processing units (GPUs), plays critical roles in making DT-RANs practical and performant \cite{oran2024digital}. It enables high-performance physics-based simulations. In particular, accurate modeling of radio wave propagation requires solving large-scale electromagnetic problems. Accelerated computing reduces runtimes for ray tracing to support large-scale scenarios. Simulating RAN protocol stacks, scheduling algorithms, and multi-cell interference in one environment further demands  accelerated computing to enable rapid design and testing of wireless algorithms. For live bidirectional synchronization between a physical RAN and its twin, latency budgets are tight. Accelerated computing is essential for supporting real-time emulation and hardware-in-the-loop \cite{villa2024colosseum}.

Besides, training of the data-driven AI/ML models for DT-RANs necessitates massive parallel computation \cite{oran2024digital}. Accelerated computing platforms allow tight integration of training with the running DT-RANs, reducing turnaround times for model updates. Within the running DT-RANs, AI/ML models should infer with low latency for timely adaptation in the physical RANs. Furthermore, DT-RAN workflows ingest massive volumes of RAN telemetry that must be cleansed, transformed, and feature-engineered before model consumption. Accelerated libraries and big-data frameworks streamline extract, transform, load (ETL) operations on the data lake \cite{lin20236g}. Lastly, accelerated computing facilitates distributed, scalable compute architectures. DT-RAN  workflows can span edge devices, on-prem servers, and public-cloud instances \cite{chen2024distributed}. Accelerated computing hardware, from laptop-scale GPUs for early research and development to multi-GPU servers and cloud-based instances, provides a edge-to-cloud continuum for DT-RANs.

\subsection{Digital Twin for O-RAN }\label{sec:O-RAN}

Despite the continuous development of wireless technology from 1G to 5G and the transition towards 6G, we still face new challenges due to the rapid growth in the number of mobile devices (e.g., 25 billion IoT devices by 2025 \cite{TranTWC22,PhuIoT21}) and the increasing demand for traffic driven by VR/AR, $360^\circ$ video, holographic video applications, and low latency requirements \cite{TranTVT20}. Consequently, it is essential to develop infrastructure and continuously adapt the architecture to meet these urgent demands. More specifically, vendors and operators have transitioned from proprietary components in Distributed Radio Access Network (DRAN) to Open RAN, which offers many new opportunities and advantages. These include interoperability between vendors, reduced operating costs, intelligent and automated platforms, and enhanced programmability. However, implementing O-RAN in practice presents many challenges. One significant challenge is managing O-RAN automatically and in real-time, rather than through offline network management, amidst a large number of mobile users, mobile stations, and dynamic traffic demands while still ensuring Quality of Service (QoS) requirements such as data rates and latency. Fortunately, the advent of digital twin technology offers an extremely effective solution for O-RAN to address these difficulties. In this section, we will provide researchers with a comprehensive overview of the evolution of network architectures from DRAN to O-RAN and explain why O-RAN is essential for 6G and beyond, i.e., subsection \ref{O-RANEvo}. We will then analyze the need for Digital Twin technology in O-RAN, i.e., subsection \ref{NDTO-RAN}, and explore use cases of Network Digital Twin (NDT) for O-RAN, i.e., subsection \ref{NDTO-RANUsecases}. These insights aim to give researchers a broad and easiest understanding of NDT for O-RAN from both academic and industry perspectives, thereby opening up new and valuable research directions.

\subsubsection{Evolution of O-RAN}
\label{O-RANEvo}
To give you a more general perspective on Open Radio Access Network (O-RAN) and why we need O-RAN for 6G and beyond, I would like to briefly describe RAN systems in the past and their development process.

a. {Distributed Radio Acess Network (DRAN):}

\begin{figure*}[t]
	\centering
	\includegraphics[height=12cm,width = 18cm]{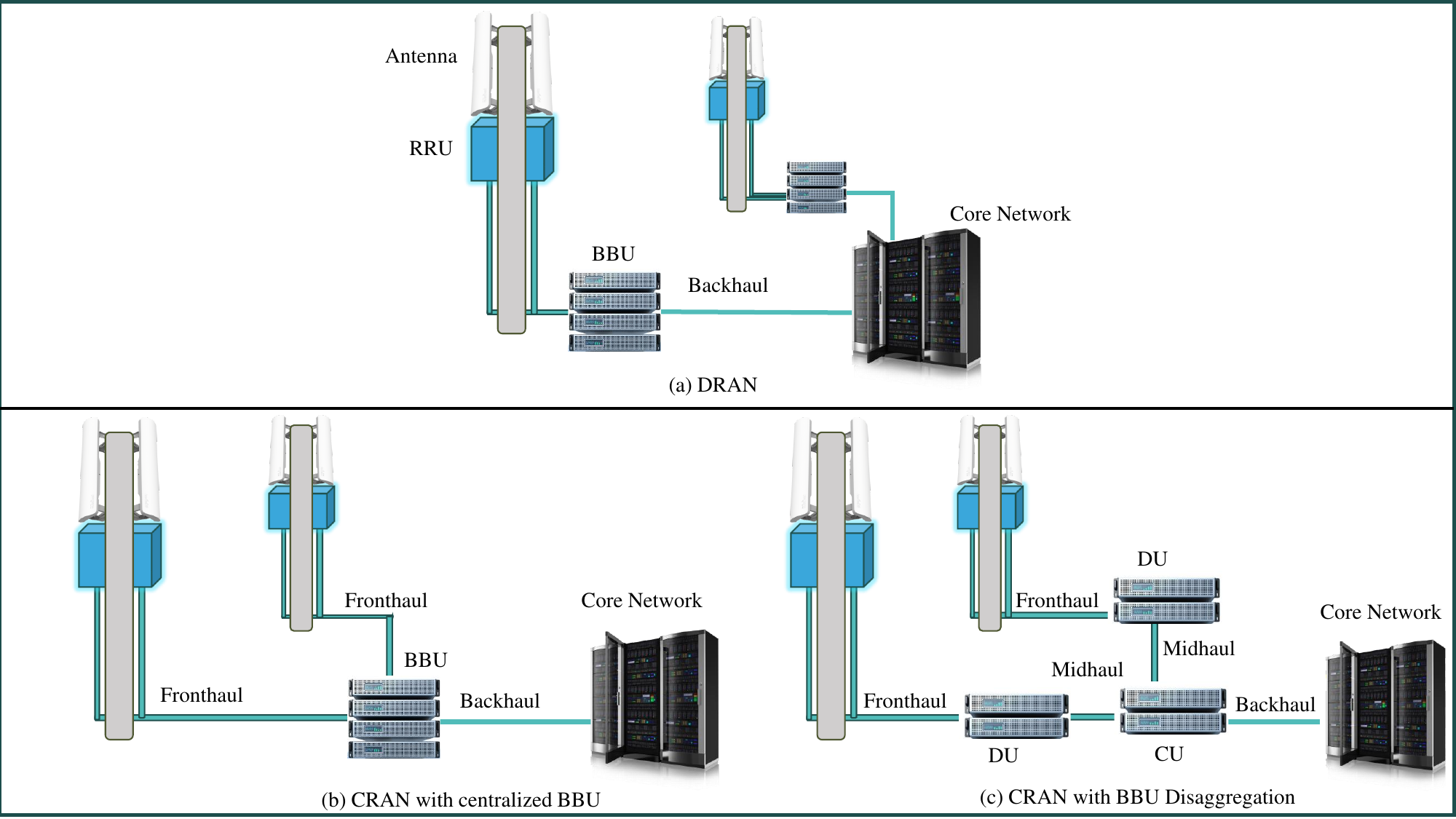}
	\caption{RAN and CRAN architectures.}
	\label{FigRANCRAN}
\end{figure*}

Firstly, we talk about RAN or distributed-RAN systems which is the earliest one. DRAN is a traditional setup where the Remote Radio Unit (RRU) and the Baseband Unit (BBU) are co-located at each mobile station. The RRU, BBU, and antenna are represented in Fig. \ref{FigRANCRAN} (a), and their functions are explained as follows. The RRU is the radio frequency (RF) processing unit responsible for transmitting and receiving radio signals. It is usually mounted near the antenna at the top of a mobile station (MS), RRU can handle the reception, transmission, filtering, and amplification of RF signals. The BBU is used to manage the whole MS, including signal processing, operating, maintenance, allocate resources while coordinating with other network elements to ensure seamless connectivity and good performance. The antenna is used to transmit and receive the RF signals. RRU, BBU, and antenna operate proprietary applications on dedicated hardware, with all functions situated in a single site location. These radio sites connect back to the core network (CN) via backhaul.

b. {Centralized Radio Access Network (CRAN):}

The next generation of RAN is called Centralized RAN (CRAN), an enhanced version of Distributed RAN (D-RAN). As shown in Fig. \ref{FigRANCRAN} (b) and (c), the BBU is not co-located with the RRU. Instead, the BBU connects to the RRU through a fronthaul interface. The new C-RAN architecture introduces several advantages and disadvantages as follows:

\textit{Advantages:}
\begin{itemize}
	\item The BBU can serve multiple MSs compared to the D-RAN, which only serves one MS.
	\item Computing resources such as CPU and RAM are shared, improving resource utilization and reducing energy consumption.
	\item Operational costs, such as rental costs for the BBU, can be reduced.
	\item Better inter-cell cooperation can be achieved due to low latency between mobile stations.
\end{itemize}

\textit{Disadvantages:}
\begin{itemize}
	\item It is still vendor-proprietary, as in D-RAN. 
	\item High data traffic from the RRU to the BBU on the fronthaul link can cause traffic congestion. This issue is resolved by disaggregating the BBU into a Distributed Unit (DU) and a Central Unit (CU). The DU supports Layer 1 (L1), such as the physical layer, and the lower Layer 2 (L2), including the medium access control (MAC) and radio link control (RLC) protocols.
\end{itemize}

c. {Virtualized Radio Access Network (VRAN):}

\begin{figure*}[t]
	\centering
	\includegraphics[height=12cm,width = 18cm]{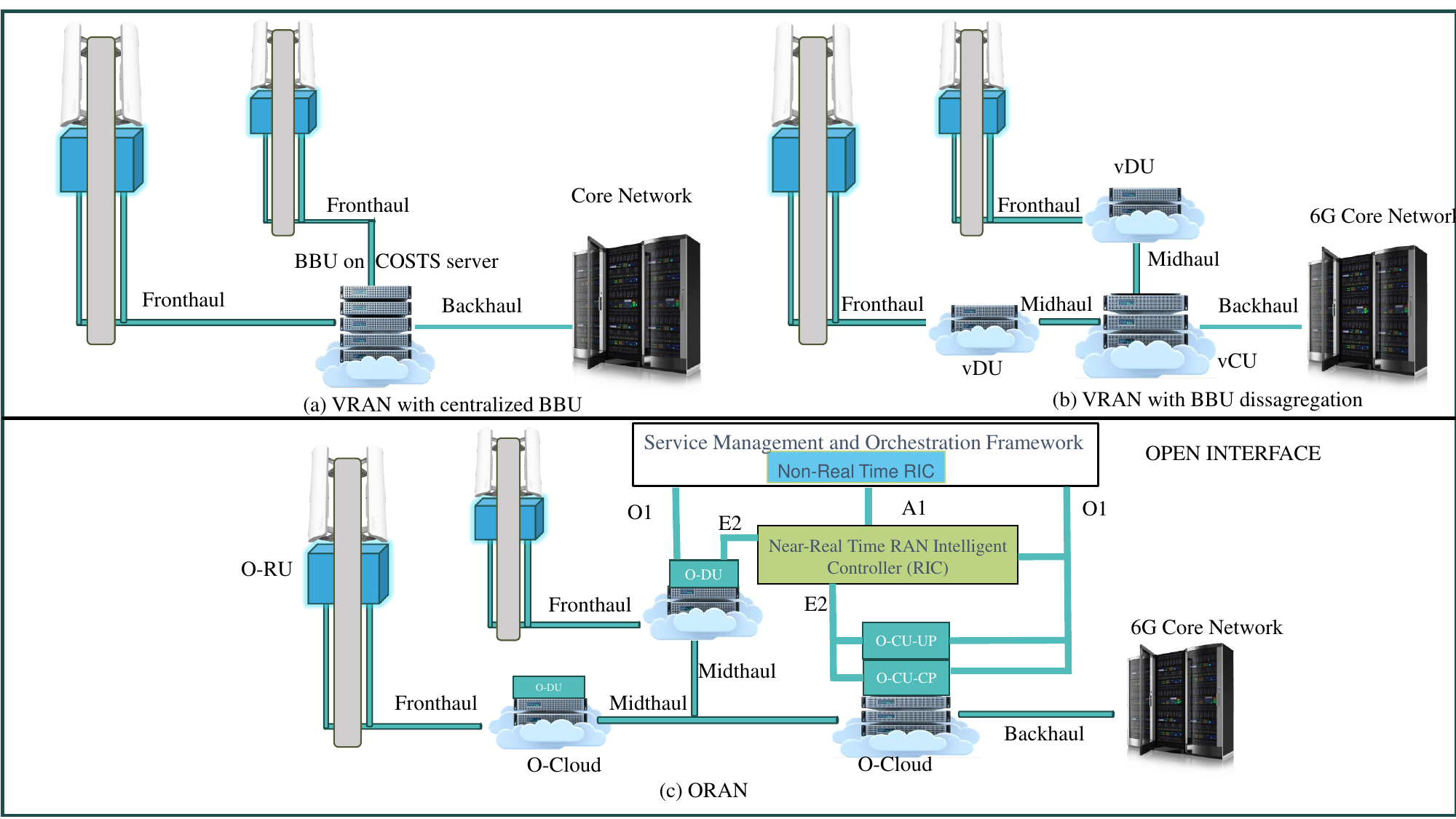}
	\caption{VRAN and O-RAN architectures.}
	\label{FigVRANORAN}
\end{figure*}

As shown in Fig. \ref{FigVRANORAN} (a) and (b), the primary difference between CRAN and VRAN is that VRAN does not use proprietary hardware like CRAN. In VRAN, the CU and DU components of CRAN are virtualized and hosted on cloud/commercial off-the-shelf (COTS) servers. Although the DU and CU are depicted in DRAN, they do not utilize open interfaces and remain vendor-proprietary, requiring the purchase of RRU, virtualized DU (vDU), and virtualized CU (vCU) from the same vendor. 

\subsubsection{Open Radio Access Network (O-RAN)}
It's evident from the discussion above that a new architecture is required to overcome DRAN, CRAN, and VRAN's shortcomings. As stated in \cite{alliance2023ORAN} Since there were few RAN suppliers available, it was difficult for new vendors to come up with creative solutions. The monolithic and proprietary nature of RAN products hindered the introduction of new features. Proprietary interfaces and licensing restrictions made it difficult for operators and independent software developers to gain control over RAN equipment and develop new features. This resulted in the development of O-RAN. The Open RAN was created in 2018 by the O-RAN Alliance. This alliance was formed by a group of major telecommunications operators, including AT$\&$T, NTT DOCOMO, China Mobile, Deutsche Telekom,  and O-RANge, with the goal of promoting open and interoperable radio access network solutions. O-RAN Alliance which is a group of mobile operators, vendors, and research institutes. Its goal is to develop open, virtualized, intelligent, and interoperable RAN systems. The network functions (NFs) and interfaces in O-RAN is described in Tables \ref{tab1} and \ref{tab2}, respectively. The benefits of O-RAN are be summarized as bellows:
\begin{itemize}
	\item \textit{Open Interfaces:} Besides 3GPP-based interfaces, O-RAN Alliance defines open internal interfaces between components in O-RAN. For example, the open Fronthaul interface between O-RU $\Leftrightarrow$ O-DU, open Midhaul between O-DU $\Leftrightarrow$ O-CU-UP/O-CU-CP, open Backhaul between core network $\Leftrightarrow$ O-CU-UP/O-CU-CP.
	\item \textit{Interoperability and reduced operating costs:} Ensuring that network components from different vendors work seamlessly together while still ensuring performance and connectivity requirements regardless of the site or environment. This property in O-RAN overcomes the limitations of previous RAN networks. Thus, it helps operators design their own network more flexibly, scaling more easily. This diverse supplier/vendor ecosystem not only leads to cost savings but also enhances network resiliency. This diverse vendor ecosystem not only leads to cost savings but also enhances network resiliency. By incorporating equipment from various suppliers, operators can mitigate the risks associated with single-vendor/single-supplier dependency and better protect their networks against potential vulnerabilities or service disruptions.
	\item \textit{Disaggregation and cloudification:} The O-RAN architecture splits the BBU into a DU and CU, fostering innovation and enabling the virtualization of these components. This approach supports cost-effective deployment and management and provides a robust framework for Open RAN security in cloud-native networks.
	\item \textit{An intelligent and automation platform:} Telecommunication network operations are becoming increasingly complex due to diverse and demanding use cases, varied frequency selections, resource allocation, virtualization, and different slicing requirements through O-RAN. Therefore, there is a need for more efficient, rapid, and stable automated system deployment. Achieving zero-touch deployment is crucial for effectively managing a network comprised of thousands of small-cell MSs. The O-RAN SMO platform is an intelligent automation solution that scales automation to simplify network complexity, improve network performance (NetPer), enhance mobile users' (MUs)' experience, and minimize operational costs.
	\item \textit{Programmability:} The evolution of the O-RAN system with RAN Intelligent Controller (RIC) introduces programmability to the RAN, enhancing its flexibility and adaptability.
	\item \textit{Energy Efficiency (EE):} When designing an O-RAN system, EE must be taken into account. The following elements should be included in EE design for O-RAN: The first step in reducing network energy consumption (NetEC) is to choose energy-efficient hardware. Although the O-RU is the main energy user, the O-DU and O-CU should also be taken into account. Furthermore, monitoring and data provision for the development of future EE algorithms depend on the measurement and reporting of the NetEC. Thirdly, using this data to help build an automated and optimization framework to improve EE is possible. In particular, NetEC may be anticipated at each time period using sophisticated machine learning (ML) techniques, which allows for well-informed decision-making.
\end{itemize}

\begin{table*}[t]
	\caption{O-RAN Network Functions.}
	\centering
	\setlength{\tabcolsep}{5pt}
	\begin{tabular}{|>{\centering\arraybackslash} m{5cm} | >{\centering\arraybackslash} m{10cm}| >{\centering\arraybackslash} m{1.5cm}|}
		\hline 
		\textbf{Network Functions} & \textbf{Description} & \textbf{References} \\  \hline\hline
		Service Management and Orchestration Framework (SMO)&  Handle management functions including transport, slicing, and RAN and CORE management. & \cite{ORANArchitecture24}\\ \hline 
Non-Real Time RAN Intelligent Controller (Non-RT RIC) &  Enhances intelligent RAN optimization, managing ML models, and providing enrichment information to the Near-RT RIC. Additionally, it supports intelligent RAN optimization control loops with latency is bigger than or equal to 1 second.  & \cite{ORANArchitecture24}\\ \hline Near-Real Time RAN Intelligent Controller (Near-RT RIC)& Perform resource allocation and optimization actions with latency from 10ms-1s& \cite{ORANArchitecture24}\\ 	\hline O-RAN Central Unit - Control Plane (O-CU-CP)& Radio Resource Control (RRC) and control plane of Packet Data Convergence Protocol (PDCP) & \cite{ORANSlicingArchitecture, ORANArchitecture24}\\ \hline O-RAN Central Unit - User Plane (O-CU-UP)& User plane of Packet Data Convergence Protocol (PDCP) and Service Data Adaptation Protocol (SDAP) & \cite{ORANSlicingArchitecture, ORANArchitecture24}\\ \hline O-RAN Distributed Unit (O-DU)& Similar to gNB-DU as following 3GPP TS 38.401. It supports medium access control (MAC), physical layer, and RLC protocol. & \cite{ORANSlicingArchitecture,ORANArchitecture24}\\ \hline O-RAN Radio Unit (ORU)& Support RF processing, low physical layer, e.g., Fast Fourier Transform (FFT)/Inverse FFT (iFFT) & \cite{ORANSlicingArchitecture, ORANArchitecture24}\\ \hline O-RAN Cloud (O-Cloud)& A computing platform hosting O-RAN NFs such as O-DU, O-CU-CP, O-CU-UP, and Near-RT RIC. & \cite{ORANSlicingArchitecture,ORANArchitecture24}\\		\hline 
	\end{tabular}
	\label{tab1}
\end{table*}

\begin{table*}[t]
	\caption{O-RAN Interfaces.}
	\centering
	\setlength{\tabcolsep}{5pt}
	\begin{tabular}{|>{\centering\arraybackslash} m{5cm} | >{\centering\arraybackslash} m{10cm}| >{\centering\arraybackslash} m{1.5cm}|}
		\hline 
		\textbf{Network Functions} & \textbf{Description} & \textbf{References} \\  \hline\hline
		A1 & Interface between non-RT RIC $\Leftrightarrow$ Near-RT RIC, support functionality $\&$ AI/ML workflows & \cite{ORANArchitecture24}\\ \hline 
		O1 & This interface is used for management between SMO $\Leftrightarrow$ O-RAN NFs & \cite{ORANArchitecture24}\\ \hline
		O2 & Interface between SMO $\Leftrightarrow$ O-Cloud & \cite{ORANArchitecture24}\\ \hline
		E2 & Interface between near-RT RIC $\Leftrightarrow$ O-CU-CP, O-CU-UP, O-DU & \cite{ORANArchitecture24}\\ \hline
		E1 & Interface between O-CU-CP $\Leftrightarrow$ O-CU-UP & \cite{ORANArchitecture24}\\ \hline
		F1-c & Interface between O-CU-CP $\Leftrightarrow$ O-DU & \cite{ORANArchitecture24}\\ \hline
		F1-u & Interface between O-CU-UP $\Leftrightarrow$ O-DU & \cite{ORANArchitecture24}\\ \hline
		X2-c & Interface between eNB $\Leftrightarrow$ eNB/en-gNB control plane (CP) information, it is defined by 3GPP & \cite{ORANArchitecture24}\\ \hline
		X2-u & Interface between eNB $\Leftrightarrow$ eNB/en-gNB for user plane (UP) information, it is defined by 3GPP & \cite{ORANArchitecture24}\\ \hline
		Xn-c & Interface between gNB/ng-eNB $\Leftrightarrow$ gNB/ng-eNB for CP information, it is defined by 3GPP & \cite{ORANArchitecture24}\\ \hline
		Xn-u & Interface between gNB/ng-eNB $\Leftrightarrow$ gNB/ng-eNB for UP information, it is defined by 3GPP & \cite{ORANArchitecture24}\\ \hline
		NG-u & Interface between gNB-CU-UP $\Leftrightarrow$ user plane function (UPF) for UP information, it is defined by 3GPP & \cite{ORANArchitecture24}\\ \hline
		NG-c & Interface between gNB-CU-CP $\Leftrightarrow$ Mobility Management Function (AMF) for CP information, it is defined by 3GPP & \cite{ORANArchitecture24}\\ \hline
		Open Fronthaul Management-Plane (Open FH M-Plane) & Management interface for controlling the O-RU & \cite{ORANArchitecture24}\\ \hline
		Open Fronthaul includes the Control User Synchronization (CUS) plane  & Real-time control, IQ sample data, synchronization between O-DU $\Leftrightarrow$ O-RU & \cite{ORANArchitecture24}\\ \hline
		
	\end{tabular}
	\label{tab2}
\end{table*}



\begin{figure*}[t]
	\centering
	\includegraphics[height=8cm,width = 15cm]{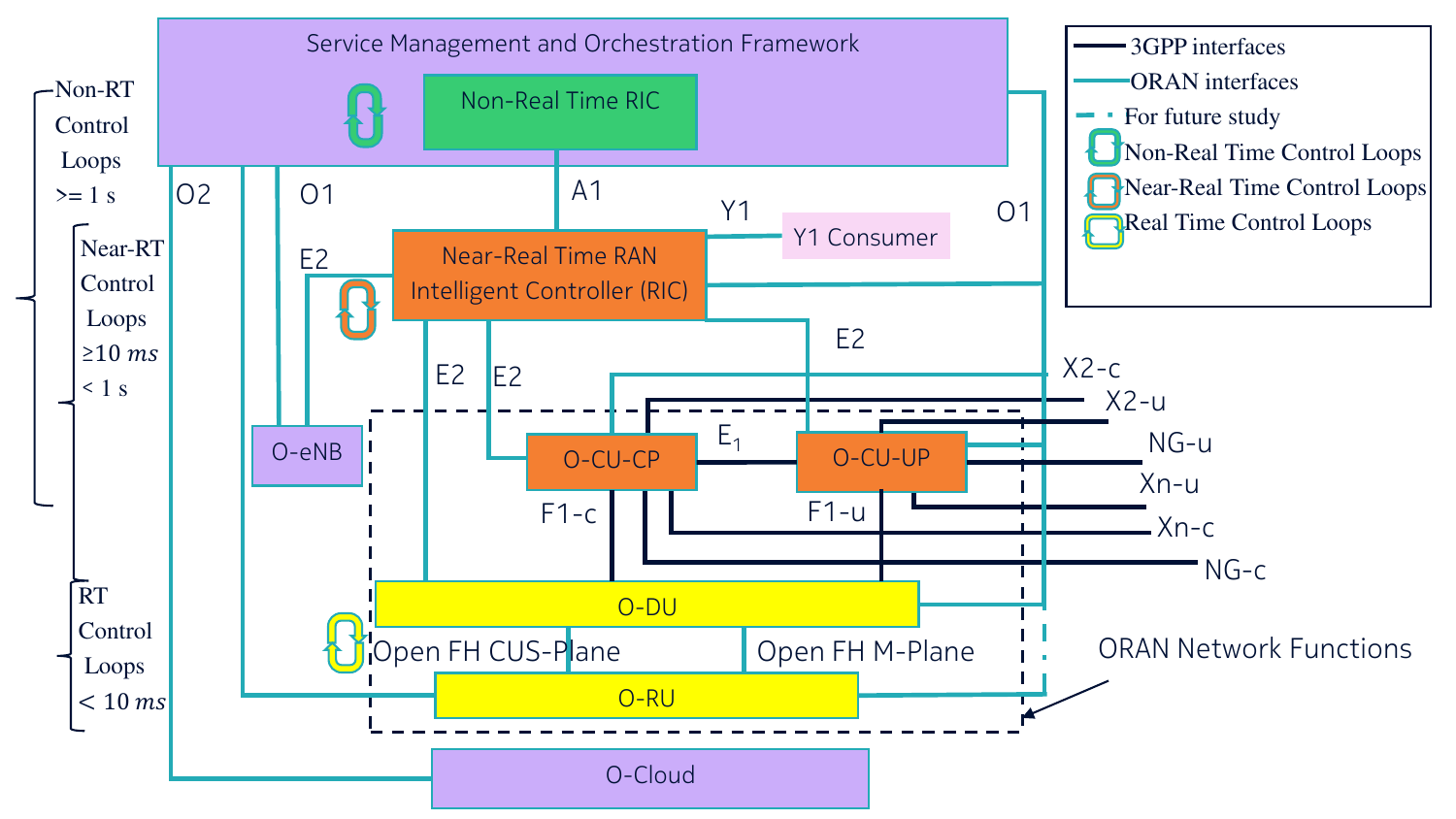}
	\caption{Logical architecture of O-RAN \cite{ORANArchitecture24}.}
	\label{FigORANArchitecture}
\end{figure*}





\subsubsection{Why is a Digital Twin necessary for O-RAN?}
\label{NDTO-RAN}
In the following subsection, we provide researchers with a clear understanding of NDT for O-RAN. Specifically, we will address the following questions: What is NDT, and why is NDT necessary for O-RAN?

a. {What is Network Digital Twin?}

The digital twin is a digital copy version of real-world entities/objects that can continuously updated to reflect its current state with a defined level of accuracy and latency requirement \cite{EricssionNDT22,almasanNDT22}. NDT includes four key elements: data, models, mapping, and interfaces \cite{NDT24} and they are described in detail as below:
\begin{itemize}
    \item \textit{Data:} an NDT should retain data (e.g., historical, real-time) about its physical network system (PNS) that are needed to represent and understand the behaviors and states of the PNS. 
    \item \textit{Models:} These techniques involve data collection in the PNS and the development of comprehensive data representations using various models. Data models, dataset models, and graphs are employed to represent the PNS, which are then instantiated to serve different applications.
    \item \textit{Interfaces:} To ensure the connection between PNS's components. There are interfaces between PNS $\Leftrightarrow$ NDT and the interface between NDT $\Leftrightarrow$ applications.
    \item \textit{Mapping:} Used to set up a real-time interactive PNS $\Leftrightarrow$ NDT or NDT $\Leftrightarrow$ NDT. This mapping provides a good understanding of the PNS's states, helping NDT to make a correct decision to optimize/maintain the PNS's performance.
\end{itemize}

b. {Why NDT necessary for O-RAN?}

In recent years, the development of network generations from 4G to 5G and 6G has introduced complex requirements that traditional network management solutions, such as network overprovisioning or admission control, struggle to meet cost-effectively. For instance, advanced communication technologies like augmented/virtual reality (AR/VR) and holographic/$360^\circ$ video streaming demand ultra-low deterministic latency. Meanwhile, modern industrial advancements, such as autonomous vehicles, require real-time adaptability to constantly changing network topologies. Additionally, the exponential growth in the number of connected devices, such as massive machine-type communication (MMTC), has made modern networks heterogeneous and dynamic. To overcome these limitations in the traditional RAN architectures, O-RAN has introduced a new architecture that offers benefits such as open interfaces, disaggregation, programmability, intelligence, and automated operation. Furthermore, it is expected to meet the diverse requirements of communication applications, including MMTC, eMBB, and URLLC. However, deploying O-RAN in the real world and meeting the above expectations remains a challenge. In particular, offline management (OFFM) in O-RAN is not feasible for the following reasons. First, due to disaggregation, OFFM results in difficulties with interoperability between vendors/suppliers and model/software updates, which reduces NetPer. Second, OFFM is challenging in a large heterogeneous network with different RRM requirements, such as latency, bandwidth, and data rate \cite{mirzaei2023network}. Fortunately, NDT has emerged as a potential solution to address many of the limitations in the O-RAN system.

NDT, as an accurate digital copy of O-RAN, can test a variety of scenarios, identifying shortcomings and risks. Then, we design different risk mitigation strategies and test them in NDT. This helps to manage large networks proactively to achieve optimal performance. These potentials of NDT facilitate quick prototyping, testing, and validation. Thus, release times are shortened while ensuring interoperability and consistent integration of O-RAN NFs. In terms of implementation, NDTs can be deployed either within the RAN Intelligent Controllers (RICs) such as \cite{10.1145/3572864.3580329} or on separate platforms, depending on the specific requirements and architecture of the network. They can operate in centralized or distributed configurations. In a centralized setup, the NDT resides within the non-Real-Time RIC (non-RT RIC), leveraging its comprehensive network view for strategic planning and AI/ML model training. Conversely, a distributed approach involves deploying NDT components closer to the edge, such as within the near-Real-Time RIC (near-RT RIC), to enable low-latency decision-making and real-time network optimization.



\subsubsection{Use Cases of Digital Twin for O-RAN}
\label{NDTO-RANUsecases}
To give researchers a clear picture of NDT for O-RAN, we present real-life NDT for O-RAN use cases from an industrial perspective. We hope that this will help academic researchers focus on these useful research directions and develop them further.

a. {Traffic Steering in Digital Twin for O-RAN:} 

With the explosion in the number of mobile users, effectively allocating resources amidst many complex factors (e.g., massive traffic demand, MU mobility, congestion, environmental influences) becomes extremely complicated and increasingly difficult. The advent of NDT brings many significant benefits that can improve this challenging problem. The benefits of NDT for O-RAN with Traffic Steering are listed below:
\begin{itemize}
    \item \textit{Advanced Real-time Traffic Monitoring and Management:} Digital twins support real-time data collection and analysis. By using data from sensors, MUs, MSs, etc., these models help predict traffic congestion, MUs' demands, MSs' states, road closures, etc. Thus, it enables quicker response times from service management orchestration.
    \item \textit{Forecast, Optimize Traffic Flow, and Alleviate Congestion:} By applying complex algorithms and ML techniques, NDT can simulate complicated traffic scenarios. Therefore, it can test and validate the effect as well as solutions for traffic management before deploying into the physical world. Through these strong capabilities, NDT helps manage/optimize traffic steering efficiently and reduce traffic congestion, significantly improving the NetPer.
    \item \textit{Ensure Safety, Promote Sustainable Urban Planning:} With the rapid development of urban infrastructure systems, optimal and effective management and allocation of resources is becoming increasingly difficult. The presence of NDT has helped significantly in that by collecting data and predicting MUs' traffic demand, vehicle traffic flows in each specific area. Thanks to that, NDT brings effective solutions to the city such as reducing the risk of accidents for autonomous cars and making quick and accurate decisions for radio resource management (RRM).
\end{itemize}

b. {Radio Resource Management in Enterprises/Factories:} 

Building communications ecosystem for businesses and factories is attracting significant interest from vendors/operators \cite{EricssionNDT22}. This market is very large and full of potential with millions of companies in each country, each requiring its own connection network. Specific needs depend on the number of employees, connectivity requirements, and existing infrastructure. For example, in a factory with thousands of different automated robots, it is very important to collect data from these robots, make accurate decisions, and ensure safety. This includes detecting incidents such as fires, explorations, electrical problems, etc. and issue warnings as quickly as possible. Furthermore, a surveillance camera system is needed to ensure security, anti-theft, and labor management. 
In one factory, various applications might be deployed, such as URLLC for robots, video streaming for camera systems, and a network system for users. NDT offers an effective solution due to its ability to collect data in real time and make automated decisions quickly. This satisfies URLLC applications with very low latency and high reliability, reported at 99.99$\%$.

c. {Network Performance Evaluation: } 

Simulations of RAN networks have become powerful tools for evaluating the performance of various RAN products and exploring solutions and methods for the next generation. Notably, integrating 3D gaming technology into NDT simulations has enhanced several aspects, such as visualization and modeling accuracy, to emulate a NDT. Ericsson \cite{EricssionNDT22} demonstrated an example of an NDT simulator compared to a real physical network. This has significantly aided researchers in conducting detailed evaluations of radio networks, including assessing the impact of MIMO interference, signal transmission paths, obstacles, and transmission delays. Based on these highly realistic simulations, the system can make quick, accurate decisions to evaluate NetPer by meeting MUs' QoS and QoE requirements while also achieving other criteria such as network energy saving (NES) and reducing allocated resources.

d. {Security in Digital Twin for O-RAN:} 

NDT for O-RAN promises to bring many benefits as mentioned above, but it is not without disadvantages. Deploying NDT for O-RAN requires collecting and processing a huge amount of data, which necessitates ensuring data privacy and security during transmission, storage, and processing. Recently, many research projects have focused on addressing security issues related to NDT for O-RAN, \cite{KumarNDTORAN23,AkramNDTORAN24,rumeshfederated,DemoNDTORAN,colosseum}. Randhir et al. \cite{KumarNDTORAN23} studied the security transmission for Internet of Vehicles (SeTIV) applying NDT in O-RAN. The NDT of SeTIV can provide the authentication and key exchange between vehicles, Road side Units (RoSiUs), and cloud servers. Then, a deep learning (DL)-based Intrusion Detection System (InDeS) algorithm was proposed to detect intrusions in the Internet of Vehicles network. Their proposed solution showed significant improvement compared to traditional solutions. In \cite{AkramNDTORAN24}, Junaid et al. proposed a novel framework namely DT-driven trust management for internet of Unmanned Aerial Vehicle Things (DTTMIUTs) in O-RAN networks focusing on monitoring forest fires. Next, they incorporate an authentication and key agreement process within the proposed framework. Then, they applied Blockchain (BloC)-based trust management and DL-based InDeS to improve the system security. Lastly, they applied a BloC-based scheme for managing the transaction. In this work, NDT was used to build a digital copy version of the internet of Unmanned Aerial Vehicle Things (IUTs). In \cite{rumeshfederated}, Rumesh et al. built and proposed a federated learning (FL) anomaly detection algorithm for NDT O-RAN. In their work, they used an O-RAN simulator to construct O-RAN components and applied NDT to train and test their FL algorithm, enhancing security. As a result, their simulations demonstrated that their anomaly detection achieved more than 99$\%$ accuracy. Unlike the works of \cite{KumarNDTORAN23,AkramNDTORAN24,rumeshfederated}, which only focused on simulation for NDT O-RAN, Peizheng et al. \cite{DemoNDTORAN} presented a demo of an NDT O-RAN system. Specifically, they proposed a near-real-time anomaly detection method using NDT. Finally, they showed a demo video of their implementation to illustrate how anomaly detection works in an NDT O-RAN system. In \cite{colosseum}, the authors built the first open-source platform for O-RAN NDT. They provided a tutorial on how to use the largest wireless network (WireNet) emulator, which was specifically built for O-RAN NDT, named Colosseum. Colosseum \cite{colosseum} evaluates O-RAN digital twin security and jamming attacks. This open-source emulator offers an opportunity to evaluate the O-RAN NDT's security, which was not previously available.

e. {Energy Efficiency in Digital Twin for O-RAN:}

Along with global warming due to many consequences caused by humans. The telecommunications system consumes a large amount of energy from MSs. Therefore, NES is becoming a topic that receives strong attention from vendors/operators, e.g., Nokia, Ericsson, Huawei, Apple, Qualcom, etc. Although there are many great benefits promised from O-RAN, the use of NDT to monitor the operation of the O-RAN system and make accurate decisions to minimize energy consumption is also receiving criticism. great concern. Temporarily turning off MSs in the O-RAN system during off-peak hours while still ensuring MUs' QoS is a quite effective solution. In \cite{DTORAN_Energy}, Anselme et al. proposed a closed-loop aided by NDT for energy saving O-RAN system. The designed two closed-loop for different network areas such as rural and urban ones. Then, they applied DRL to optimize the allocated resources using their proposed energy model.

\subsection{Digital Twins for Intelligent Surface}

Intelligent Reflecting Surfaces (IRS) are planar surfaces made up of numerous small and passive components that can deliberately reflect incoming signals to improve wireless communications \cite{TinIoT2024,long2025deep}. In this case, the use of DT can provide virtual representation and optimization for IRS. For the virtual representation, the DT can accurately reflects the physical IRS in real-time, encompassing its configuration, environment, and operational condition. Meanwhile, for the optimization, the DT has the ability to simulate different configurations of the IRS and analyze their effects on the network. This enables the precise adjustment of the reflecting parts for best performance. To further enhance the performance of IRS using DT, the ML/AI-based approaches can be used to analyze data in real-time and make dynamic adjustments to the elements of the IRS in order to maximize the reflection of signals. Furthermore, they have the ability to anticipate changes in the communication environment, and consistently optimize the IRS settings for improved network performance (e.g., maximizing signal-to-noise ratio, minimizing interference).

For example, the work in~\cite{Oliveri2022Towards} discusses a DT 
framework for reflect-array unit cells (RA-UC) through electromagnetic-assisted ML methods. In particular, the DT framework predicts the the values of scattering matrix components for both single- and dual-polarization elements in the RA-UC. This utilizes artificial neural networks that are built upon augmented radial basis functions (A-RBFs), support vector regression (SVR) techniques, and statistical learning processes that rely on ordinary kriging (OK). The training process focuses on understanding the correlation between the scattering matrix coefficient and the UC descriptors. Various data encodings, such as real and imaginary values, magnitude and phase values, and normalized values, are also utilized. To assess the trade-offs between the accuracy of co-/cross-polar scattering fidelity and the time saved in synthesizing UC-DT models, an analysis is conducted.  This analysis takes into account the high-performance RA cells designed for various operational conditions, such as high-frequency applications, high-gain scenarios, wide-band designs, and dual-polarization radiation. Based on the numerous performance results, the proposed framework can achieve the prediction accuracy by considering the number of geometrical/physical descriptors and the complexity of the wave manipulation phenomena caused by the UC. Additionally, All DT implementations achieve a time savings of over 99\% compared to classical full-wave modeling.

In another work, the authors in~\cite{Cui2023Digital} introduce a novel system, i.e., RIS-assisted, uplink, user-centric cell-free (UCCF) system, using DT. This aims to maximize the sum-rate by simultaneously optimizing the access point and user association (AUA), power control, and RIS beamforming. Specifically, the propose system first separates the AUA from the power control and RIS beamforming (PCRB) according to the distinct characteristics of their variables, resulting in a reduction of the solution space. Here, the AUA is utilizing a newly developed position-adaptive binary particle swarm optimization (PABPSO) approach. Then, a DRL approach based on two twin-delayed DDPG models, with advanced state pre-processing layers, are developed for the PCRB. In this case, the DT is utilized to train the learning framework by replaying recorded channel estimations. For the system model, a number of user equipments (UEs) are served simultaneously by access points (APs) with the help of a number of RISs. The DT is utilized as the intermediate layer between the hardware, e.g., the APs and edge server, and software components, e.g., the control algorithm that addresses the AUA, power control, and RIS beamforming. In this case, the DT can direct the hardware to carry out control decisions established by the algorithm in the physical environment. It can also evaluate how the environment responds to these decisions, such as measuring the data rates that can be achieved by the UEs. Moreover, the DT can store and reproduce the measurements of parameters taken in the real-world setting, such as channel estimations. Extensive simulations indicate that the proposed system can provide 25\% more episode reward and 12\% higher physical sum-rate under dissimilar maximum transmit powers than those of other baseline methods. This shows that using RISs greatly improves the overall data transmission rate of UCCF systems. 

\section{Digital twin for transport network }\label{sec:optical} 


Optical networks offer large capacity and long-distance data transmission due to their high bandwidth, low loss, and strong resistance to interference. However, existing optical communication networks face significant challenges in accurately monitoring, managing online, controlling in real-time, and facilitating rapid maintenance due to complex deployment environments. These challenges can be mitigated by integrating optical networks with   DT, which can improve network optimization, enable predictive maintenance, and support autonomous operations\cite{Wang2021optical}.

\begin{figure}[htb]
	\centering
	\includegraphics[height=7.5cm,width = 9.5cm]{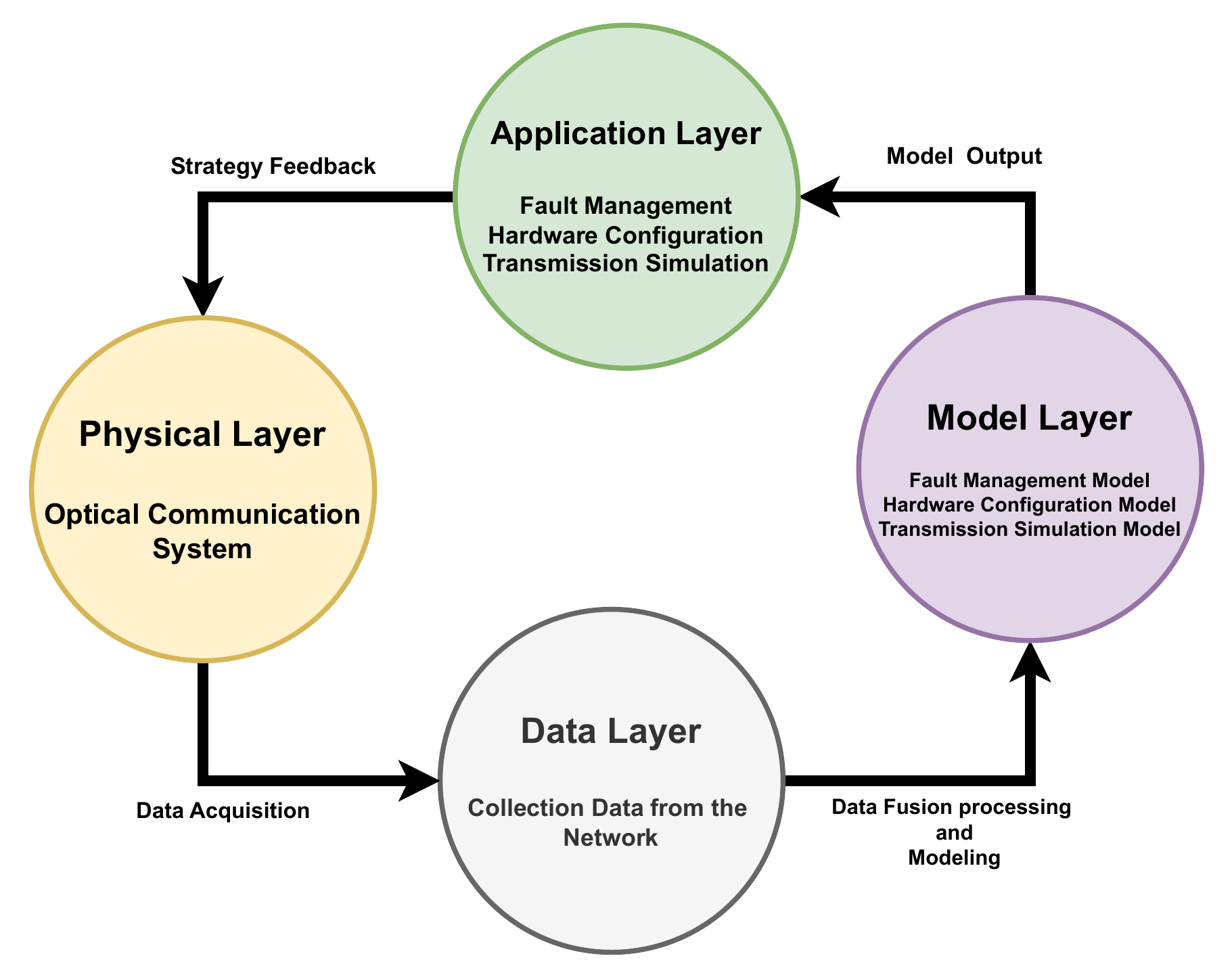}
	\caption{Digital twin framework for optical communication system\cite{Wang2021optical}.
}
	\label{opticalDT}
\end{figure}

Fig. \ref{opticalDT} shows a framework for using DT in optical communication system.  Firstly, data is collected from the physical layer, which refers to the optical system and may consist of optical devices, network elements, hardware modules, fiber links, etc. Secondly, fusion processing is applied to the collected data to obtain useful information in the data layer. Afterward, a model is created from this useful data in the model layer, and based on the created model, the application layer can perform equipment fault prediction, diagnosis, location, and repair, hardware resource optimization, network efficiency improvement, transmission process dynamic simulation, real-time display, and numerical analysis. Finally, based on the results of the comprehensive analysis, response actions are taken in the application layer, and feedback is provided to the physical layer from the top down\cite{Wang2021optical}.

While having a general understanding of how the DT can be applicable in optical systems from Fig. \ref{opticalDT}, several studies have explored different approaches to implementing DT in optical communication systems, each focusing on specific network elements, signal transmission, and operational challenges. In \cite{Sequeira2023IQ,Devigili2024IQ}, a DT is used for estimating the quality of transmission (QoT) and managing failures efficiently when a lightpath connects two end locations equipped with a transponder node through an optical network consisting of several Reconfigurable optical add/drop multiplexer nodes (ROADMs) and optical links. The DT utilizes the extracted information of Optical in-phase and quadrature (IQ) from the optical signal and applies deep neural network (DNN) models to exploit IQ constellation features for soft-failure detection, identification, and severity estimation. Furthermore, \cite{Vilalta2023validDT},   provided detailed information on the use of  DT  in optical networks to facilitate the validation of optical light-path behavior before implementation in the physical network. While in  \cite{Zhang2023large}, the DT model is developed based on the Gaussian noise model and a deep neural network (DNN) to perform efficient and accurate QoT estimation for optimization and control of physical layer devices, real-time responses to deterioration and fault alarms in networks, as well as network routing and resource reallocation.

Accurate physical layer models facilitate the development of intelligent, self-driving optical networks. For example, the dynamic wavelength-dependent gain characteristics of erbium-doped fiber amplifiers (EDFAs) determine the power spectrum, which is important for estimating the optical signal-to-noise ratio as well as the magnitude of fiber nonlinearities. For this, a gray-box EDFA gain modeling scheme is proposed, making it more feasible to build a customized digital twin of each EDFA in optical networks, which is essential, especially for next-generation multiband network operations\cite{Yichen2023EDFA1,Yichen2023EDFA2}.  On the other hand, in ROADMs, soft failures may occur in both inter-node and intra-node components, such as wavelength-selective switches and fiber spans. While machine learning (ML) is a promising solution for soft-failure localization, data scarcity presents a significant challenge for its implementation. A digital-twin-assisted meta-learning framework to achieve effective soft-failure localization is proposed in [56] to address this issue. This framework utilizes digital twins to create mirror models and provides multiple training tasks. As a result, meta-learning can be effectively employed to localize soft failures\cite{Wang2024Rodam}.


\begin{table*}[]
	\centering
	\caption{Summary of recent works on Digital Twin for Optical Network}
	\label{TableDTforOpticalReview}
	\begin{tabular}{|l|l|l|l|}
		\hline
		Reference & Technology/Scenario & Problem/Purpose & ML/AI Method \\ \hline
		\cite{Qu2022FedTwin} & \multicolumn{1}{c|}{\multirow{5}{*}{Device configuration method}}                 & Flat and Maximized received SNR per channel & NN-based EDFA mirror modeling \\ \cline{1-1} \cline{3-4} 	
		\cite{Aloqaily2023Reinforcing} &  & SNR optimization for the multi-span DT system   
        & Gradient descent \\ \cline{1-1} \cline{3-4} 
		\cite{Lv2023Blockchain} &  & Maximize the mean and standard deviation of GSNR & Heuristic algorithm \\ \cline{1-1} \cline{3-4}       
		\cite{Prathiba2024Fortifying} &  & maximize system capacity in MB system
        & Ml-based power profiles optimization \\ \cline{1-1} \cline{3-4}
		\cite{Mu2023Digital} &  & Flat and maximized received power/SNR & Autoencoder-based multi-span optimizer \\ \hline

	\end{tabular}
\end{table*}

\section{5G and Beyond Core Network DT }\label{sec:5gcore}
We discussed NDT for O-RAN in section \ref{sec:O-RAN}, about its potential, advantages, why we need NDT for O-RAN and use cases. It would be a big mistake not to mention NDT for 5G CORE Networks and Beyond (NG-CORE) because NG-CORE performs many important functions such as authentication, security, registration, mobility, access management, etc. Therefore, in this section, we would like to describe to researchers what NG-CORE is, its functions \ref{sec:coreintro}, why we need NDT for NG-CORE \ref{sec:coreNDT}, and topics related to NDT for NG-CORE \ref{sec:coretopics}. 

\begin{figure*}[t]
	\centering
	\includegraphics[height=8cm,width = 18cm]{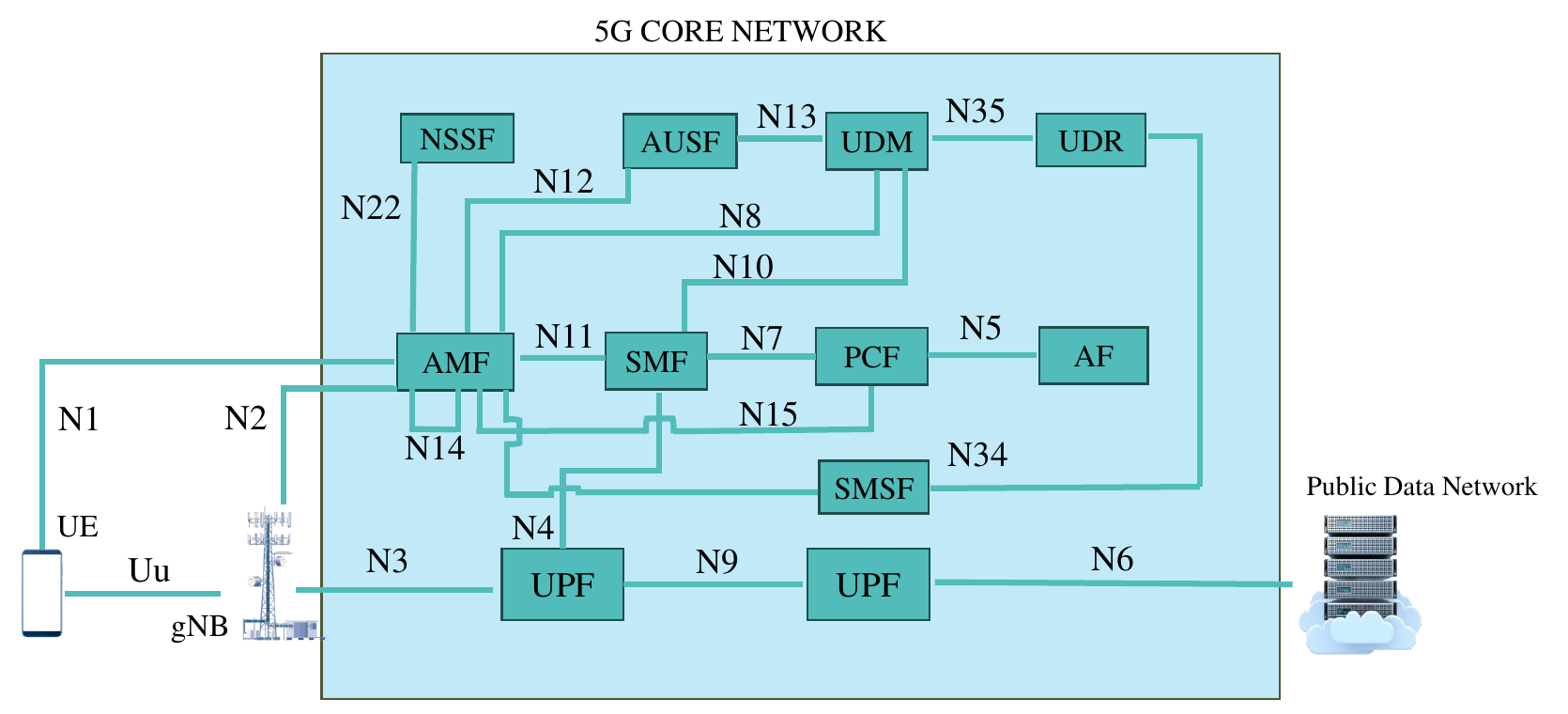}
	\caption{5G system architecture.}
	\label{Fig6GCore}
\end{figure*}
\subsection{Introduction about 5G CORE Network and Its Components} \label{sec:coreintro}
In Fig. \ref{Fig6GCore}, we show the 6G system architecture including a UE, 1 gNB/RAN \cite{kaada2024multi}, a 5, and a public data network (PDN) \cite{Nokia6GSystem, Dell6GCore,ETSI6GCore}. The 5 has many different NFs and each one has its own purpose. Details of each NF are described as follows:
\begin{itemize}
    \item \textit{User Plan Function (UPF):} This is an important NF in 5. It is also responsible for managing the user data during the user data transmission process. More specifically, it provides the connection point between RAN and the PDN as shown in Fig. \ref{Fig6GCore}, i.e., encapsulates/decapsulate the general packet radio service (GPRS) Tunnelling Protocol (GTP). It also responsible for packet routing, quality of service (QoS) handling, and MS's data traffic reporting for billing.
    \item \textit{Access and Mobility Management function (AMF):} AMF is one of the control plane (CP) NF of the 5, it is responsible for the mobile device (MD) registration, access management, mobility management, policy implementation (i.e., it ensures the adequately allocated resource to the MU to satisfy the MU's QoS), session management, MU's subscriber data management, network slicing (NetSlice), policy control, location management, optimized network, ensuring the security of 5G MUs. AMF connects with MU, gNB, and many other NFs such as Policy Control Function (PCF), Network Slice Selection Function (NSSF), Authentication Server Function (AUSF), Unified Data Management (UDM), Unified Data Repository (UDR), and Session Management Function (SMF).
    \item \textit{Session Management Function (SMF):} SMF is a CPNF in 5. It is responsible for the data session establishment, maintenance, and release between the MU and PDN. Moreover, it supports for the NetSlice, whereas multiple slicing with different latency and data rate requirement are created on the top of one physical network hardware. More specifically, it manages these different sessions and slicing. Specifically, it can be deployed in a standalone NF or combined with other NFs. Furthermore, SMF can connect with many other NFs such as UPF, SMF, PCF, and UDM.
    \item \textit{Policy Control Function (PCF):} PCD acts as a CPNF in 5. Its function is to guarantee the network operates properly and meets the various service requirements of the MUs. Moreover, it provides customizable policies to business customers. For example, it define different policies corresponding to various applications such as MMTC, eMBB, and URLLC. As shown in Fig. \ref{Fig6GCore}, PCF connects to different other NFs such as AMF, SMF, and Application Function (AF).
    \item {AF: } AF is an important AF in QoS assignments, application policies, and traffic management for 5G networks. It executes policies related to allocated resources and application behavior. Moreover, it is used for prioritization such as prioritizing real-time applications/services over non-RT data transmission. Furthermore, AF also support for the billing process by monitoring the amount of application usage data. As shown in Fig. \ref{Fig6GCore}, AF connects to PCF for policy management.
    \item \textit{Short Message Service Function (SMSF):} SMSF is a NF that is used for the short message service (SMS)/texting between MS $\leftrightarrow$ MS in 2G/3G/4G/5G. Its functions are listed as follows: Manage the subscriber's information (i.e., inserts/removes subscriber's information), processing roamer (i.e., manage Security Edge Protection Proxy (SEPP) and Public Land Mobile Network (PLMN) interconnections). As shown in Fig. \ref{Fig6GCore}, SMSF connects to UDR and AMF.
    \item \textit{Network Slice Selection Function (NSSF):} It can be used to support AMF to select the correction NetSlice providing to the MS. Moreover, it is used to allocate a specific AMF in the case that the allocated AMF is unable to support all NetSlice for a specific MS. Furthermore, NSSF contains the network slice selection assistance information (NSSAI), identifier of NetSlice instance, i.e., Nsild,  roaming information, i.e., RoamingIndication, configuration information of the single NSSAI (S-NSSAI), i.e., ConfigureSnssai, slicing information for registration process, i.e., SliceInforForRegistration. As shown in Fig. \ref{Fig6GCore}, NSSF connects to AMF.
    \item \textit{Authentication Server Function (AUSF):} The functions of AUSF are authentication (i.e., identify the MU subscriber (MUSub) when it accesses to the 5G network, the AUSF is responsible for verifying this subscriber and make sure that this one connects to the network properly.), authorization (i.e., AUSF checks whether this subscriber can access a specific service or not), security (i.e., establish the security for the MUSub), mobility management (i.e., AUSF connects to the AMF to clarify the MU mobility and handover decisions), session management (i.e., AUSF supports for the session establishment, maintenance, and release), MU's data management (i.e., AUSF connects to UDM for MU's profiles). As shown in Fig. \ref{Fig6GCore}, AUSF can connect to UDM, and AMF.
    \item \textit{Unified Data Management (UDM):} Functions Of UDM are listed as follows: a) manage the MU's data, b) MU identification, i.e., store $\&$ manage the MU's Subscription Permanent Identifier (SUPI), c) Persistent service/session support, e.g., continuing to assign tSMF/data network name (DNN) to ongoing sessions, d) SMS $\&$ subscription management. As shown in Fig. \ref{Fig6GCore}, UDM connects to AMF, SMF, and UDR.
    \item \textit{Unified Data Repository (UDR):} UDR provides a unified cloud database for MUSub, authentification, authorization, and application. Functions of UDR are listed as follows: a) Manage $\&$ store Subscriber Identity Module (SIM) card, b) Announce other NFs about the changing of MUSub information. As shown in Fig. \ref{Fig6GCore}, UDR connects to SMSF and UDM. 
\end{itemize}

In the paragraph below, we describe the connection interface between NFs in NG-CORE:
\begin{itemize}
    \item \textit{N1:} The interface between MU/User Equipment (UE) $\leftrightarrow$ AMF.
    \item \textit{N2:} The interface between gNB $\leftrightarrow$ AMF.
    \item \textit{N3:} The interface between gNB $\leftrightarrow$ UPF.
    \item \textit{N4:} The interface between SMF $\leftrightarrow$ UPF.
    \item \textit{N5:} The interface between PCF $\leftrightarrow$ AF.
    \item \textit{N6:} The interface between UPF $\leftrightarrow$ PDN.
    \item \textit{N7:} The interface between SMF $\leftrightarrow$ PCF.
    \item \textit{N8:} The interface between AMF $\leftrightarrow$ UDM.
    \item \textit{N9:} The interface between UPF $\leftrightarrow$ UPF.
    \item \textit{N10:} The interface between SMF $\leftrightarrow$ UDM.
    \item \textit{N11:} The interface between AMF $\leftrightarrow$ SMF.
    \item \textit{N12:} The interface between AMF $\leftrightarrow$ AUSF.
    \item \textit{N13:} The interface between AUSF $\leftrightarrow$ UDM.
    \item \textit{N14:} The interface between AMF $\leftrightarrow$ AMF.
    \item \textit{N15:} The interface between AMF $\leftrightarrow$ PCF.
    \item \textit{N22:} The interface between AMF $\leftrightarrow$ NSSF.
    \item \textit{N35:} The interface between UDM $\leftrightarrow$ UDR.
\end{itemize}

\subsection{Why Do We Need NDT for NG-CORE?}
\label{sec:coreNDT}

NG-CORE becomes a crucial part of 5G and future networks as it needs to perform many important functions such as authentication, security, registration, mobility, management, etc. The rapid increase in the number of MUs, the need for different service requirements, and the need for increasing network complexity make the operation, maintenance, and development of NG-CORE increasingly difficult and complex. In order to design a NG-CORE system that works smoothly in practice, we need to build a testing environment as close to reality as possible. However, this becomes highly expensive and difficult with so many different MUs and components from many various suppliers/vendors, not to mention how to design algorithms that work effectively for a large network, e.g., thousands of MSs and thousands of MUS. This is truly a difficult problem \cite{zheng2022research}. Fortunately, NDT has become an extremely effective and inexpensive solution to solve the difficulties mentioned above. NDT is able to reconstruct a replica of a real NG-CORE, and it can update the digital information of the NG-CORE in real time and periodically. Therefore, we identified the problem in NG-CORE quickly, and from there we could analyze the situation, provide a timely solution, and deploy the solution as quickly as possible into a physical NG-CORE system. Right after, we would like to describe the advantages of NDT for NG-CORE:
\begin{itemize}
    \item \textit{Reduce cost:} It is clear that by building a virtual replica of the real world, NG-CORE, it will be much cheaper than purchasing real components from multiple suppliers/vendors. In reality, a lab with hundreds of gNBs and a large 5-GCOREB system is extremely expensive and even difficult to deploy. Building an NDT system using only software and the cloud for storage is now quite easy and cheap for all research centers or universities. This not only helps researchers have more conditions to do practical research and access new technology quickly, but it also reduces cost barriers. More specifically, we can emulate problems that may occur in practice, through which we can design algorithms and test them quickly and effectively, which can then be deployed in practice. NDT also helps to test new features, devices, and components before bringing products to market. Therefore, it helps significantly reduce testing and trial operation costs.
    \item \textit{{Advanced network management and optimization:} } NDT helps monitor the physical NG-CORE system in real time. Therefore, it proactively recognizes problems as early as possible and provides solutions before they impact the real-world NG-CORE system. Furthermore, the appearance of NDT helps manage things automatically instead of managing manually, which is slow and ineffective. Therefore, we can build a complex zero-touching system that operates completely automatically, updating states to NDT periodically. Based on that, algorithms can be tested to optimize system performance before running on the real system. NDT can also emulate different traffic demand scenarios quickly, thereby providing solutions to optimize traffic efficiently. This is to ensure NG-CORE operates accurately and smoothly, avoiding excessive congestion and overloading which will affect MUs's QoS and QoE.
    \item \textit{Fault/Anomaly prediction:} To detect errors or anomalies, NDT needs to build the basic or normal behavior of the NG-CORE system based on regularly received data from history and current operating status. NDT will then collect data regularly from NG-CORE based on common key performance indicators (KPIs) such as data rate, latency, and packet error rate monitored in real-time to detect or predict the error/abnormal behavior. Machine algorithms can be applied to detect the bias between current states and the baseline. Then, they predict the potential anomalous behavior based on the pattern in the data. Therefore, they make decisions to check more frequently to prevent the problem from escalating more seriously. Machine algorithms can be applied to detect the bias between current states and the baseline. Then, it predicts the potential anomaly behavior based on the pattern in the data. Therefore, IT can decide to check more frequently to prevent the problem from escalating more seriously. Then, NDT provides solutions to some problems, such as rerouting or reallocating resources, or it can issue warnings and notify humans to intervene to solve problems more effectively.
    \item \textit{Effective network planning and deployment:} The emergence of NDT has greatly supported network planning and deployment. More specifically, NDT can simulate the entire NG-CORE system, allowing operators and suppliers to freely research and test it before putting it into actual operation. Instead of managing manually like before, NDT offers the potential for an almost automatic, zero-touch management system with 99.99$\%$ automation \cite{ZerotouchNDT23}. Automating management brings many great benefits, such as allocating resources faster and more efficiently without encountering small errors caused by humans. In addition, it also reduces operating costs significantly because one person can undertake many tasks. Furthermore, network scalability also becomes easier because NDT supports managing a large number of MSs and MUs in a large area. NDT also helps us monitor and maintain the system effectively by updating the network state. Therefore, analyze the real-time behaviors of the NG-CORE and detect errors to provide solutions in the most timely manner.
    \item \textit{Improved security:} The integration of NDT into NG-CORE has greatly improved security in many ways. By monitoring NG-CORE continuously and regularly, NDT can detect potential threats to the security of NG-CORE, thus taking timely prevention measures. First, NDT can emulate different security attack cases, based on that, security researchers can analyze potential risks and develop algorithms to prevent and minimize risks. In addition, with real-time monitoring of NG-CORE, unusual manifestations that potentially affect security will be detected early, thus being promptly prevented before the attack spreads. Furthermore, regular and long-term data collection helps NDT have a large data set and diverse attack cases in the past, so it can simulate attack cases that may occur in the future effectively, to proactively visualize possible situations and come up with solutions in advance.
\end{itemize}

\subsection{Topics Related to NDT NG-CORE}
\label{sec:coretopics}
In this subsection, We would like to describe the research works that have been done related to NDT for NG-CORE as follows:
\begin{itemize}
    \item \textit{Design NDT for NG-CORE:} We all know that NDT is a digital copy version of an object, e.g., NG-CORE, but how to build an NDT and what properties of NG-CORE does it copy to solve a specific problem. For example, we can build NDT as a physical or IP network copy version of NG-CORE. Below we would like to introduce some of the research that has been and is being done to build NDT for NG-CORE. In \cite{SanzCore23}, Sanz et al. Sanz et al. proposed a new design of NDT for the NG-CORE system. This work proposed a method of seven phases to regularly update the NG-CORE's NFs information to the NDT. Different from literature works that did not connect the physical system to the NDT, this work builds a real tested environment using GNS3 that could connect the NDT for NG-CORE using free5gc. In \cite{TaoCore24}, Zhenyu et al. proposed two different NDT architectures to build the 5G CP. More specifically, they implemented DL models, named 5GC-former and SeQ2Seq to rebuild a virtual copy of NG-CORE's NFs. This is the first research work to rebuild NG-CORE's NFs based on deep learning instead of programming. In particular, their simulation results showed that the two models 5GC-Seq2Seq and 5GC-former reached 99,997$\%$ and 99,999$\%$ F1-score, i.e., a metric to measure the accuracy of samples. In \cite{wang2024digital}, In \cite{wang2024digital}, Jiadai et al. NDT proposal for the NG-CORE network slicing framework, provides a virtual environment for DRL agents to interact, thus reducing wrong decisions that occur in real-world environments. In \cite{GeiBlerCORE23}, Stefan et al. build a mobile virtual network operator core network (MVNOCoreSim). This was the first protocol-level simulator for IoT networks. The proposed simulator mimicked the signaling message, timeout, and interaction of IoT users (IoTUs) in the physical real system. Furthermore, they showed a 2G/3G MNVO NG-CORE describing the signaling signals obtained from a real MNVO network. They also showed a case study of bottleneck detection and resource allocation in MNVO. In addition, an NDT is built to present the NG-CORE NFs.
    \item \textit{Flying vehicles connectivity:} In, \cite{Bilen5GCORE22} Bilen et al. introduced a novel proof-of-concept (PoC) of NDT in the Wifi NG-CORE system. Their NDT model was used to select appropriate connectivity links and traffic-based NG-CORE. As in-flight connectivity (IFCon) faces various difficulties due to the inherent characteristics of airplane in-flight topologies and flying NG-CORE links (F5GCOL). Therefore, we need to continuously check F5GCOLs to ensure continuous connectivity for flights. This is especially important to avoid unwanted accidents such as flight MH370, which has gone missing and has still not been found. Therefore, NDT is applied to continuously monitor and manage IFCon transmission lines to proactively analyze possible risks, based on which timely prevention plans can be made. Additionally, NDT can simulate, analyze, and predict using ML algorithms to provide stable transmission and reduce packet error rates.
    \item \textit{Security improvement:} To build an almost completely automated management system for the NG-CORE network, Yagmur et. al. \cite{YigitDDOSCORE22} has built an NDT system using online learning (Olearn) to detect distributed denial-of-service (DDoS) attack problems quickly and effectively. First, they designed a DDOS attack detection system using NDT for the Internet service provider (ISP) NG-CORE network. Then, they applied Olearn to continuously update the model to improve the model learning process quickly and make accurate predictions. They concluded that their framework provides 97$\%$ accurate results for DDOS attack detection. In \cite{zheng2022research}, Yifend et al. built a NDT for NG-CORE networks to detect fault and then self-recovery automatically. This work was based on the real network architecture from the China Mobile Research Institute for NDT including three domains, three layers, and three closed loops. They built a framework consisting of a physical layer, a network layer, an application layer, an operation management system, and a security system. They focused on the NG-CORE network for error detection and recovery via four steps, i.e., perception $\rightarrow$ analysis $\rightarrow$ decision $\rightarrow$ deployment.
\end{itemize}



\section{Cloud Computing/Edge Computing DT }\label{sec:cloud}
Cloud and edge computing have the potential to significantly enhance DT technology by providing robust data processing and real-time analytics capabilities. Particularly, cloud computing offers the necessary infrastructure to handle large-scale data storage and intensive computations, ensuring that DTs can process and analyze vast amounts of information from their physical counterparts. Edge computing complements this by processing data closer to the source, reducing latency and enabling faster response times.

\subsection{Cloud Computing DT}

Cloud computing is the technology that delivers computing services over the internet, allowing organizations to access these services remotely from cloud providers. Cloud computing, by providing scalability, flexibility, and computational resources, can empower DT technology in a wide range of applications. 

A noticeable application is elderly healthcare, discussed in \cite{mohamed2023leveraging}, \cite{liu2019novel}. In~\cite{mohamed2023leveraging}, a DT approach is proposed for healthcare systems, particularly focusing on a cloud-based solution for elderly care. The proposed framework uses cloud-based DTs to replicate the health status of patients, continuously updating them with real-time data from wearable sensors and medical devices. The cloud platform performs advanced analytics to monitor health conditions, predict potential issues, and provide personalized care recommendations. The implementation improved elderly healthcare services by enabling continuous monitoring and early detection of health problems, leading to better patient outcomes and reduced healthcare costs. Taking another approach, in~\cite{liu2019novel}, the authors propose a cloud-based framework, namely CloudDTH, for elderly healthcare services using DT technology. The framework creates a DT of the healthcare ecosystem by integrating physical objects (patients, medical devices), virtual models, and healthcare data. The system leverages key technologies such as IoT for data collection, cloud computing for processing, and advanced analytics for insights. Among them, cloud computing serves as the foundational technology for the CloudDTH framework, providing the scalable infrastructure, processing power, and data management capabilities necessary to implement DT simulations for elderly healthcare services. It also enables centralized storage and analysis of large volumes of healthcare data, supports real-time monitoring and remote diagnosis, facilitates the delivery of various healthcare services, and allows for efficient resource allocation and management. A case study is conducted to demonstrate the feasibility of the proposed approach in two key areas, i.e., real-time monitoring with personalized medication reminders for individual patient and hospital bed allocation scenarios. Besides, ~\cite{zheng2021towards} introduces PSim-DTH, a method for secure and efficient healthcare monitoring using similarity queries on a DT cloud. The approach involves first creating a formal model for similarity-based healthcare monitoring within the DT cloud environment. To organize healthcare data, they use a partition-based tree (PB-tree) indexing structure. For privacy, the paper employs matrix encryption to develop a privacy-preserving similarity range query (PSRQ) algorithm that works with the PB-tree. The PSim-DTH scheme is built upon the PSRQ algorithm. Extensive security analysis and performance evaluation show that the PSim-DTH scheme is both privacy-preserving and efficient.

Apart from health care, \cite{han2022cloud}, \cite{zhang2022practical}, and \cite{stergiou2022digital} discuss DT applications in power management. In~\cite{han2022cloud}, the authors explore the use of cloud and edge computing to host DTs for managing distributed energy resources. Specifically, by splitting the DT framework between cloud and edge layers, the system benefits from real-time data collection and processing at the edge, while the cloud handles more intensive computations and data storage. This architecture enables optimized coordination and control of distributed energy resources, enhancing the efficiency and reliability of energy distribution systems. Experiment results show significant improvements in energy management, with better resource utilization and reduced communication overhead. In~\cite{zhang2022practical}, the authors investigate the practical adoption of cloud computing in power systems, highlighting its role in enhancing DT technology for grid management. By integrating power system components with cloud platforms, DTs can model and monitor grid behavior in real-time. The cloud platform facilitates vast data collection and analysis from grid components, supporting applications like predictive maintenance, fault detection, and demand forecasting. Simulation results show that cloud computing significantly enhances the efficiency, reliability, and resilience of power grids, leading to more effective management and reduced downtime. ~\cite{stergiou2022digital} introduces a DT and cloud-side-end collaboration for intelligent battery management system. The edge node bridges the physical and digital worlds, transmitting battery data to create DT models and returning state estimation and health predictions to the vehicle. The cloud-based big data platform leverages extensive storage and computational resources for analysis and feedback, supporting additional functions like service, operation, maintenance, product tracking, and fault retracing. A four-layer networked architecture is also developed, leveraging cloud computing and storage for advanced algorithms, real-time data-driven DT models with dynamic mapping, and integrating real-time processing with cloud intelligence for sophisticated battery management. Experiment results demonstrate the high accuracy of the proposed approach in predicting the state of the real battery based on the DT counterparts.  

Besides, DT applications in industrial production is discussed in ~\cite{mikhailov2022new}, ~\cite{huang2021digital}, and ~\cite{pan2021digital}. In~\cite{mikhailov2022new}, the authors introduce DT as a Service (DTaaS) for Industrial Internet of Things (IIoT) systems, leveraging cloud computing to create scalable and flexible solutions for monitoring and optimizing industrial processes. The proposed architecture integrates cloud-based DTs with IIoT sensors and devices, providing real-time data analysis and simulation. The cloud platform supports various services, including predictive maintenance and operational optimization. Simulation results show that DTaaS approach can enhance scalability, flexibility, and efficiency in industrial process management, reducing downtime and improving overall performance. Similarly, ~\cite{huang2021digital} presents a new framework that uses DT technology to detect anomalies in industrial systems in real time and even predict them. This framework follows the idea of edge AI (or edge intelligence) by using DTs within a flexible network that spans both the industrial edge and the cloud. This setup allows for fast and accurate anomaly detection. By placing DTs at the edge, the system can process data efficiently using the computing and storage available on local devices. They built a working model using a LiBr absorption chiller, which showed that their framework and its technologies are practical. Their study also indicated that the proposed method can identify anomalies early on.~\cite{pan2021digital} introduces a multi-level cloud computing-enabled DT system designed for real-time monitoring, decision-making, and control of a synchronized production logistics system. This IoT-driven production logistics synchronization (PLS) system uses real-time information to precisely capture the activities of the physical layer and assess how they influence the overall system's operation. Depending on the severity of these dynamics-slight, moderate, or severe-edge computing, fog computing, and cloud computing are employed to form a dynamic multi-level distributed decision-making system, optimizing effectiveness and cost-efficiency. A industrial case study is also conducted where an optimization model is developed for production and storage. The results of the case study demonstrate the effectiveness of the proposed approach. 

\begin{figure}[!]
	\centering
	\includegraphics[scale=0.6]{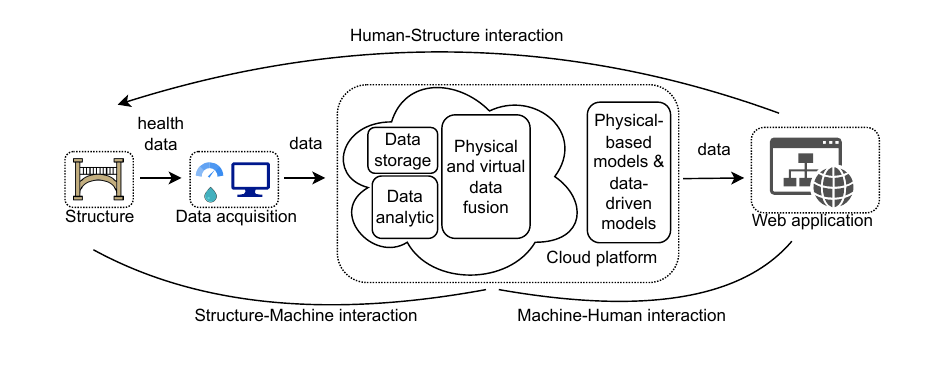}
	\caption{Cloud-based DT model.~\cite{dang2021cloud}}
	\label{CloudBasedDT}
\end{figure}

In~\cite{dang2021cloud}, a cloud-based DT framework for structural health monitoring is developed, as illustrated in Fig.~\ref{CloudBasedDT}. Particularly, monitoring large structures is typically expensive and labor-intensive. DT technology aims to overcome these challenges by creating a high-fidelity digital replica of the physical entity. To this end, the framework combines physical structure, digital models, and cloud computing to efficiently monitor real-time data and enable proactive maintenance. For damage detection, the framework employs a deep 1D Convolutional Neural Network (CNN), which achieves a 92\% accuracy rate in identifying damaged truss connections.

Taking another approach, ~\cite{dai2022adaptive} integrates DT technology into VEC networks to enable adaptive network management and policy scheduling. First, the framework and key issues of VEC networks are introduced. Then, the DT concept is presented, and an adaptive DT-enabled VEC network is proposed. In this architecture, DTs facilitate adaptive management using two closed loops between the physical VEC networks and their digital versions. Furthermore, a DT-based approach addresses the VEC offloading problem, employing vehicle and roadside unit (RSU) digital models. A Deep Reinforcement Learning (DRL)-based offloading scheme is developed to minimize total latency, with numerical results confirming the algorithm's effectiveness.

\subsection{Edge Computing DT}

Edge computing, unlike traditional cloud computing, enables IoT devices to process at the edge of the network, thereby bringing data closer to the data source. Edge computing can enhance DT in aspects such as reducing latency, improving reliability and resource efficiency \cite{9830363}. Multiple DT applications integrated with edge computing have been proposed. 

Among those applications, \cite{karobi2024ecoedgetwin}, \cite{lu2021adaptive}, and \cite{yang2023edge} present DT in 6G networks. Particularly, both \cite{karobi2024ecoedgetwin} and \cite{lu2021adaptive} apply mobile edge computing (MEC) to overcome challenges of 6G network such as the need for higher data throughput and reduced latency. In \cite{karobi2024ecoedgetwin}, the authors introduce EcoEdgeTwin model, which consists of a physical infrastructure layer presenting the actual network components such as edge servers and mobile users, and a DT Integration Layer presenting the DTs of those network components. Moreover, from developed core mechanisms, performance model, an optimization problem is formulated to balance latency, energy consumption, utility function, and to maximize the Quality of Experience (QoE). To that end, an Advantage Actor-Critic (A2C) algorithm is developed for optimizing task offloading and energy management. Simulation results show that, compared to a benchmark model, EcoEdgeTwin has lower energy consumption and higher QoE values by roughly 15\% and 25\%, respectively. On the other hand, in \cite{lu2021adaptive}, the authors proposed a three-layer model: a radio access layer contains mobile devices with limited computing resources, a DT layer where base stations, equipped with MEC servers, execute computation tasks and maintain DTs, and a cloud layer contains cloud servers with large computing and storage resources. Moreover, an A2C-based DRL approach is proposed to minimize average system latency when selecting base stations as DT servers. Additionally, to quickly adapt the DT model when users move between servers, a transfer learning model is used. Simulation results show that the proposed model can reduce the total system cost by up to 99\%, compared to baseline models. Also leveraging MEC, the authors in \cite{yang2023edge} present a framework for energy-efficient video analytics for DT construction in 6G networks. Particularly, the framework uses mobile devices to capture video frames, and edge nodes for computation offloading. Moreover, a mixed integer nonlinear programming (MINLP) problem is formulated to minimize energy consumption while maximizing video analytics accuracy, subject to latency constraints. To this end, a Multi-Agent Deep Deterministic Policy Gradient (MADDPG) algorithm, which allows for centralized training and decentralized execution of the video analytics tasks, is proposed to solve the optimization problem. Simulation results show that, using a reward function, MADDPG can outperform different benchmark algorithms by 10\%-40\%.

Besides 6G network, how edge computing supports DT in human-related applications is the focus of \cite{wang2024human}, \cite{xiang2023realizing}, and \cite{martinez2019cardio}. In \cite{wang2024human}, the authors propose a mobile edge computing MEC-based DT deployment scheme in the smart home domain with a focus on human activity recognition and sensor update inference. Particularly, this scheme leverages a cloud platform to collect data from sensors to create DTs, and edge servers as intermediaries to preprocess and manage sensor data locally. Moreover, a DRL-based data upload timing model is developed to improve the DT fidelity i.e., how accurately the virtual model represents actual sensor changes, while reducing communication costs. Simulation results show that this approach maintains a higher accuracy by 6.51\% and 1.58\% on different datasets compared to a baseline approach. Another application i.e., Human Digital Twin (HDT) is presented in \cite{xiang2023realizing}. To that end, the authors introduce an immersive communication framework for HDT leveraging edge computing and tactile Internet. The framework, namely IC-HDT-ECoTI, aims to enhance HDT applications by providing strong interactions and extremely immersive QoE between physical twins (PTs) \cite{xiang2023realizing} and virtual twins (VTs) \cite{xiang2023realizing}. In particular, the system architecture of IC-HDT-ECoTI has an Edge Interaction Domain, which acts as a bridge between a Physical Master Domain \cite{xiang2023realizing} and a HDT Domain \cite{xiang2023realizing}. The Edge Interaction Domain enables data offloading to edge servers, thereby reducing communication bandwidth and service delay. Simulation results show that, compared to a baseline framework, IC-HDT-ECoTI can reduce the delay by up to roughly 50\%, and offer higher QoE for HDT in practical implementation. An application of DT for human organ replication is introduced in \cite{martinez2019cardio}. Specifically, the authors propose an architecture, namely Cardio Twin, for ischemic heart diseases (IHD) detection by replicating a DT of the human heart. Moreover, Cardio Twin is designed to run on edge devices such as smart phones to provide continuous monitoring and data collecting, high responsiveness, and privacy. Additionally, AI or ML models can be used to detect abnormalities and make decisions based on data collected from the edge devices. Simulation results show that Cardio Twin can detect the IHD with an accuracy of mostly 85\% using a CNN model.

\begin{figure}[!]
	\centering
	\includegraphics[width=\columnwidth]{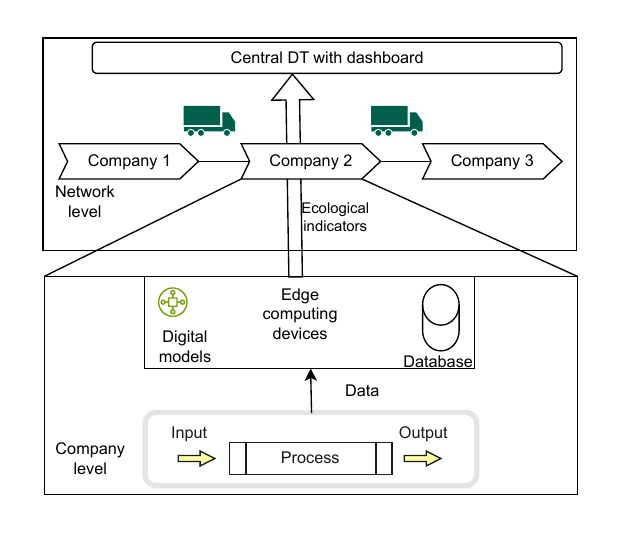}
	\caption{Edge-based DT framework in cross-company production networks.~\cite{glatt2021edge}}
	\label{EdgeBasedDT}
\end{figure}

DT applications in monitoring industrial process of manufacturers are discussed in \cite{ouahabi2021distributed}, and \cite{glatt2021edge}. Particularly, in \cite{ouahabi2021distributed}, the authors propose a distributed DT three-layer architecture for shop floor monitoring based on edge-cloud collaboration. The first layer, namely physical layer, collects data from shop floor sources such as sensors, RFID tags, videos, images. The second layer, namely the edge layer, performs data preprocessing, executes local DT tasks, thereby improving real-time capabilities and reducing cloud computing load. Lastly, the cloud layer stores processed data, performs global DT tasks, and manages model replication, training, and optimization for both edge and cloud layers. Moreover, a microservice architecture is recommended to develop the DT systems to provides benefits such as elasticity, modularity, reducing risk of cascading failures, better resource orchestration, and potential for containerization. Possible manufacturing application scenarios of DT, such as equipment health management, production scheduling, production control and optimization, and quality control, are also discussed. Addressing another challenge, in \cite{glatt2021edge}, the authors propose an edge-based DT framework to assess and ensure ecological sustainability in cross-company production networks. In particular, the framework consists of a company level and a network level, as illustrated in Fig.~\ref{EdgeBasedDT}. In the company level, edge devices at each manufacturer computes sustainability indicators for their specific processes, thereby keeping sensitive data within each company. In the network level, non-sensitive sustainability indicator data from all companies are aggregated by a central DT, thereby providing an overview of the entire supply chain. This approach could reduce latency and bandwidth usage compared to cloud-based alternatives, especially for data-intensive applications like video processing.

In \cite{girletti2020intelligent}, the authors propose an Edge-based DT solution for robotic systems which aims to move the DT's computational needs from being on the robot or in a cloud to being distributed across the network. In particular, the system relies on an Intelligent 5G Edge and Fog Infrastructure, which combines cloud-to-things continuum to spread computation at closer distant to the robot, 5G connectivity for real-time data transfer between robots and edge resources, and distributed intelligence capabilities to incorporate AI and ML throughout the infrastructure. Additionally, the DT is based on the Robot Operating System, which consists of two main components: a computational stack to define the robot's behaviour and control, and an analytics stack leveraging AI/ML for movement prediction and optimization. Simulation results show that the synchronization accuracy between a physical robot and its DT is mostly 98\% when the controlling application is deployed on edge nodes, compared to unpredictable movement when the application is deployed on cloud. 

In \cite{van2022edge}, the author proposes a novel DT scheme supported metaverse leveraring MEC and ultra-reliable low-latency communications (URLLC). Particularly, the IIoT devices are connected to an access point (AP) via URLLC links, allowing DTs to be replicated in the metaverse. Moreover, edge servers are used to optimize task offloading and to provide caching capabilities. The end-to-end latency optimization problem is divided into three sub-problems (caching policy, offloading policy, joint communication and computation) and is solved using an alternating optimization (AO) based algorithm. Simulation results show that the proposed scheme can improve the quality of the DT with lower latency and higher reliability in a metaverse. 

In edge computing, AI/ML-based DT represents a significant advancement to monitor, manage, and optimize physical systems and processes in edge networks through using system replica either at the central server or edge server. In this case, edge devices can collect and pre-process data from various sensors and IoT devices and then deploy AI/ML models using DT to analyze the data locally. The central server can also implement the centralized learning process using the DT to help the edge devices in monitoring and optimizing the real networks.

For example, \cite{Dong2019Deep} presents a DL-assisted mobile edge computing scheme that involves training the DL algorithm offline at a central server using a DT of the actual network environment. The objective is to decrease the normalized energy consumption through the optimization of user association, resource allocation, and offloading likelihood, while meeting the quality-of-service requirements. Here, the mobility management entity (MME) can manage the user association in real-time from the pre-trained deep neural network (DNN), while each access point (AP) can orchestrate the rest. To this end, the DT can dynamically monitor the fluctuations in real networks and adjusts the DNN accordingly, taking into account that real networks are dynamic. This DT creates a replica adopting the network topology, the queueing and channel models, and basic principles of the utilized real networks. From the simulation results, the proposed framework can save normalized energy consumption up to 89\% compared to other conventional methods while decreasing the computational complexity. Another DL-based edge-fog-cloud computing scheme to enhance the security of cloud storage is discussed in~\cite{Lv2022Edge}. Specifically, smart machines as the edge devices collect and pre-process data including data storage and transmission. Then, a DT is employed to develop a virtual replica in the physical world using 3Dmax at the cloud server, enabling the simulation of the data-driven behavior of the machines. This DT can simulate a network intrusion detection leveraging a DNN model. Through the experiments using a two-layer cloud database model and Amazon Relational Database Service, the proposed scheme can obtain success rate between 0.9 and 1.0 of the database load/frequency.

The authors in~\cite{Sanchez2024Building} then adopt a DT mechanism to create a digital duplicate of Espinardo's campus at the University of Murcia in Spain utilizing ETSI next-generation service interfaces-linked data open standard. The initial stage involves the utilization of the current sensor network deployment and the on-going development of the Fog-Edge-Cloud computing infrastructure, which is built upon a 5G private network. Specifically, the DT-based smart campus includes the DT of 23 buildings working as edge nodes, each with chosen sets of data attributes. This DT is implemented by integrating sensor data with external open data sources, analytics models, and the information processed by these analytics. Using linear ML models, i.e., logistic regression (LR), decision tree (DT), random forest (RF), and support vector machine (SVM), for building occupancy prediction of 19 buildings in the campus, the proposed mechanism can obtain minimum root mean squared error (RMSE) and mean absolute error (MAE) between $3 \sim 53$ and $1 \sim 14$, respectively when decision tree is used.

\begin{figure}[t]
	\centering
	\includegraphics[scale=0.45]{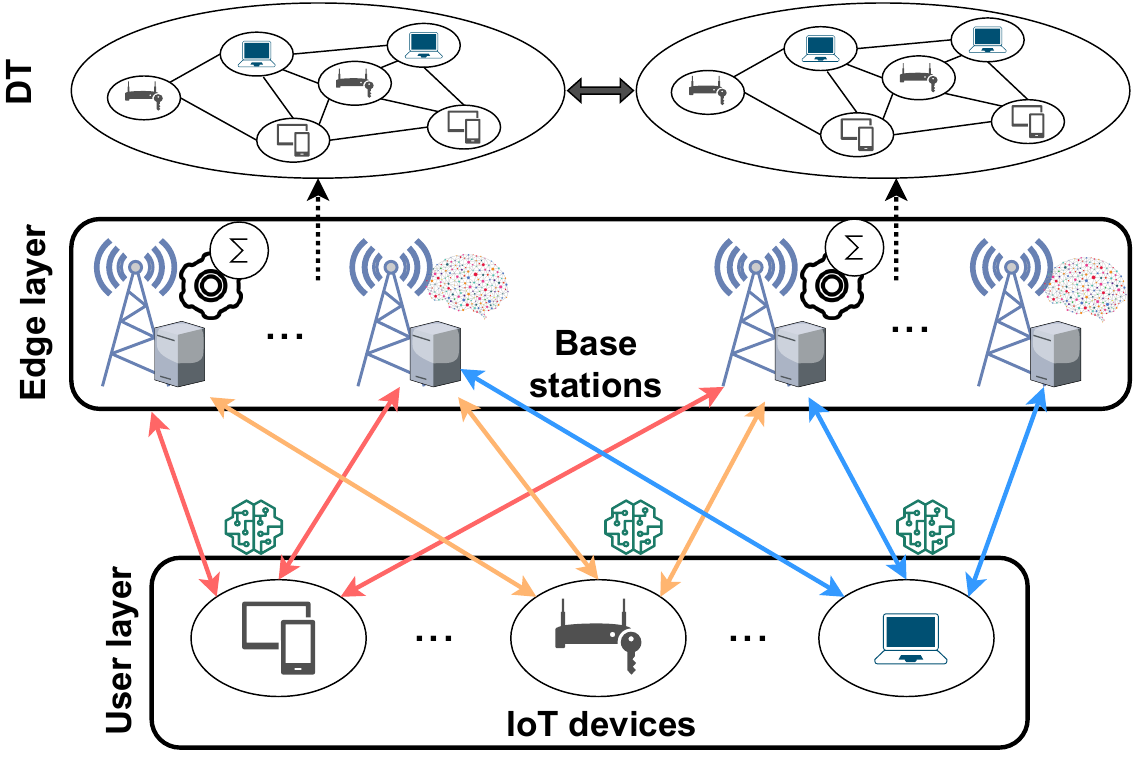}
	\caption{The FL-based framework with user, edge, and DT layers.}
	\vspace{-0.5em}
	\label{fig:ai_ml2}
\end{figure} 

To provide a collaborative learning without data sharing, the work in~\cite{Lu2021Comm} presents the architecture of DT edge networks (DITEN), which combines DT technology with edge computing to create a streamlined connection between IoT devices and cyber systems for IIoT applications. In this case, an FL framework with asynchronous model update formulated as an optimization problem is utilized to address the communication overhead problem. The problem is subsequently divided into two subproblems and solved using DNN model. As illustrated in Fig.~\ref{fig:ai_ml2}, the DITEN has three layers including user layer with IoT devices, edge layer with mobile edge computing server embedded at base stations, and DTs which are deployed within the base stations. From the simulations, the DITEN can obtain more than 90\% of accuracy for various number of FL participants and a lower time cost by 60\% compared to the baseline algorithm. Then, the authors in~\cite{Yang2023Hyper} extend the above work using a hierarchical FL approach that incorporates a hypernetwork (HN) algorithm, i.e., HFedHN, to mitigate the impact of non-IID data across devices. This HN is a lite version of a main neural network that generates weight parameters and uses the main network's weight structure as input and outputs them. Specifically, the system utilizes HNs in the lower layer to produce the parameters of local model for the devices. In the higher layer, the HNs are updated by combining their model parameters. The system then separates the number of parameters sent by the upper and lower levels, resulting in improved communication efficiency and reduced calculation costs without sacrificing accuracy. Using five classification datasets, the proposed system can outperform other baseline FL methods through achieving more than 90\% of test accuracy.

In another work, a DT-based multi-FL service mechanism to tackle the dynamicity in resources and mobile users for mobile edge computing networks is studied in~\cite{Zhang2024Digital}. Particularly, a DT is first applied to improve the scheduling/selection of devices and allocation of mobile edge computing resources, aiming at maximizing the overall utility of FL services. Here, the DT can provide seamless synchronization between the mobile edge computing networks and the twin of multiple FL services. This is realized by applying LSTM model to forecast future network circumstances according to the accumulated knowledge stored in the DT. Using the network insights from the DT and the behavior of FL participants, the FL device selection and bandwidth allocation strategy during the FL training process can be determined. To further intelligently allocate bandwidth, a DRL approach with the Deep Deterministic Policy Gradient (DDPG) algorithm is introduced. This DDPG utilizes the knowledge gained from the network analysis conducted by the DT and actively determines the allocation of bandwidth in each communication round, allowing it to adjust to the changing conditions of the mobile edge computing networks. The performance evaluations show that the system utility of the proposed approach can be improved by 49.8\% compared to a baseline one.

The works surveyed in this section is summarized in Table.~\ref{TableCloudEdgeDTReview}

\begin{table*}[]
	\caption{Summary of Cloud/Edge computing approaches in DT applications.}
	\label{TableCloudEdgeDTReview}
\centering
\resizebox{\textwidth}{!}{%
\begin{tabular}{|l|l|l|l|}
\hline
Reference                                       & Architecture                              & Objectives                                                     & Approaches                                                                                                                                                                                                                                               \\ \hline
\cite{mohamed2023leveraging}   & \multirow{8}{*}{Cloud Computing}          & Elderly healthcare services                                    & \begin{tabular}[c]{@{}l@{}}Cloud-based DTs to replicate the health status of patients\\ Cloud platform performs advanced analytics\end{tabular}                                                                                                          \\ \cline{1-1} \cline{3-4} 
\cite{liu2019novel}            &                                           & Elderly healthcare services                                    & \begin{tabular}[c]{@{}l@{}}Create a DT of the healthcare ecosystem by integrating physical objects, virtual models, and healthcare data\\ Leverage IoT for data collection, cloud computing for processing, advanced analytics for insights\end{tabular} \\ \cline{1-1} \cline{3-4} 
\cite{dang2021cloud}           &                                           & Health monitoring                                              & \begin{tabular}[c]{@{}l@{}}Create a high-fidelity digital replica of the physical entity\\ Use cloud computing to efficiently monitor real-time data and enable proactive maintenance\end{tabular}                                                       \\ \cline{1-1} \cline{3-4} 
\cite{zheng2021towards}        &                                           & Health monitoring                                              & Formalize a healthcare monitoring model over the DT cloud platform                                                                                                                                                                                       \\ \cline{1-1} \cline{3-4} 
\cite{zhang2022practical}      &                                           & Grid management                                                & \begin{tabular}[c]{@{}l@{}}Integrate power system components with cloud platforms\\ Use DTs to model and monitor grid behavior in real-time\end{tabular}                                                                                                 \\ \cline{1-1} \cline{3-4} 
\cite{mikhailov2022new}        &                                           & DTaaS for IIoT systems                                         & Integrates cloud-based DTs with IIoT sensors and devices to provide real-time data analysis and simulation                                                                                                                                               \\ \cline{1-1} \cline{3-4} 
\cite{pan2021digital}          &                                           & Production logistics system control                            & Form a dynamic multi-level distributed decision-making system, optimizing effectiveness and cost-efficiency                                                                                                                                              \\ \cline{1-1} \cline{3-4} 
\cite{dai2022adaptive}         &                                           & Adaptive network management and policy scheduling              & \begin{tabular}[c]{@{}l@{}}Use DTs to facilitate adaptive management\\ Develop DRL-based offloading scheme to minimize total offloading latency\end{tabular}                                                                                             \\ \hline
\cite{karobi2024ecoedgetwin}   & \multirow{8}{*}{Edge Computing}           & DT in 6G network                                               & Develop anA2C algorithm for optimizing task offloading and energy management                                                                                                                                                                             \\ \cline{1-1} \cline{3-4} 
\cite{lu2021adaptive}          &                                           & DT in 6G network                                               & \begin{tabular}[c]{@{}l@{}}Develop an A2C-based DRL approach to minimize average system latency\\ Use a transfer learning model is to adapt the DT model when users move between servers\end{tabular}                                          \\ \cline{1-1} \cline{3-4} 
\cite{yang2023edge}            &                                           & Video analytics for DT construction in 6G networks             & Develop a MADDPG algorithm to solve MINLP optimization problem                                                                                                                                                                                           \\ \cline{1-1} \cline{3-4} 
\cite{xiang2023realizing}      &                                           & HDT                                                            & Leverage edge computing to enable data offloading to edge servers, thereby reducing communication bandwidth and service delay                                                                                                                            \\ \cline{1-1} \cline{3-4} 
\cite{martinez2019cardio}      &                                           & DT of the human heart                                          & \begin{tabular}[c]{@{}l@{}}Use edge devices to continuously monitor and collect data\\ Develop AI or ML models to detect abnormalities and make decisions\end{tabular}                                                                                   \\ \cline{1-1} \cline{3-4} 
\cite{glatt2021edge}           &                                           & Ecological sustainability in cross-company production networks & Use a central DT to provide an overview of the entire supply chain                                                                                                                                                                                       \\ \cline{1-1} \cline{3-4} 
\cite{girletti2020intelligent} &                                           & DT for robotic systems                                         & Leverage edge computing to distribute computation closer to robots                                                                                                                                                                                       \\ \cline{1-1} \cline{3-4} 
\cite{van2022edge}             &                                           & DT for metaverse                                               & Use edge servers to optimize task offloading                                                                                                                                                                                                             \\ \hline
\cite{han2022cloud}            & \multirow{5}{*}{Edge-cloud collaboration} & Distributed energy resources management                        & \begin{tabular}[c]{@{}l@{}}Use edge computing for data collection and process\\ Use cloud for more intensive computations and data storage\end{tabular}                                                                                                  \\ \cline{1-1} \cline{3-4} 
\cite{stergiou2022digital}     &                                           & Intelligent battery management system                          & \begin{tabular}[c]{@{}l@{}}Leverage edge nodes to bridge the physical and digital worlds\\ Use cloud-based big data platform for storage and computation\end{tabular}                                                                                    \\ \cline{1-1} \cline{3-4} 
\cite{huang2021digital}        &                                           & Health monitoring and anomaly prediction in industrial systems & Use edge/cloud network to achieve high-performance anomaly detection                                                                                                                                                                                     \\ \cline{1-1} \cline{3-4} 
\cite{wang2024human}           &                                           & Human activity recognition and sensor update inference         & \begin{tabular}[c]{@{}l@{}}Leverages a cloud platform to collect data from sensors to create DTs\\ Use edge servers to preprocess and manage sensor data locally\end{tabular}                                                                            \\ \cline{1-1} \cline{3-4} 
\cite{ouahabi2021distributed}  &                                           & Shop floor monitoring                                          & \begin{tabular}[c]{@{}l@{}}Use edge nodes to preprocess data and execute local DT tasks\\ Use cloud computing to store data and perform global DT tasks\end{tabular}                                                                                     \\ \hline
\end{tabular}%
}
\end{table*}

\section{Digital Twin for Applications }\label{sec:aiml}


The evolution of 6G and beyond networks is driving the development of a broad range of applications that increasingly rely on DT technology to achieve intelligent, adaptive, and efficient operation. In this context, AI/ML plays a critical role as core enablers for applications in NDTs for 6G and beyond networks including blockchain-based frameworks, anomaly detection schemes, healthcare, manufacturing, and vehicular networks. To this end, AI/ML-aided DT can provide significant benefits in optimizing network performance and improving user experience through network environment simulation and performance prediction, reducing downtime using proactive maintenance, enhancing system security by anomaly detection approach, and increasing cost efficiency through resource optimization and automated management. Integrating AI/ML-based digital twins into 6G networks can completely revolutionize their design, operation, and maintenance. Additionally, this integration guarantees that 6G and beyond networks are more resilient \cite{tran2024deep}, efficient, and user-centric, as discussed in the following existing scenarios.

\subsection{Blockchain}

Blockchain-enabled AI/ML framework has been considered as an advanced integrated technology that can enhance the efficiency, security, and functionality for digital twin (DT) networks. Specifically, the integration of blockchain and AI/ML approach, e.g., FL, within digital twin frameworks offers several benefits. First, blockchain's secure and immutable ledger guarantees that data fed into AI/ML algorithms in digital twin remains trustworthy and private. Second, AI/ML algorithms have the capability to analyze large quantities of encrypted data stored on the blockchain to provide more accurate predictions/decisions. Third, AI/ML-driven decisions may be used to automate processes in digital twin networks using smart contracts on the blockchain in which the transaction would be securely stored. Finally, the decentralized nature of blockchain makes the digital twin system more resilient to failures and cyberattacks, while AI/ML algorithms make it more adaptable to real-world developments.

For example, the authors in \cite{Qu2022FedTwin} propose FedTwin that combines blockchain technology with dynamic asynchronous FL, aiming at providing privacy and decentralization for digital twin (DT) networks. This work addresses several challenges including the existing DT networks which utilize centralized computation, the presence of data fabrication and privacy leakage, and lack of incentive scheme. Particularly, a blockchain-based consensus algorithm based on Proof-of-Federalism is designed at all training DTs. Here, the blockchain-based consensus process refers to the local model training of FL that stores the local trained parameters in blocks and then sends them to the network of blockchain. In this phase, a generative adversarial network with differential privacy is applied at each DT to preserve the privacy of local models. Additionally, each DT must execute a smart contract that uses the Isolation Forest algorithm to detect faked local model parameters and eliminate them. Then, in the aggregation phase, the training DTs can generate global model parameters using the selected local models via an enhanced Markov decision process (to comply with a dynamic asynchronous aggregation mechanism) and save them in a candidate block. The candidate block with the most selected global parameters is then cross-verified and added to each DT's ledger for the next FL process. The consensus process of FL repeats until a defined threshold is reached or the FL convergence is achieved. To this end, the training DTs can obtain rewards as incentives for participating in the blockchain consensus and FL process. This FedTwin can achieve local test accuracy up to 96\%, 5-8\% higher than that of other baseline FL methods, after 50 rounds using CNN and MLP algorithms.

The extended blockchain approach using a hierarchical FL-based DT framework for Industrial IoT (IIoT) is investigated in~\cite{Aloqaily2023Reinforcing}. This work integrates the DT into a cyber-physical system (CPS) to offer flexible and dynamic configurations for collaborative CPS. Specifically, a two-stage FL training process using authentication and verification of the blockchain is proposed to provide accuracy for the IIoT devices' behaviors. IIoT-based machines in the factory are clustered with a cluster coordinator for each cluster in the first stage. This cluster coordinator leads its cluster for the local training processes at each local machine. It also creates some DTs that help some incapable machines, e.g., machines with limited computation resources, to perform the local training processes. The trained local models from all machines and DTs are then aggregated by the cluster coordinator as the cluster global model. In the second stage, the training processes occur between the cluster coordinators, validator nodes, and the cloud server. In this case, the validator nodes will cross-verify and validate the received global models from cluster coordinators with the shared global model before aggregating them using FedAvg. This verified aggregated model is stored at a block and added to the blockchain. Since all blocks are uploaded to the blockchain, this allows smart contracts to issue transactions between clusters with different criteria. From the experiments using CNN, it is shown that the FL can obtain similar accuracy with the centralized CNN at 90\% with small block size at 2MB.

\begin{figure}[t]
	\centering
	\includegraphics[scale=0.48]{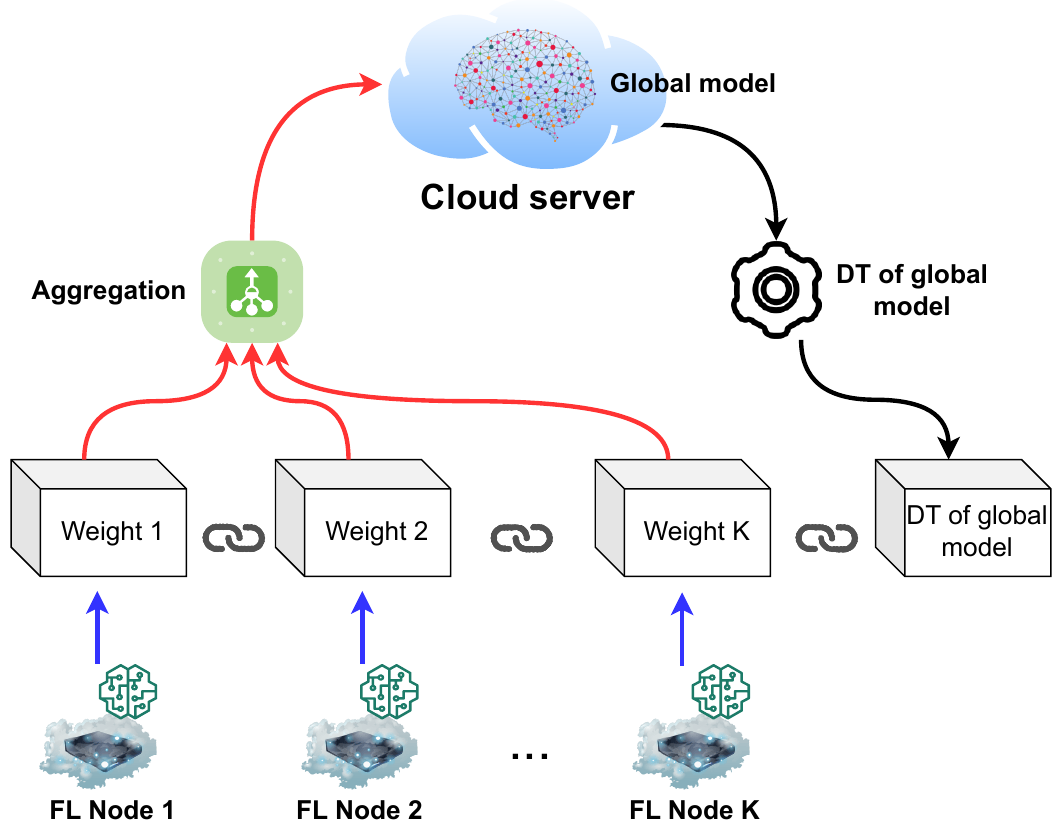}
	\caption{DT-based SFL framework with blockchain technology.}
	\vspace{-0.5em}
	\label{fig:ai_ml1}
\end{figure} 

The use of blockchain to protect DT networks and user privacy from malicious communication activities using distributed learning is also discussed in~\cite{Lv2023Blockchain} and~\cite{Prathiba2024Fortifying}. In~\cite{Lv2023Blockchain}, the authors focus on designing secure FL for data protection and reliability in the IoT environment while implementing digital resource allocation using a broad learning with federated continuous learning (BL-FCL) algorithm. Specifically, a three-layer DT service model is developed to realize the DT network with the blockchain. Here, the model primarily utilizes blockchain and smart contract technologies to guarantee the security of data storage and connectivity within the blockchain. This data blockchain is exchanged between the data owner and the data transmitter. Through the scenario of data poisoning attack detection, the proposed framework can achieve a higher prediction accuracy by 20\%-60\% compared with the conventional federated averaging approach. Then, in~\cite{Prathiba2024Fortifying}, the authors present a blockchain-based secure FL (SFL) using DT to address data dan model poisoning as well as Sybil attacks, i.e., a type of attack by creating many fake IDs, in decentralized IIoT environments. Specifically, the DT works as a secure scheme that protects the trained models during aggregation step. To enhance the security of IIoT machines, non-fungible tokens (NFTs) that are stored on a blockchain to certify authenticity and ownership are utilized to mitigate Sybil attacks through FL distributed nodes. As illustrated in Fig.~\ref{fig:ai_ml1}, the typical FL process happens with the local training and aggregation processes. Here, the blockchain works as a protected and fixed ledger in the FL process.  This stores each model updates' transaction including weights from the local training process at each FL node. Then, in the aggregation process, the DT can duplicate the global model as the reference. In this case, the system can regularly compare the updated global model with the DT as to detect the data as well as model poisoning and return to the current DT state when such anomalies occur. From the experiments, the proposed SFL can reach loss reduction up to 200\% compared to the baseline FL approaches and achieve 97\% accuracy In IIoT environments.

For mobile networks, the use of AI/ML-based DT with blockchain can optimize and improve network performance as well as user experience with the existence of data privacy and security protection. In~\cite{Mu2023Digital}, a communication-assisted sensing situation using federated transfer learning (FTL) in DT-enabled time-varying mobile networks is investigated. In this context, the transfer learning reduces the transfer communication overhead and the DT facilitates prototype, testing, and optimization, enabling efficient virtual modeling and practical application guidance in the mobile networks. Specifically, two communication-assisted sensing schemes, i.e, centralized and decentralized of FTL, are employed in the DT, aiming at enhancing mobile network communication efficiency. To this end, the centralized FTL is first executed to perform feature extraction and train data on the central server. To ensure data safety, a decentralized FTL with blockchain is then used. The feature extraction module is created by combining the blockchain sharing and local models. This blockchain technology is combined with a new detection method that relies on the similarity of models. For the blockchain mechanism, each participant corresponds to a single miner where the local model's parameters are uploaded to the miner. Subsequently, the miners disseminate the parameters as transactions to the remaining miners, which are then stored in their respective trading pools, and broadcast transactions to other participants in the blockchain. When a miner receives a transaction information broadcast from another miner, it checks the transaction's validity and puts the parameters in the trading pool. Upon the generation of a new block, the miner downloads the new block to update the global model. Then, the participant initiates another round of local model training. Meanwhile, for the detection method, if there is a discrepancy in model similarity between the present time and the next time, it indicates the presence of a malicious node in the system. In such cases, the malicious node is removed from the FL training. Using a MSTAR dataset for SAR ground target recognition in non-IID manner, the proposed framework can provide a higher average accuracy 97.5\% and faster convergence than conventional FL with federated averaging method.

\subsection{Health System}

The DT has been considered as a promising solution to revolutionize the operation of digital healthcare systems. Specifically, the DT in health applications can provide a virtual replica or model through enabling proactive and personalized patient care, simplifying health monitoring, optimizing health delivery, and fostering innovation in medical research and development. In AI/ML area, the DT can be implemented in health system to build a detailed model of the patient or structure health operation. This model captures not only the current state but also predicts future states based on different scenarios and interventions. The DT can also continuously monitors and updates the health real-time monitoring based on incoming data from the actual patients or structures. This allows for early detection of health anomalies. Moreover, the DT can be used to simulate different treatment strategies, predict outcomes, and optimize care plans tailored to individual patients/structures.

In \cite{Elayan2021Digital}, the authors study DT and DL-based heart disease and heart problem detection using an electrocardiogram (ECG) heart rhythms model. Specifically, three-step framework is developed consisting of processing and prediction step, monitoring and correction step, and comparison step. In the first step, the framework collects patient data by using IoT sensors and then stores the real-time data including patient body and health information to the temporary cloud server. After the pre-processing and DL process using CNN and LSTM, the system can predict and detect the health anomalies. The prediction/detection results are then saved at the result database. In the second step, the healthcare professionals can verify and validate the given prediction/detection results as well as provide feedback and optimize the current model. In the final step, the comparison between the real prediction/detection and the DT of similar case patients are incorporated to enhance the DT model for the next use to continuously monitor and detect health anomalies. Through performance evaluation using ECG datasets, the framework can achieve the accuracy of 96\% and 97\% using CNN and LSTM, respectively. To increase both speed and accuracy for disease assessment, the authors in~\cite{Yu2023FMCPNN} present a factorization machine combine product-based neural network (FMCPNN)-based framework. This framework can enhance product-based neural network (PNN) performance from poor capability for generalization due to insufficient interaction between low-order features by employing additional second order interaction. The simulations results show that the proposed framework can provide the lowest loss at 0.36 and the highest accuracy at 84\% compared with other baseline ML and DL methods.

Next, the authors in~\cite{Tai2022Digital} propose a DT-based Internet of Medical Things (IoMT) framework for telemedical simulation to remotely perform lung cancer surgery. This framework combines a generative adversarial network (GAN), mixed reality, and 5G cloud computing system. In particular, the lung cancer data are first collected and pre-processed from patients using IoMT devices via 5G networks. To build accurate prediction model, a new robust auxiliary classifier GAN (rAC-GAN)-based smart network is developed. This prediction accuracy is then enhanced during the remote surgery using DT. This can be achieved through utilizing a real-time operating room viewpoint, where a surgical navigation image is projected onto the surgeon's helmet through the application layer, employing 5G-based mixed reality. Based on the performance results, the proposed IoMT framework can provide accuracy of 92\%, 20\% better in average compared with KNN, SVM, and DNN approaches. Then, the work in~\cite{Khan2023ANovel} introduces a new Respiration DT (ResDT) model that utilizes Wi-Fi Carrier State Information (CSI) and ML algorithms to monitor and classify patient respiration into binary and multi-class categories. Particularly, the respiration data is first collected using an ESP32 Wi-Fi sensor equipped with Wi-Fi CSI. Then, the Patient's Breaths Per Minute (BPM) is determined by analyzing sensor data using various signal processing techniques, including denoising (i.e., smoothing and filtering) and dimensionality reduction, i.e., Principal Component Analysis (PCA), SVM, Empirical Mode Decomposition (EMD), and EMD-PCA. Various filters and dimensionality reduction techniques are compared to accurately estimate BPM. Based on the experiments, the PCA can achieve more effective dimensionality reduction, resulting in 87.5\% accurate BPM values. Additionally, the fine tree algorithm is the best option in the ResDT model by providing accuracy of 96.9\% and 95.8\% for multi-class classification and binary-class classification, respectively, outperforming the other supervised classification algorithms.

\begin{figure}[t]
	\centering
	\includegraphics[scale=0.42]{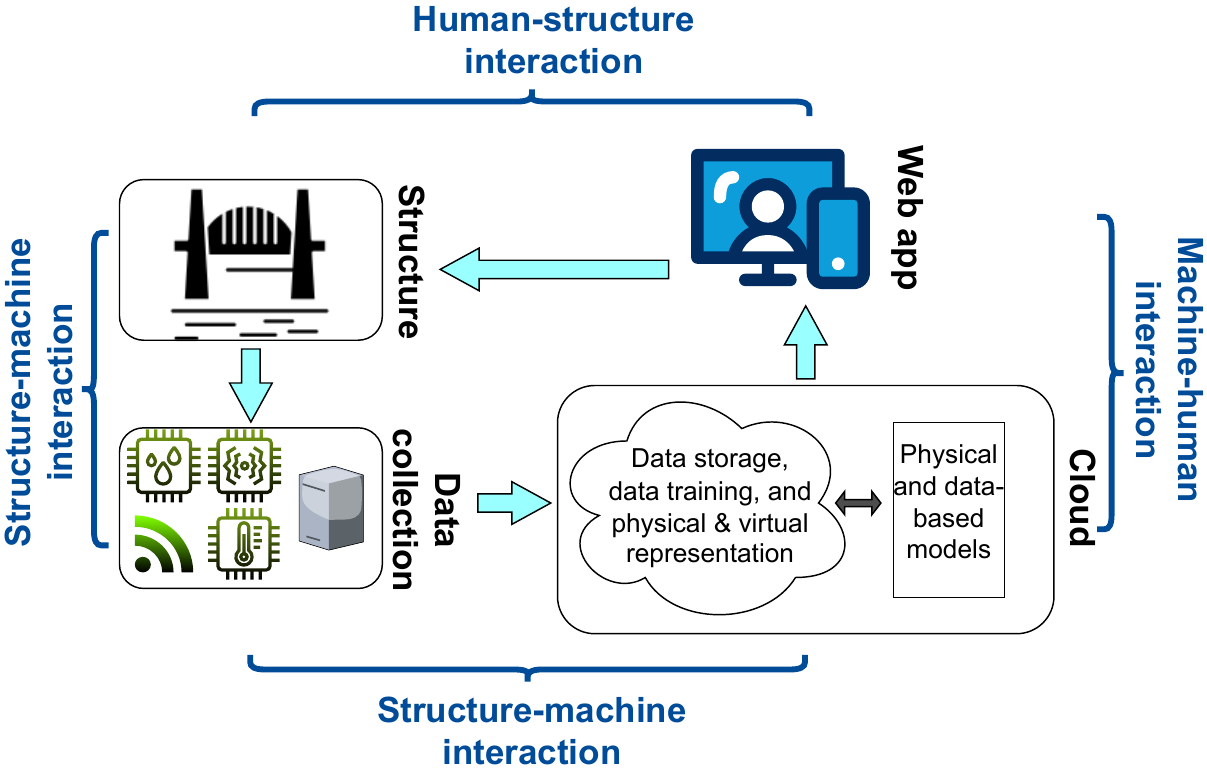}
	\caption{The DT-based structural health monitoring framework with structure-machine-human interaction.}
	\vspace{-0.5em}
	\label{fig:ai_ml3}
\end{figure} 

Another work that presents DL-based structural health monitoring using DT in the cloud server is investigated in~\cite{Dang2022Cloud}. Here, the system orchestrates data interaction among physical infrastructure, digital model, and human interventions through cloud computing and web interface. To accurately detect damages on the structures, several phases including structure-machine interaction, machine-human interaction, and human-structure interaction as illustrated in Fig.~\ref{fig:ai_ml3}. For the structure-machine interaction, data collection using IoT sensors, accelerometers, and cameras is executed for the DL process. In the machine-human interactions, the collected data are trained using CNN models at the cloud server. In this case, the DT-based virtual structure according to the physical structure is built to replicate the current behavior and predict the future behavior of the real structures. For the last phase, the web interface is deployed to provide valuable information for the structure professionals regarding the current status of structures based on the DL process. From the experiments, it is demonstrated that the CNN model can provide the damage detection accuracy up to 92\% in the scenario of model and real bridge structures.   

\subsection{Manufacturing}

The use of AI/ML-aided DT network can help the manufacturing systems including IIoT to optimize operations, enhance productivity, improve quality, and reduce downtime. Through a virtual model that replicates the physical manufacturing processes, machinery, and factory environment in real-time, the manufacturing systems can conduct real-time simulation, prediction, and decision making, with the help of AI/ML approaches.

For example, the authors in~\cite{Mortlock2022Graph} introduce a cognitive DT for Industry 4.0 which enable manufacturers to strategically leverage implicit information derived from the experience of current industrial systems and provide greater independent decision-making and control, while enhancing performance on a large scale. To do so, a graph learning is used to improve cognitive capabilities in the manufacturing DT, especially throughout the product design phase of manufacturing. This graph learning model has the ability to conduct cognitive methods including learning, perception, attention, memory, reasoning, and problem solving through three-step processes, i.e., graph formation, graph operations, and learning objective. In this case, the graph formation handles perception, memory, and reasoning, while the graph operations control attention and learning processes. Finally, the learning objective performs the problem solving through data prediction. It is observed that the proposed DT with a structural graph CNN (SGCNN) can obtain the accuracy at 91\% for the sub-graph classification.

Another use of DT to manage complex machinery for modern manufacturing industry is proposed in~\cite{Ren2022Machine}. The primary goal is to create a DT for complex machinery and oversee its whole lifespan through a mathematical model that is influenced by both models and data. By using this approach, one may have a more comprehensive understanding, closely monitor, accurately track, and effectively control the present condition of complex machinery. In this work, three layers are developed, i.e., data layer, edge-cloud computing layer, and service layer. Specifically, data layer can perform data collection from physical entities and data transmission. The edge-cloud computing layer consists of modeling algorithm and computing reasoning modules for edge-cloud collaboration. Finally, the service layer contains digital virtual entity that offers direct services to reality based on the reasoning results form the previous layer. Using the diesel locomotive dataset with the combination of ML approaches, i.e., Lasso model, SVR, and XGBoost, the proposed framework can achieve the lowest root mean-squared error and mean absolute error at 0.71 and 0.43, respectively.

Additive manufacturing is also considered as a potential manufacturing process that can be visualized using ML/AI-based DT. In~\cite{Sampedro20233D}, the authors study a DT approach that utilizes the advanced ensemble 3D-AmplifAI algorithm to monitor faults in 3D printer machines a.k.a additive manufacturing. Specifically, the DT system constantly analyzes current temperature levels in real-time and identifies any malfunctions to proactively prevent potential harm to the printer. The 3D-AmplifAI technique uses an ensemble method to integrate many ML models, resulting in improved fault detection capabilities for 3D printers. To this end, the Unity-developed DT environment acts as the intermediate scheme that links the physical printer to the virtual realm. Compared to other individual ML models, the 3D-AmplifAI system can outperform the accuracy at 82.35\%. The above work is expanded in~\cite{Jyeniskhan2023Integrating} through incorporating ML models, Unity, OctoPrint, and Raspberry Pi for real-time control and monitoring. Particularly, the system employs transfer learning, i.e., EfficientDet-Lite, with bi-directional feature pyramid network (BiFPN) to detect defects, attaining an Average Precision (AP) score of 92\%. Meanwhile, the Unity client user interface is designed to enable control and visualization, making it convenient for monitoring the additive manufacturing process.

\begin{figure}[t]
	\centering
	\includegraphics[scale=0.53]{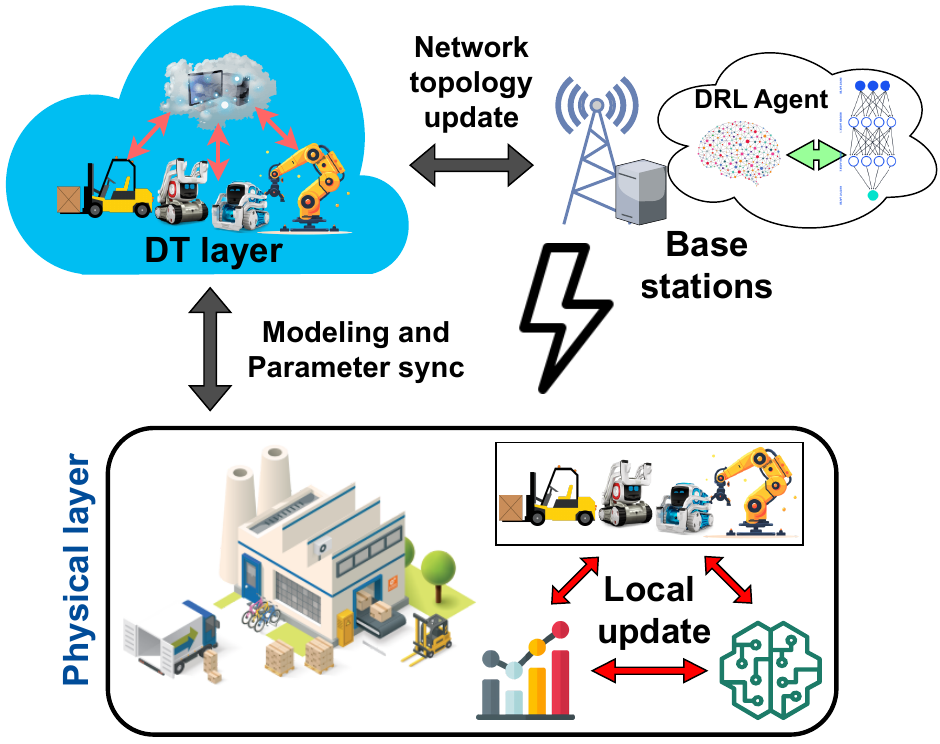}
	\caption{The DT-enabled IIoT framework with FL and DRL approaches.}
	\vspace{-0.5em}
	\label{fig:ai_ml5}
\end{figure} 

To address the issue of resource-constrained IIoT manufacturing machines, the authors in~\cite{Qiao2024Resources} propose a DT-enabled IIoT (DTENI) framework using FL approach in wireless networks as illustrated in Fig.~\ref{fig:ai_ml5}. This framework first utilizes DTs to capture the specific features of industrial machines, allowing for real-time processing and intelligent decision-making. To optimize the efficiency of the FL training processes, the wireless parameters such as CPU frequency, bandwidth, and transmission power are then theoretically analyzed. Using the analysis, the authors formulate the stochastic loss minimization problem of the FL model within a certain resource limitation. By further using the model-free learning capabilities of DRL, the DTENI-assisted DRL is employed for the adaptive adjustment of wireless settings to effectively solve the stochastic optimization problem. Simulation findings show that the proposed framework can significantly reduce communication costs by up to 74.23\%, 69.51\%, and 60.94\% compared to the three baseline schemes.



\subsection{Security}

The virtual replica of DT can also be used to improve threat or anomaly detection, response, and overall security management. To this end, the additional AI/ML approaches can further predict potential threats and provide insights for proactive security measures using real-time synchronization from the DT, i.e., via real-time data from sensors, cameras, access logs, and other security devices to maintain an up-to-date representation.

Anomaly detection is one of the security measures which identifies patterns in data that deviate from the expected or normal behavior due to cyber threats. For example, the authors in~\cite{Castellani2021Real} propose a DT-based anomaly detection system in industrial environments using weakly supervised learning. Specifically, the system utilize a DT to create a training dataset that replicates the normal functioning of the machinery. Additionally, a limited number of abnormal measurements from the actual machinery are included. Here, the abnormal data from the actual machinery and the normal data from the DT are trained using a clustering-based method, i.e., cluster centers (CC) and a neural network structure utilizing Siamese Autoencoders (SAE). Leveraging the trained model, the proposed system is then evaluated against several cutting-edge anomaly detection algorithms using a real-world dataset from a facility monitoring system. Additionally, the study examines the impact of hyperparameters associated with feature extraction and network design. Through the simulations, the proposed method using SAE can obtain the AUC at 87.6\%, the highest AUC compared with that of other unsupervised and weakly-supervised learning methods. Then, the authors in~\cite{Li2023ANovel} introduce an end-to-end anomaly detection which aims to achieve prompt and accurate real-time anomaly detection to address the difficulty of anomaly detection in DT. This is due to the fact that current techniques typically involve multiple stages and need laborious training and detection procedures. In this work, the anomaly detection utilizes a multidimensional deconvolutional network of the CNN model and attention mechanism to identify essential features. The deconvolutional network is used to convert the low-dimensional features of the original data into high-dimensional features, enabling the succeeding network to effectively represent the original data. Meanwhile, the attention mechanism has the capability to acquire knowledge about the key regions that differentiate normal data from abnormal data, thereby significantly enhancing the performance of the model. From the performance evalution, the system can provide 94\% and 92\% of F1-score when SWaT and WADI datasets are used, respectively.

To tackle insider threats, the authors in~\cite{Wang2023DTITD} introduce a novel framework for insider threat detection using DT and DL models with self-attentions. Specifically, the framework first performs insight analysis of users' behavior and entities. To enhance this analysis, contextual word embedding techniques using the Bidirectional Encoder Representations from Transformers (BERT) model is employed. Then, the sentence embedding technique using the Generative Pre-trained Transformer 2 (GPT-2) model is utilized to augment the data and overcome significant data imbalance. Lastly, a temporal semantic representation of users' behaviors to construct user behavior time sequences is adopted. To further quickly identify insider threats, DL models with self-attentions are implemented using simplified transformer model called DistilledTrans, Robustly Optimized BERT Approach (RoBERTa), and a hybrid approach that combines pre-trained models (BERT, RoBERTa) with either a CNN or a Long Short-term Memory (LSTM) network model. Based on the experimental results using the dataset CERT r6.2 augmented by contextual embedding words, the framework with DistilledTrans can achieve the highest accuracy at 99.72\%. To tackle the adversarial attacks in the wireless communication systems, e.g., denial-of-service (DoS) attacks, the authors in~\cite{Boche2024On} propose extension of DL-based DTs, i.e., neuromorphic twins, that can detect the DoS attacks using Blum-Shub-Smale machines, addressing the incapability of Turing machines. In particular, there exists two DTs that are deployed in the system. The first DT represents the physical network containing end users and their corresponding base stations. The second DT is used to provide a digital representation of the attack strategy space. Both DTs are then utilized to detect DoS attacks in the system. Using jamming with full-knowledge and partial-knowledge, the proposed system can detect whether the DoS is possible or not.

The integration of intrusion detection with DL-based DT can also be utilized to enhance network defense capabilities and mitigate security threats in service computing systems. Specifically, the authors in~\cite{Lv2024Secure} propose a network intrusion detection technique using a DL model in a DT. To this end, a trust mechanism utilizing Keyed-Hashing-based Self-Synchronization (KHSS) is first implemented in the cloud service system. This mechanism accurately predicts the security level and identifies intrusions based on the existing malicious attacks, guaranteeing the consistent functioning of the network security defense system. Then, the simulations confirm the viability of the Deep Belief Networks (DBN) model with the cloud trust model. The DBN method enhances the accuracy of identifying unknown samples by 4.05\% in comparison to the Support Vector Machine (SVM) technique. In this case, The DBN algorithm correctly recognized 818 more attacks than the SVM algorithm out of the 20,100 bits of data in the test dataset. Moreover, the KHSS$+$DBN model accurately predicts cloud security states, with a minimal error rate of 1\%$\sim$2\% when compared to the actual states. The authors in~\cite{Kumar2023Blockchain} also introduce a DT-based blockchain and DL system to provide decentralized data processing and learning for intrusion detection in an IIoT network. Specifically, the system first generate a DT model that simulates and reproduces security-critical processes of IIoT. Then, the blockchain technology employs smart contracts to guarantee the data integrity and validity for data transmission. Meanwhile, the DL method is developed to detect any unauthorized access to genuine data obtained from the blockchain. Here, the authors use Long Short Term Memory-Sparse AutoEncoder (LSTMSAE) technique to obtain spatial-temporal representation. The extracted features are further utilized by the proposed Multi-Head Self-Attention (MHSA)-based Bidirectional Gated Recurrent Unit (BiGRU) algorithm to acquire knowledge of distant features and effectively identify attacks. From the simulations, the system can achieve more than 98\% of accuracy or up to 27\% higher than that of other baseline ML methods.

In addition to the above anomaly, threat, DoS attack, and intrusion detections, AI/ML-based DT can be applied for anti-counterfeiting scenarios, e.g., copy detection. In~\cite{Belousov2024AMachine}, the authors introduce a novel system to model a printing-imaging channel by utilizing an ML-based DT for copy detection patterns (CDPs). Specifically, the DT-based anti-counterfeiting system can help to avoid attackers in optimizing the process of estimating digital templates from physical samples in order to conduct copy attacks and creating adversarial samples for the physical domain. Here, the DT is constructed using the TURBO framework, i.e., a generative model with a physical latent space, which is based on information theory. This TURBO framework utilizes variational approximations of mutual information for both the encoder and decoder to facilitate the bidirectional transmission of information. To this end, the system only requires training data consisting of digital templates supplied to a printing device and data obtained from an image device. In the physical channel, an industrial printer first reproduces a digital template on the surface of a digital item in the form of a CDP. Then, a mobile phone user captures an image of CDP. The obtained image is a deteriorated representation of the original digital template. Here, the DT of the printing-imaging channel utilizes DNN to estimate the CDP of the obtained image from the original digital template in a direct and inverse manner. Through using normalized and non-normalized data, the proposed system with TURBO can achieve the lowest mean-squared error at 0.003$\sim$0.004 and 0.004$\sim$0.005, respectively.

\subsection{Vehicular Networks}

Vehicular networks including high-mobility networks and low-mobility networks can take advantages from AI/ML-based DT systems by optimizing and enhancing vehicular network performance, safety, and operational efficiency. The use of AI/ML-based DT can involve a virtual model that replicates the real-world vehicular network's structure, components (vehicles, roadside units, traffic signals), and behavior in real-time. It also can continuously synchronize with real-world data from vehicles, infrastructure, and environmental sensors to maintain an up-to-date representation.

The authors in~\cite{Zhao2023ELITE} introduce an RL-based hierarchical routing mechanism for Intelligent Digital Twin-based Software-Defined Vehicular Networks (IDT-SDVN), entitled intelligent digital twin hierarchical (ELITE) routing. Specifically, there are two main processes are considered in the scheme, i.e, the policy training and generation in the virtual network as well as deployment and relay selection in physical networks. During the policy learning process, several parallel agents are used in DT networks to create numerous single-target policies. For the policy generation process, the learned policies are integrated and new policies are created to meet complicated communication needs. Next, during the deployment process, the most appropriate generated policy is chosen based on the current network conditions and message types. The controller then determines the road path by utilizing the chosen policy and subsequently transmits it to the requester vehicle. Finally, the relay selection process is used to determine relay vehicles in a step-by-step process along the chosen path. Here, the whole vehicular network is considered as the environment, where agents refer to the on-road vehicles, the SDVN controller, or virtual agents that move among the vehicles. The accumulation of reward/experience is derived from the actual routing of data or the network awareness. Through using the RL-based routing via Q-learning algorithm, the agents do not require prior knowledge of the environment and the RL can effectively adjust to highly dynamic virtual networks due to its capability to respond to temporary fluctuations in network conditions. The simulation performances using OpenStreetMap and generated traffic flows by Simulation of Urban Mobility (SUMO) reveal that the proposed scheme can achieve a higher packet delivery ratio by 10\%, a lower end-to-end delay by 20ms, and a lower communication overhead up to 700 messages for various number of vehicles compared to other conventional schemes.

To enable efficient model training and real-time processing in fast mobile networks, the authors in~\cite{Zhou2023Digital} develop a three-layer, i.e., end-edge-cloud, federated reinforcement learning (FRL)-based DT framework. This framework incorporates a DT system that is structured with both edge and cloud components. Particularly, a dual-RL scheme is implemented to optimize client node selection (via deep-Q network) and global aggregation frequency (via DDPG) in the FL process using a cooperative decision-making strategy. This scheme is supported by a two-layer DT system deployed in the edge-cloud that enables real-time monitoring of mobile devices and environmental changes. Then, a model pruning and federated bidirectional distillation (Bi-distillation) scheme is locally employed with the generative adversarial network (GAN) for simple model training. Simultaneously, a global model splitting scheme with a lightweight data augmentation mechanism is applied to optimize the aggregation weights separately based on the encoder and classifier. These two mechanisms work together to effectively reduce overall communication cost and improve the non-IID problem. Using two different real-world datasets, i.e., CIFAR-10 and MNIST, the proposed scheme can obtain up to 18\% and 7\% accuracy improvemet in the non-IID scenario (i.e., high-speed mobile networks) as well as 16\% and 7\% accuracy improvement in the IID scenario for CIFAR-10 and MNIST datasets, respectively.

\begin{table*}[]
	\centering
	\caption{Summary of DT Networks for Applications}
	\label{TableAIMLReview}
	\begin{tabular}{|l|l|l|l|}
		\hline
		Reference & Technology/Scenario & Problem/Purpose & ML/AI Method \\ \hline
		\cite{Qu2022FedTwin} & \multicolumn{1}{c|}{\multirow{5}{*}{Blockchain}}                 & Privacy-preserving and decentralized DT & Asynchronous FL \\ \cline{1-1} \cline{3-4} 	
		\cite{Aloqaily2023Reinforcing} &  & IIoT devices monitoring   
        & Two-stage FL with CNN \\ \cline{1-1} \cline{3-4} 
		\cite{Lv2023Blockchain} &  & Data privacy protection on IoT devices                  
        & BL-FCL \\ \cline{1-1} \cline{3-4}       
		\cite{Prathiba2024Fortifying} &  & Data and model poisoning attacks on distributed IIoT
        & SFL \\ \cline{1-1} \cline{3-4}
		\cite{Mu2023Digital} &  & Communication-aided sensing & FTL \\ \hline
		
		
		\cite{Elayan2021Digital} &  \multicolumn{1}{c|}{\multirow{5}{*}{Health System}}         & Heart disease and problem detection & DL with CNN and LSTM \\ \cline{1-1} \cline{3-4} 
		\cite{Yu2023FMCPNN} &  & Disease diagnosis                            
        & FMCPNN \\ \cline{1-1} \cline{3-4} 
		\cite{Tai2022Digital} &  & Telemedicine simulation on IoMT devices                      
        & rAC-GAN \\ \cline{1-1} \cline{3-4} 
		\cite{Khan2023ANovel} &  & Patient respiration monitoring and decision       
        & PCA, SVM, EMD, and EMD-PCA \\ \cline{1-1} \cline{3-4} 
		\cite{Dang2022Cloud} &  & Structural health monitoring 
        & DL with CNN \\ \hline
        
		
		\cite{Mortlock2022Graph} &  \multicolumn{1}{c|}{\multirow{5}{*}{Manufacturing}}         
        & Cognitive functionalities for manufacturing system & SGCNN \\ \cline{1-1} \cline{3-4} 
		\cite{Ren2022Machine} &  & Complex equipment management                              
        & Lasso, SVR, and XGBoost \\ \cline{1-1} \cline{3-4} 
		\cite{Sampedro20233D} &  & Fault monitoring for additive manufacturing                 
        & 3D-AmplifAI with ensemble method \\ \cline{1-1} \cline{3-4} 
		\cite{Jyeniskhan2023Integrating} &  & Defect detection for additive manufacturing   
        & BiFPN   \\ \cline{1-1} \cline{3-4} 
		\cite{Qiao2024Resources} &  & Resource allocation on IIoT devices  
        & FL and DRL    \\ \hline
		
		\cite{Castellani2021Real} &  \multicolumn{1}{c|}{\multirow{7}{*}{Security}}             & Anomaly detection in industrial environment & Siamese Autoencoders \\ \cline{1-1} \cline{3-4} 
		\cite{Li2023ANovel} &  & Anomaly detection in industrial control systems   
        & Multidimensional deconvolutional network \\ \cline{1-1} \cline{3-4} 
		\cite{Wang2023DTITD}  &  & Insider threat detection                             
        & DL with CNN and LSTM \\ \cline{1-1} \cline{3-4} 
		\cite{Boche2024On} &  & DoS attack detection                             
        & Neuromorphic twins-based DL \\ \cline{1-1} \cline{3-4} 
		\cite{Lv2024Secure} &  & Intrusion detection 
        & DL with DBN \\ \cline{1-1} \cline{3-4} 
		\cite{Kumar2023Blockchain} &  & Intrusion detection in IIoT networks                  
        & DL with LSTMSAE \\ \cline{1-1} \cline{3-4} 
		\cite{Belousov2024AMachine} &  & Anti-counterfeiting system                  
        & DL with DNN  \\ \hline
  
        \cite{Zhao2023ELITE} & \multicolumn{1}{c|}{\multirow{2}{*}{Vehicular Networks}}              & Hierarchical routing scheme & RL with Q-learning \\ \cline{1-1} \cline{3-4} 	
		\cite{Zhou2023Digital} &  & High-speed mobile networks & RL-based FL \\ \hline
  
	\end{tabular}
\end{table*}

\section{Digital twin for NTN }\label{sec:ntn} 
With the recent significant advancements in nongeostationary orbit satellites, it is now feasible to construct the 3rd generation partnership project (3GPP) non-terrestrial networks (NTNs) \cite{wigard2023ubiquitous,nguyen2024emerging}. These networks encompass satellite communication networks, high-altitude platform systems (HAPS), and air-to-ground networks. NTNs are a feasible choice for providing communication services to remote and rural areas where conventional land-based networks frequently encounter difficulties in efficiently and dependably covering such areas.

This makes NTNs play a crucial role in offering worldwide coverage for various applications that demand high availability and resilience \cite{nguyen2023security,nguyen2022security,nguyen2022outage}. They also help bridge the digital divide by providing new opportunities for innovation, economic growth, and social progress. NTNs will provide these exciting improvements, but they will also result in increased complexity due to the large number of network entities and user nodes communicating in a highly dynamic and diverse environment. Consequently, the design and management of NTN have become excessively complex and costly.

Meanwhile, DT technology has emerged as a highly promising method for simulating intricate and dynamic systems. Therefore, integrating this technology into NTNs can result in numerous advantages and benefits, while also helping to address the technical challenges related to the convergence of the air-to-ground networks. By precisely and comprehensively simulating and testing new technologies prior to their implementation in space, it can expedite the advancement of NTNs.


\subsection{DT for Satellite communication system}  
\begin{figure}[t]
	\centering
	\includegraphics[scale=0.3]{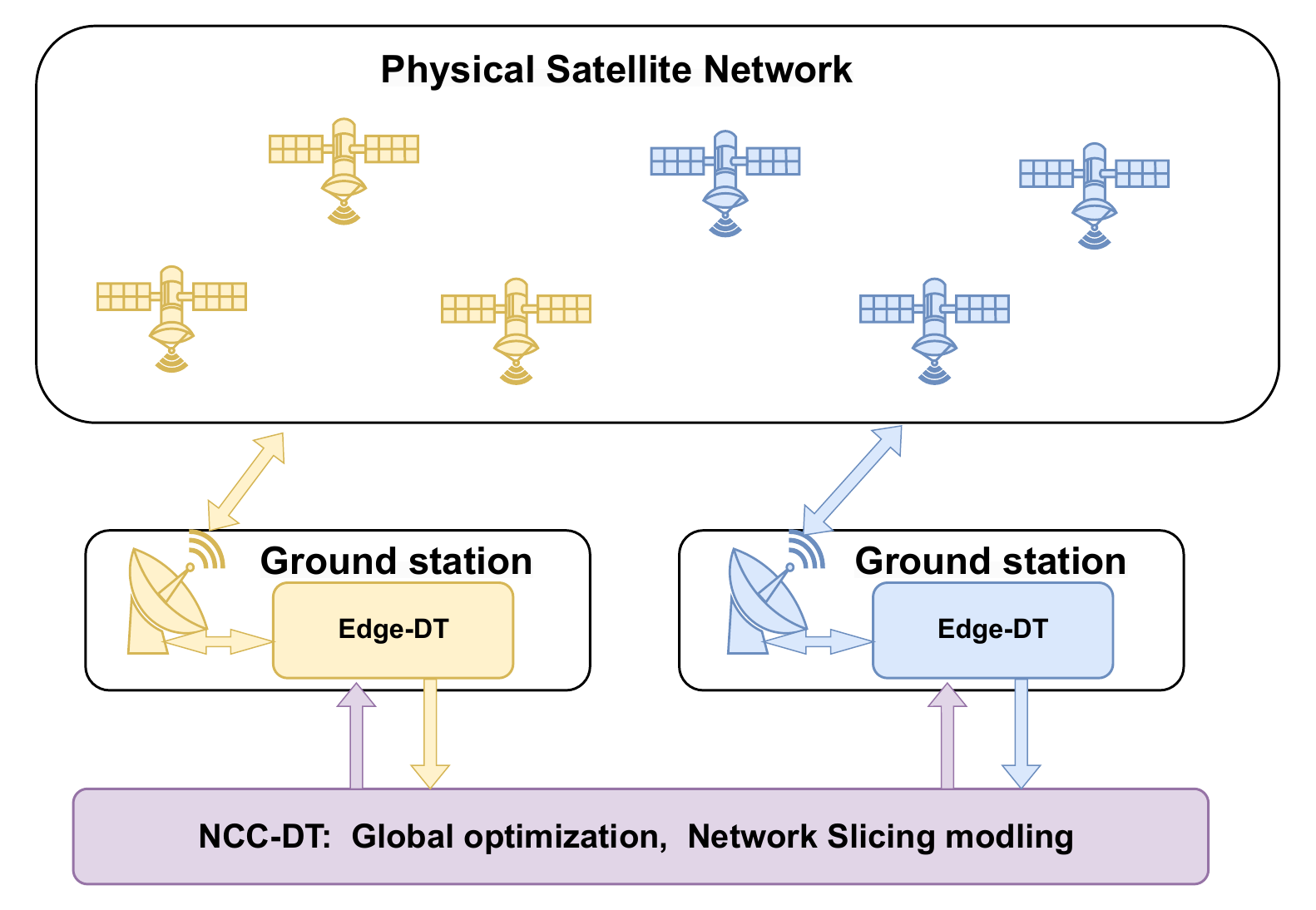}
	\caption{The function classifications of the DT system\cite{Zhou2023HDTSatnet}.}
	\label{classDT}
\end{figure}

Satellite networks are a dynamic topology that includes space and ground communication networks. The dynamic nature of the network has challenges in design, emulation, deployment, and maintenance. For these challenges, DT can play a role in establishing a virtual replica of the physical system, providing emulation, validation, prediction, and troubleshooting for the network. In this context,  a hierarchical DT network has been proposed for satellite communication networks,   with key components edge-DTs, central DTs, communications, and controllers as shown in Fig.\ref{classDT}.   The edge-DTs collect information from the physical network, such as device status, mobility information, radio resource status, and Quality of Experience (QoE). With this data, each edge-DT builds a model for the physical entity, which enables the local controller to perform real-time operations such as fault diagnosis, beam scheduling, radio resource allocation, and data processing. While, the central DTs,  deployed in the centralized network control center (NCC), collect data from edge DTs to construct the global network topology of the satellite network, such as slicing management\cite{Zhou2023HDTSatnet}. 

The speed of low-orbit satellite networks is much faster than that of ground terminals, leading to frequent satellite handovers. This can result in data loops among the satellites before the information is transferred to the end user. For this,  the Genetic Algorithm (GA) is proposed to
directly solve the problem with satellite handovers. Then, a DT is used to map the satellite networks to virtual space, and it allows for planning the routing to ensure that the content required by the users is always maintained and avoids data loops in the satellite network\cite{Zhao2022interlink}. Furthermore, dynamic contact window scheduling using digital twins with a network flow graph based algorithm is used to solve inefficient task scheduling caused by the dynamic change of task priority and satellite position over time has been proposed in \cite{Fan2023contact}.

On the other hand, satellite orbit prediction plays a crucial role in enhancing awareness of space situations, including collision warnings and observation scheduling. However, several factors, such as measurement errors, estimation inaccuracies, and unmodeled orbit perturbations, negatively impact the conventional orbit prediction method, leading to low-accuracy results. To address this issue, a DT framework is proposed to improve the accuracy of conventional dynamic orbit prediction models which consists of a digital twin system and physical objects (satellite network). The physical objects component provides real-time orbit data to the DT control center. The DT system uses a temporal convolutional neural network technique within an error prediction model, which aids in correcting the parameters of the orbit prediction model to enhance its accuracy\cite{Xu2023MlSatIot}.

\subsection{DT for flight objects }

\subsubsection{UAV}

UAVs, commonly known as drones, are increasingly utilized in a wide range of areas such as surveillance, delivery services\cite{tran2022throughput}, agriculture, and disaster response \cite{tran2022satellite}. AI/ML-powered DTs for UAVs offer a potent means of optimizing as well as augmenting UAV performance and operational effectivity, especially at computational-and-storage-constrained UAVs. Through using real-time data and dynamic optimization, they can tackle a range of challenges in cutting-edge UAV operations that are elaborated in the following. 

\begin{figure}[t]
	\centering
	\includegraphics[scale=0.45]{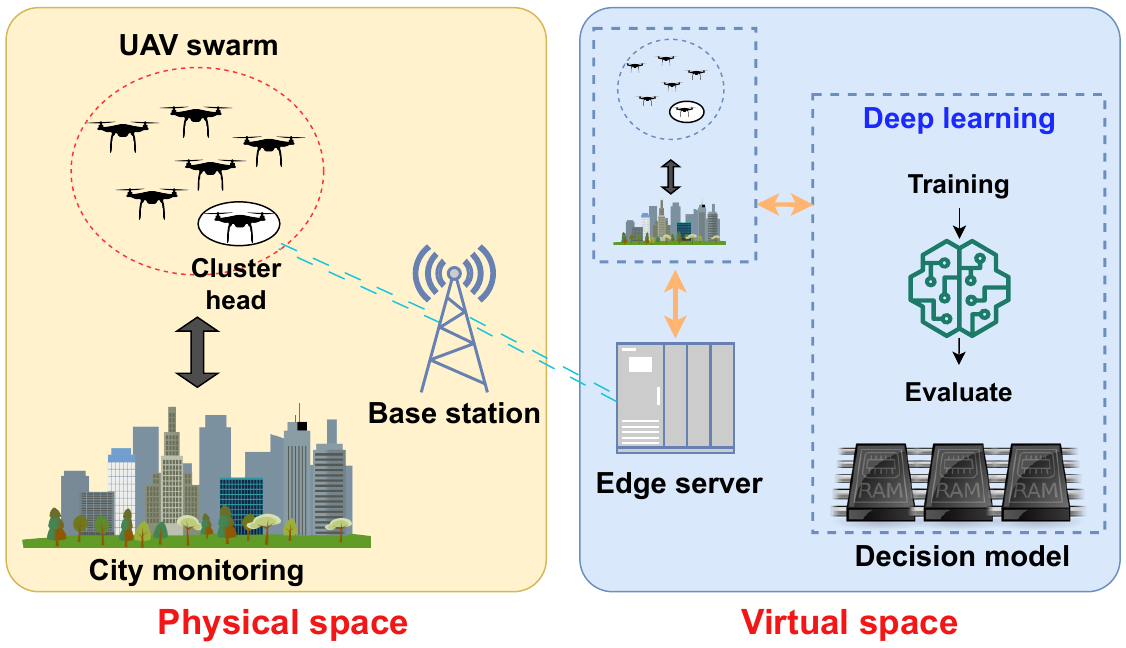}
	\caption{The DT-based framework for intelligent cooperation of UAV swarm.}
	\vspace{-0.5em}
	\label{fig:ai_ml8}
\end{figure} 

In~\cite{Lei2021Toward}, a DT-based framework for intelligent cooperation of UAV swarm is presented. Particularly, as shown in Fig.~\ref{fig:ai_ml8}, the DT model is first created within the framework to accurately represent the physical entity of a swarm of UAVs, and closely observe its whole life cycle. In this case, the DT acquires environmental state information from the physical entity and produces a dataset for different purposes through simulation and modeling. Once a sufficient amount of data in the sample dataset is obtained, a decision model is deployed that incorporates a DNN algorithm to achieve the global optimal strategy and control the actions of swarm of UAVs. The DNN utilizes the state inputs from the DT to generate an optimal prediction and then sends the result to the DT. Subsequently, the DT disseminates the final decisions to assist the physical entity. Using a case study of intelligent UAV swarm in time-varying condition, the framework can obtain a comprehensive score (i.e., a weighted sum of throughput, average packet delay, and packet loss ratio) up to 80 or up to 29\% higher than that of carrier sensing multiple access with collision avoidance (CSMA/CA), dynamic time division multiple access (DTDMA), and self-organized time division multiple access (STDMA) in all stages.

In another work, the use of dynamic DT and FL for air-ground networks, e.g., UAVs, is discussed in~\cite{Sun2022Dynamic}. Specifically, a UAV acts as the aggregator, while the ground clients work together to train the model using the network dynamics gathered by a DT in the middle of air and ground networks. To dynamically modify the selection of the most suitable clients and their level of participation, a Stackelberg game-based incentive mechanism for the FL is applied under the consideration of the varying DT deviations and network dynamics. Here, the DT of the UAV works as the leader and determines the preferences for clients. Meanwhile, the ground clients act as followers and select the global training rounds. Through the simulations using MNIST dataset, the proposed scheme can obtain by more than 90\% model accuracy. Additionally, when clients' data quality and heterogeneity increase, the energy consumption can be reduced by 25\% compared with the static incentive approach. The above work is then extended in~\cite{Sun2023Lightweight} through utilizing lightweight DT to address the challenges of constrained energy capacity and computation in UAVs for 6G networks. Specifically, this work utilizes an FL approach to share the DT modelling workload across many ground devices. Here, the DT on the UAV creates digital maps that represent the important conditions of the network, such as communication link quality and resource availability. It then uses these maps to develop an ML model that helps with intelligent network management. To enhance the effectiveness of DT modelling, a distributed incentive mechanism based on the Stackelberg game that encourages high-performing ground devices to participate the FL process is developed. From the simulation results, it is shown that the learning accuracy achieves more than 90\% with comparable energy efficiency of 0.0012.

To further enhance the use of resources and the efficiency of DRL model in a multi-UAV system, a DT-enabled task assignment strategy system is studied in~\cite{Tang2023Digital}. In particular, the system is segmented into two distinct phases: the initial task assignment and task re-assignment. During the initial phase, an airship uses a genetic algorithm to divide a task into several subtasks depending on the shortest distance. These subtasks are then assigned to UAVs. In the second phase, the DT is utilized to facilitate the airship in acquiring knowledge from task features and producing the Q-value of the estimated value network of DRL for UAVs through pre-training of the DT. The Q-value can be directly utilized in the DQN for UAVs to minimize the training episode. In this case, the DQN algorithm is used to train the task re-assignment solution. The simulation results reveal that the proposed DT with DQN can increase the task completion ratio and energy efficiency by 30\% and 19\%, respectively, compared with other baseline schemes. Then, the authors in~\cite{Hazarika2024Hybrid} introduce a hybrid ML technique that combines asynchronous FL (AFL) with multi-agent DRL to collaborately optimize task completion rate, energy consumption, and delay parameters in DT-assisted and UAV-aided Internet-of-Vehicles (IoV) network. Specifically, a DT network is first developed for vehicle-to-vehicle (V2V) and vehicle-to-infrastructure (V2I) task offloading. In this case, the DT network consists of three separate DTs, i.e., task vehicles, service vehicles, and roadside units, which are deployed in the UAVs. For the tasks, they can be either processed locally or offloaded via V2V or V2I modes, where task vehicles are in charge of task offloading and service vehicles are responsible for processing the offloaded tasks. Then, an optimization problem with the goal of optimizing the system efficiency while decreasing both the delay and the total energy consumption is formulated. To solve this optimization problem, the multi-agent DRL with MARS algorithm is utilized to allocate resources within the DT-assisted IoV network. The MARS algorithm is trained to employ the hybrid AFL and maximize the total utility of the system. From the simulation results, it can be observed that the MARS algorithm can provide the highest mean utility or up to 80\% higher that that of other baseline methods. Furthermore, the proposed framework can achieve the highest task completion rate at more than 90\%, the lowest energy consumption at 11kWh, and the lowest mean delay at 80ms.

Training trajectory and selection models on board UAVs increase their energy consumption. To save energy and improve the efficiency of UAVs, DT is used to replicate the physical environment and assist the UAV in planning the optimal trajectory. For this, a Dueling DQN with Prioritized Experience Replay (PER) function is used to train the model, and the UAV can get the newest model to adjust its trajectory with less energy consumption\cite{Zhao2024traject}.

In the context of multi-UAV operations, flocking motion methods are usually designed for specific environments. However, the real environment is often unknown and stochastic, which significantly reduces the practicality of these methods. A DT-enabled DRL method has been proposed in \cite{Shen2022flock} for this problem, where the DT of the multi-UAV system is built in the central server to train the DRL model. After the training stage, the DRL model can be quickly deployed to real-world UAVs through the connections between the physical entity and the digital model. On the other hand, a DT-aided task assignment for multi-UAV systems has been proposed in \cite{Tang2023task}.  It initially assigns tasks using GA. Then, during task reassignment, the UAVs' behavioral decisions are made by DQN based on DT to achieve task reassignment and improve task completion under time constraints. Similarly, a  DT with a weighted particle swarm optimization algorithm to establish a multi-objective task assignment optimization has been proposed in \cite{Deng2023task2}.  

A conventional DT consists of monitoring, imitation, and feedback control, which may be inefficient for real-time imitation in high-dynamics scenarios such as UAV-based target tracking. For this, a federated DT framework has been proposed, which includes three key elements: cooperative sensing, DT model aggregation, and DT inspection. The cooperative sensing algorithm minimizes data redundancy while speeding up the data collection process in mobile targets. This framework can quickly aggregate local DT models using an attention-based mechanism, enhancing the accuracy of mobile imitations. The DT inspection provides a method for real-time synchronization, addressing various factors such as differences in sensor capabilities, dynamic communication environments, and unavoidable imitation latency\cite{Zhou2024FdDtUav}.

\subsubsection{In-Flight Connectivity} In-flight connectivity (IFC) can be established by accessing the internet via multiple core network links, including air-to-ground (A2G), air-to-satellite (A2S), and air-to-air (A2A) connections at different slots throughout the flight. However, the highly dynamic nature of aeronautical networks, caused by the aircraft’s high velocity, can affect the efficiency of core network selection. This may lead to connectivity delays and increased packet loss. Additionally, the diverse traffic requirements of users can further complicate core network selection. A DT for IFC has been proposed to ensure continuous connectivity via traffic-based core network selection models. Using the K-means algorithm, the connectivity-based core network selection model determines potential core network links for aircraft. On the other hand, the traffic-based selection model uses a deep learning approach to address the diverse traffic requirements of passengers\cite{Bilen2022Flight}.

\section{Digital Twin for Quantum Networking }\label{sec:quantum}

Quantum networking (QN) significantly diverges from traditional networking by utilizing principles of quantum mechanics. Unlike classical networks that use bits as the fundamental unit of information, restricted to states of 0 or 1, quantum networks employ quantum bits (qubits). Qubits can exist in multiple states simultaneously due to superposition, enabling faster and more secure data transmission. DARIUS \cite{darius} defines a digital twin in quantum networking as a virtual model that replicates components of the QKD system, including the Quantum Transmitter (QTx), Quantum Receiver (QRx), and the fiber channel. This model dynamically adjusts system parameters to optimize performance, particularly by monitoring the State of Polarization (SOP) of photons in the channel and adapting to fluctuations introduced by environmental conditions.

Ongoing research and development of applications for 6G and beyond networks have led to increased demands for advanced technologies, including quantum networking \cite{Chau2023}. Quantum networking leverages quantum mechanics to enhance communication by using quantum bits (qubits), which can exist in multiple states simultaneously, allowing potentially faster and more secure data transmission \cite{Amir2022}. However, implementing quantum networks poses significant challenges, such as qubit decoherence, high error rates, and the need for precise synchronization over long distances \cite{Khan2022}. Additionally, limited quantum processing power, scalability issues, and noise in Noisy Intermediate-Scale Quantum (NISQ) systems further complicate implementation \cite{Amir2022}. Developing specialized quantum computers and error-correcting mechanisms is necessary to address these challenges. Integration with digital twin frameworks and blockchain technology also presents obstacles, as classical blockchains are vulnerable to quantum attacks, and quantum-resistant solutions must be developed \cite{Khan2022}. The complexity of managing large quantum systems and the high computational burden of sampling from high-dimensional distributions highlight the need for continued research and innovation in this field \cite{Amir2022, van2024dynamic, Khan2022}.

For example, in a smart city scenario, autonomous vehicles require real-time communication with each other and traffic management systems to ensure safety and efficiency. Traditional networks may face latency and security challenges in such high-stakes environments. However, integrating QN within the 6G framework allows for instantaneous and secure information exchange between vehicles, significantly reducing the risk of accidents and improving traffic flow. This example highlights how quantum networking, when combined with 6G advancements, can transform communication and data transmission.

This section explores the application of Digital Twin (DT) technology in quantum networking, reviewing recent advancements and solutions proposed in the literature. By leveraging DTs, network performance, security, and user experience in quantum communication systems can be enhanced. Integrating DTs in quantum networking represents a significant step toward realizing the full potential of next-generation networks, ensuring that 6G and beyond networks are more resilient, efficient, and user-centric.

\subsection{Challenges in Quantum Networking}
\label{sec:quantumchal}

Jaschke et al. identified several critical challenges in benchmarking quantum processors, particularly within Rydberg atom-based systems, that impact quantum networking. One major issue involves \textit{crosstalk during simultaneous gate operations} due to long-range van der Waals interactions, which degrade gate fidelity. This interference is especially problematic for Noisy Intermediate-Scale Quantum (NISQ) devices, where parallel operations are essential for efficiency but risk compromising computational accuracy \cite{jaschke2022ab}. Additional challenges include \textit{decoherence and state decay}, as quantum states are inherently fragile, with accelerated decoherence during parallel operations. These phenomena increase error rates, complicating the maintenance of fidelity over extended or complex circuit depths. Furthermore, scalability constraints arise as larger qubit arrays introduce complex interactions that necessitate precise control over circuit depth and parallelization, making scalability without performance loss a difficult objective.

In Quantum Key Distribution (QKD) networks, intrinsic challenges further impact security and efficiency, as highlighted by DARIUS \cite{darius}. Imperfections in optical components, such as polarization controllers and beam splitters, lead to photon loss and increased Quantum Bit Error Rate (qBER). Environmental factors, including fiber stress, further distort photon polarization, leading to misalignments in basis selection between quantum transmitters (QTx) and receivers (QRx), complicating secure key exchange. Additionally, detecting eavesdropping is challenging, as legitimate environmental fluctuations and malicious interventions both affect qBER similarly, making it difficult to distinguish between them without interrupting the QKD process. Ahmad et al. expanded on QKD challenges, focusing on \textit{State of Polarization (SOP) distortion} caused by environmental conditions that misalign QTx and QRx bases, increasing qBER \cite{ahmadian2022designing}. Non-ideal optical components, such as beam splitters and polarization controllers, further degrade system accuracy due to photon loss and random polarization shifts. Eavesdropping detection remains problematic, as environmental disturbances and malicious activities introduce similar SOP changes, complicating secure QKD.

Quantum networking for federated learning in healthcare also presents unique challenges. Privacy in data sharing is a major concern, as traditional machine learning relies on centralized data collection, raising issues around patient data privacy. Long training times and high communication costs present additional hurdles, as extensive communication delays hinder timely updates of diagnostic models, which are critical for healthcare \cite{qu2023dtqfl}. Moreover, Variational Quantum Neural Networks (VQNNs) in DTQFL are impacted by \textit{quantum noise and model degradation}, especially within NISQ devices, reducing diagnostic accuracy over time. Quantum networking in Metaverse applications also involves unique challenges. Wang et al. highlight \textit{resource constraints at base stations} that process high data volumes from UAVs, often resulting in data loss and delays. Communication delays and transmission losses due to non-line-of-sight (NLOS) conditions further disrupt real-time synchronization required for Metaverse applications. Adapting UAVs to dynamic environments presents additional complexity, as traditional reinforcement learning models require retraining for redeployment, delaying decision-making \cite{wang2023efficient}. 

Integrating quantum networking within the Metaverse introduces challenges around \textit{scalability and stability} of quantum communication systems, which are essential for low-latency, secure connections. Quantum communication networks reliant on entangled states face limitations from environmental noise and the difficulty of maintaining long-distance entanglement \cite{tuli2024integration}. Hardware limitations in NISQ devices further constrain large-scale quantum processing, complicating efforts to use quantum computing for simulations in the Metaverse. Cybersecurity concerns also arise, as quantum computing may compromise traditional cryptographic methods, necessitating quantum-enhanced security frameworks. Middleware challenges add complexity, requiring solutions for data exchange between quantum and classical systems without latency or data loss.

For quantum networking in field applications, such as satellite-based sensors, Chau et al. emphasize challenges in \textit{frequency stability in compact designs}, as maintaining high stability despite reduced size, weight, power, and cost (SWaP-C) demands precise control. Quantum sensors in field applications must also be resilient to extreme environmental conditions, including temperature variations and radiation exposure \cite{Chau2023}. In QKD networks supporting post-quantum cryptography, challenges arise in ensuring \textit{forward security}, as frequent key updates are resource-intensive, especially for IoT devices \cite{nouma2023post}. Implementing cryptographic agility to transition smoothly to post-quantum standards while maintaining compatibility with existing infrastructure is also essential for future quantum networks. Quantum networking for climate modeling faces scalability and synchronization challenges in handling large-scale data. Current quantum systems are limited in processing extensive datasets typical of climate applications \cite{otgonbaatar2023quantum}. Additionally, synchronizing quantum and classical computations is complex, as classical models require high-speed processing, while quantum computations lack comparable efficiency. Hardware limitations, such as error rates and coherence issues, further limit quantum contributions to complex climate models.

Semantic communication within the Metaverse introduces challenges in quantum networking due to the need for high-fidelity, low-latency transmission. Semantic noise, resulting from discrepancies between transmitted and received information, distorts meaning and hinders accurate virtual object reconstruction \cite{Khalid2023}. Resource inefficiency due to high machine learning demands and knowledge inconsistencies across devices further complicate semantic extraction, while semantic attacks threaten data fidelity, adding privacy and security risks. Scaling QKD networks to support secure data transmission presents further challenges, particularly in \textit{efficient resource management}. High costs and limited availability of quantum devices constrain QKD deployments, while compatibility with security standards, such as those set by the European Telecommunications Standards Institute (ETSI), is essential for maintaining robust QKD systems \cite{martin2024service}. Additionally, quantum networks for consumer applications face \textit{resource allocation and scalability} challenges, as classical and quantum resources must be managed in real-time to meet fluctuating demands \cite{cao2024softwarized}.

Finally, quantum networking in manufacturing introduces data security challenges, especially against quantum attacks on blockchain-based systems securing data. Traditional cryptographic methods are vulnerable to quantum threats, requiring quantum-resilient data management systems \cite{Khan2022}. Processing power limitations and high error rates in current quantum systems also complicate data handling within complex systems, affecting the accuracy of simulations \cite{Amir2022}.

\subsection{Digital Twin Solutions to Quantum Networking Challenges}
\label{sec:quantumSolutions}

Digital Twins (DTs) have been employed across various quantum networking applications to address challenges such as scalability, fidelity, and security in quantum networks. This section synthesizes solutions from the literature, illustrating how DTs provide practical tools to mitigate these issues.

Jaschke et al. demonstrated that DTs could mitigate \textit{crosstalk and decoherence} challenges in Rydberg atom-based systems by offering a virtual environment to model and optimize gate configurations before deployment \cite{jaschke2022ab}. By simulating long-range interactions and adjusting gate distances and operational parameters within the DT, the system can maintain high fidelity and minimize errors due to crosstalk and state decay. This approach enhances scalability, as DTs can model increasing qubit arrays without impacting real-system resources.

In Quantum Key Distribution (QKD) systems, DARIUS employs a DT to replicate real-world QKD operations, enhancing the \textit{detection of eavesdropping and environmental impacts on photon polarization} \cite{darius}. The DT provides real-time monitoring and adjustment of QKD system parameters to correct for polarization shifts due to fiber stress or optical component imperfections. This allows the system to differentiate between natural fluctuations and potential malicious attacks, thereby improving the Quantum Bit Error Rate (qBER) without interrupting secure key exchanges. Ahmad et al. further illustrate how DTs in QKD can address \textit{SOP distortion and optical imperfections} \cite{ahmadian2022designing}. The DT models the optical components and environmental conditions, simulating their impact on SOP and photon polarization. Through these simulations, the DT allows real-time adjustments to the Electronic Polarization Controller (EPC), reducing qBER and supporting secure key exchanges despite external influences.

For federated healthcare applications within quantum networking, DTQFL uses DTs to enhance \textit{privacy, efficiency, and accuracy} in VQNNs \cite{qu2023dtqfl}. By creating a digital representation of patient data and network resources, DTs enable decentralized federated learning that protects data privacy while optimizing quantum network resources to manage long training periods and reduce communication costs. The DT's real-time simulations also improve the robustness of VQNNs by accounting for quantum noise, thus maintaining diagnostic reliability over time.

In the Metaverse, DTs address \textit{resource allocation and synchronization issues} in quantum networks supporting high data volumes from UAVs. Wang et al. employ a DT framework to simulate resource demands and optimize data management at base stations, reducing delays and improving communication reliability \cite{wang2023efficient}. The DT dynamically adapts to environmental changes and user demands, ensuring real-time updates that are crucial for immersive Metaverse experiences. For quantum communication applications in the Metaverse, Tuli et al. leverage DTs to improve \textit{scalability and stability of quantum communications} \cite{tuli2024integration}. DTs enable virtual testing of entanglement-based communication protocols, which are sensitive to environmental noise and long-distance maintenance issues. By optimizing these protocols within the DT, quantum networks can enhance low-latency, high-security data transmission in Metaverse applications. Additionally, DTs facilitate seamless middleware solutions for hybrid quantum-classical data exchanges, reducing latency and improving the reliability of integrated systems. Chau et al. present DTs as a solution to the \textit{frequency stability and environmental resilience} challenges in portable quantum sensors \cite{Chau2023}. The DT monitors temperature fluctuations and other environmental factors in real time, allowing for adaptive control over laser frequency and enhancing stability in harsh conditions. This real-time feedback improves system resilience and expands the potential for field-deployed quantum sensors.

In post-quantum cryptographic applications, Nouma et al. use DTs to provide \textit{cryptographic agility and forward security} \cite{nouma2023post}. The DT models IoT devices within the quantum network, facilitating real-time key updates and efficient cryptographic protocol transitions. This agility allows quantum networks to stay ahead of quantum attacks while optimizing resource usage in IoT applications.

In climate modeling applications, DTs address the \textit{scalability and synchronization challenges} faced by quantum networks handling large datasets \cite{otgonbaatar2023quantum}. The DT bridges quantum and classical computations, optimizing data processing and reducing the burden on quantum hardware. Through hybrid simulations, DTs enable effective resource management, supporting complex climate models that leverage quantum processing.

For semantic communications in the Metaverse, DTs mitigate \textit{semantic noise, resource inefficiency, and security concerns} by simulating semantic data processing and adapting ML resource demands \cite{Khalid2023}. The DT enables synchronization across devices, ensuring consistent semantic extraction and reducing the risk of semantic attacks that could compromise virtual reconstructions. This approach improves resource efficiency and data fidelity, essential for secure Metaverse interactions.

In QKD networks, Martin et al. highlight DTs as crucial for achieving \textit{scalability and efficient resource management} \cite{martin2024service}. By virtually emulating QKD protocols and aligning with established security standards, the DT enhances the scalability and robustness of QKD systems, overcoming limitations in available quantum devices.

For consumer quantum applications, DTs enable \textit{dynamic resource allocation and real-time adaptation} in networks managing both classical and quantum resources \cite{cao2024softwarized}. Through virtual modeling, the DT allocates resources adaptively, meeting variable demands and ensuring synchronization between quantum and classical components in consumer networks. In manufacturing, Khan et al. leverage DTs to enhance \textit{quantum-resilient data management} by using hash-based digital signatures in the Twinchain system, which securely manages data against quantum threats \cite{Khan2022}. The DT improves the resilience of blockchain-based security by simulating quantum-proof cryptographic measures, maintaining data security in distributed networks.

Finally, DTs in quantum simulations address \textit{computational limits and hardware constraints}, as illustrated by Amir et al. \cite{Amir2022}. DTs simulate error correction processes and optimize hardware configurations, reducing error rates and enhancing coherence for more reliable quantum simulations.

Overall, these studies demonstrate the essential role of DTs in overcoming challenges within quantum networks. By simulating, adapting, and optimizing quantum systems in real time, DTs contribute to the robustness, scalability, and security of quantum networking across various applications.

\begin{table*}[!h]
	\caption{Summary of Challenges in Quantum Networking, Digital Twin Solutions, and Applications}
	\centering
	\setlength{\tabcolsep}{4pt} 
	\begin{tabular}{|>{\centering\arraybackslash} m{2.5cm} | >{\centering\arraybackslash} m{3.5cm} | >{\centering\arraybackslash} m{4cm} | >{\centering\arraybackslash} m{3.5cm} | >{\centering\arraybackslash} m{3cm}|}
		\hline 
		\textbf{Reference} & \textbf{Quantum Networking Challenges} & \textbf{Digital Twin Solutions} & \textbf{Application or Use Case} & \textbf{DT Implementation Focus} \\  \hline\hline
		
		Jaschke et al. \cite{jaschke2022ab} & Crosstalk during simultaneous gate operations; Decoherence and state decay; Scalability constraints. & Virtual environment to model and optimize gate configurations, minimizing crosstalk and improving fidelity. & Quantum processor benchmarking in Rydberg atom-based systems. & Simulating gate configurations and optimizing scheduling for larger qubit arrays. \\ \hline
		
		DARIUS \cite{darius} & Photon loss and increased qBER due to optical imperfections and environmental factors; Eavesdropping detection challenges. & Real-time monitoring and adjustment of QKD parameters to correct polarization shifts. & Quantum Key Distribution (QKD) systems. & Monitoring photon polarization and detecting security breaches in QKD. \\ \hline
		
		Ahmad et al. \cite{ahmadian2022designing} & SOP distortion from environmental factors; Non-ideal optical components; Eavesdropping detection issues. & Modeling optical components and environmental effects, enabling real-time adjustments to reduce qBER. & Quantum Key Distribution (QKD) systems. & Simulating SOP changes and adjusting polarization controllers in QKD. \\ \hline
		
		Qu et al. \cite{qu2023dtqfl} & Privacy in healthcare data sharing; Long training periods; Quantum noise in VQNNs. & Creating digital representations for privacy-preserving federated learning, with real-time simulations for resource optimization. & Federated learning in healthcare networks. & Maintaining data privacy and VQNN accuracy in federated learning models. \\ \hline
		
		Wang et al. \cite{wang2023efficient} & Resource constraints at base stations; Communication delays and transmission losses; UAV adaptation in dynamic environments. & Simulating resource demands and managing data flow, adapting to environmental changes for real-time updates. & Metaverse applications with UAV data handling. & Resource management and real-time communication reliability in UAV-based networks. \\ \hline
		
		Tuli et al. \cite{tuli2024integration} & Scalability and stability in quantum internet; Environmental noise; Middleware for quantum-classical data exchange. & Virtual testing of entanglement protocols and middleware integration for seamless data exchange. & Metaverse applications for secure, low-latency communication. & Improving low-latency, secure data transmission in quantum-enhanced Metaverse applications. \\ \hline
		
		Chau et al. \cite{Chau2023} & Frequency stability in compact sensors; Environmental resilience; Modularity in laser systems. & Monitoring environmental conditions to maintain laser frequency stability. & Portable quantum sensors for field applications. & Environmental adaptation and laser stability in portable quantum sensor networks. \\ \hline
		
		Nouma et al. \cite{nouma2023post} & Forward security and cryptographic agility; Resource constraints in IoT devices. & Modeling IoT devices to support real-time key updates and cryptographic protocol transitions. & Post-quantum cryptographic security in IoT networks. & Cryptographic agility and resource efficiency in post-quantum networks. \\ \hline
		
		Otgonbaatar et al. \cite{otgonbaatar2023quantum} & Scalability and synchronization for large climate datasets; Hybrid quantum-classical processing issues; Hardware limitations. & Hybrid simulations bridging quantum and classical computations for resource management. & Climate modeling with quantum-enhanced networks. & Managing quantum and classical data processing for large-scale environmental models. \\ \hline
		
		Khalid et al. \cite{Khalid2023} & Semantic noise; Resource inefficiency in ML; Knowledge-base inconsistencies; Risks of semantic attacks. & Simulating semantic processing to ensure consistency and prevent semantic attacks. & Metaverse applications with semantic communication. & Resource efficiency and data fidelity in semantic communication. \\ \hline
		
		Martin et al. \cite{martin2024service} & Scalability and resource management in QKD; Limited quantum device availability; Compatibility with security standards. & Emulating QKD protocols to align with security standards, optimizing scalability and resource use. & Quantum Key Distribution (QKD) systems in secure communication networks. & Scalability and standard compliance in QKD systems. \\ \hline
		
		Cao et al. \cite{cao2024softwarized} & Resource allocation and scalability in consumer networks; Synchronization of classical and quantum resources. & Adaptive resource allocation for real-time synchronization across quantum and classical components. & Consumer applications with dynamic resource demands. & Dynamic resource management in quantum-enabled consumer applications. \\ \hline
		
		Khan et al. \cite{Khan2022} & Risk of quantum attacks on blockchain; Vulnerability of cryptography to quantum decryption; Quantum-resilient data management needed. & Simulating quantum-proof cryptographic measures to secure blockchain against quantum threats. & Manufacturing data management with blockchain. & Quantum-resilient security for blockchain in manufacturing. \\ \hline
		
		Amir et al. \cite{Amir2022} & Processing power limits and high error rates; Reliability issues in quantum simulations for digital twins. & Simulating error correction and optimizing hardware configuration to reduce errors. & Quantum simulations in complex system applications. & Error correction and hardware optimization for reliable quantum simulations. \\ \hline
	\end{tabular}
\end{table*}

\subsection{Future Work in Quantum Networking and Digital Twin Integration}

The reviewed studies highlight several promising directions for future research aimed at advancing Quantum Networking (QN) through the integration of Digital Twins (DTs). Key areas for future work include enhancing scalability, improving real-time adaptability, advancing hardware for network stability, and bolstering security measures specific to quantum communication networks.

One major area for future research is \textit{scalability in quantum networks}. As DTs are increasingly used to model complex quantum networking environments, additional research is needed to create scalable DT frameworks that can support larger qubit arrays and more complex QN architectures without a loss in fidelity or increased error rates \cite{jaschke2022ab}. Advanced DT algorithms for simulating extensive quantum circuits and managing the effects of crosstalk and decoherence will be crucial for expanding QN applications in fields like climate modeling and the Metaverse \cite{otgonbaatar2023quantum, tuli2024integration}. In terms of \textit{real-time adaptability}, DTs must evolve to enable continuous, dynamic adjustments in QN systems, especially in response to environmental or operational fluctuations. Improved adaptability will enhance performance in sensitive applications, such as Quantum Key Distribution (QKD) and UAV-based networking for the Metaverse, where QN operations must adjust quickly to maintain secure and synchronized connections \cite{ahmadian2022designing, darius, wang2023efficient}. 

Another key area for future work involves \textit{hardware and resource optimization} within QN environments. Research efforts should focus on optimizing quantum hardware and reducing noise, improving coherence times, and developing cost-effective, compact components suitable for real-world networking environments \cite{Chau2023, Khan2022}. Additionally, DTs that support hybrid quantum-classical computation will play a significant role in balancing resource usage and enabling broader applications for QN in areas such as healthcare and consumer networking \cite{qu2023dtqfl, cao2024softwarized}. Strengthening \textit{security and cryptographic resilience} in DT-integrated QN systems is critical as quantum computing poses risks to traditional cryptography. Future DTs should support post-quantum cryptography frameworks that incorporate forward security and cryptographic agility, allowing QN systems to maintain resilience against quantum-based cyber threats \cite{nouma2023post}. Standardized DT frameworks that align with established security protocols, particularly for applications in QKD and healthcare, will be essential \cite{martin2024service}.

Lastly, achieving \textit{interoperability between quantum and classical systems} remains a significant research challenge. Effective DT implementations should support seamless, latency-free data exchange between quantum and classical elements within QN environments, as applications increasingly rely on hybrid processing. This need is especially pressing in fields like climate modeling and advanced manufacturing, where DTs are used to optimize data flows and ensure synchronized quantum-classical integration \cite{Amir2022, Khan2022}. Future work on middleware solutions and QN architectures that facilitate smooth quantum-classical interactions will be vital for expanding DT-enhanced quantum networking.

Together, these future directions will advance the integration of Digital Twins in Quantum Networking, fostering scalability, security, and adaptability across diverse applications and addressing the unique challenges of quantum-enhanced communication networks.

\section{Challenges, Open Issues, and Future Research Directions }\label{sec:challenge}

As the development of end-to-end NDTs for 6G networks progresses, it becomes evident that several critical technical challenges remain unsolved. Addressing these open issues is essential to realizee full potential of network digital twins in beyond-5G environments. In this section, we outline key research challenges and promising future directions, focusing on real-time data synchronization, scalability, high-fidelity modeling, explainable AI, security and privacy, and resource efficiency. Each subsection discusses the limitations of current approaches and highlights opportunities for further investigation.

\subsection{Real-Time Synchronization and Data Fusion}
One fundamental challenge for network digital twins is achieving \textit{real-time} synchronization between the physical network and its virtual twin. The digital twin must continuously ingest and fuse data streams from millions of dynamic and distributed sources (IoT sensors, base stations, user devices, satellites, etc.) to mirror the live state of a 6G network. Ensuring timely updates with minimal \textit{Age of Information} is non-trivial, as network latency and jitter can cause the twin's state to lag behind reality \cite{vaezi2023delay, zheng2023vehicular}. Recent studies have begun addressing this; for example, optimizing twin placement and data pipelines can minimize latency and information age in updates \cite{alkhateeb2023realtime}. In vehicular networks, game-theoretic scheduling of synchronization messages has been proposed to achieve low-latency, consistent updates between vehicles’ twins and roadside units \cite{zheng2023vehicular}. Developing robust synchronization frameworks, advanced AI-driven fusion algorithms, unified time-coordination protocols, and predictive synchronization techniques represents promising directions moving forward.

\subsection{Scalability and Complexity}
6G networks will be extremely large-scale and complex, which poses a major scalability challenge for digital twins. The 6G vision includes massive machine-type communications, ultra-dense deployments of small cells, and the integration of terrestrial, aerial, and satellite networks into a single system \cite{lin2023from}. Modeling and simulating such a vast, heterogeneous environment in a unified digital twin is immensely difficult. As the number of nodes and links grows into the billions, the computational and memory requirements to maintain an up-to-date network state explode. Addressing these scalability challenges will likely involve hierarchical architectures, modular twin designs, AI-assisted model abstractions, and efficient distributed computing solutions to handle the inherent complexity.

\subsection{High-Fidelity Modeling of Dynamic Environments}
Another challenge is achieving high fidelity in modeling the 6G network’s physical environment, which is extraordinarily dynamic. Wireless networks experience continual changes in propagation conditions, user mobility, interference, and traffic patterns. Accurately capturing these in a digital twin requires sophisticated models that border on real-world physics. High-fidelity channel models (e.g., ray tracing for mmWave/THz bands, deterministic propagation models for specific environments) must be continually updated as the environment evolves \cite{alkhateeb2023realtime, lin2023from}. Integrating real-time environmental data, predictive modeling techniques, and hybrid modeling strategies combining deterministic and machine learning approaches will be essential to enhance modeling accuracy and responsiveness.

\subsection{Explainable AI}
AI-driven digital twins in 6G networks must be transparent and interpretable, ensuring human operators can understand and trust automated decisions \cite{wang2024XAI}. Current black-box AI models lack transparency, creating trust and regulatory issues. Implementing domain-specific explainable AI methods and integrating interpretability into digital twin architectures is a key research direction. Initial efforts have outlined frameworks for applying methods like SHAP or LIME to interpret decisions of network control algorithms \cite{wang2024XAI, lin2023from}. Important next steps include embedding interpretability frameworks directly within twin architectures, leveraging hybrid AI approaches for greater transparency, and developing robust fairness and anomaly detection methods for AI-driven decisions.

\subsection{Security and Privacy}
Network digital twins raise significant security and privacy concerns. By design, a digital twin maintains a live digital replica of a physical network, including sensitive information about users, devices, and network state. This effectively becomes an attractive target for adversaries: compromising the digital twin could allow detailed intelligence gathering about the real network or malicious manipulation. Protecting this infrastructure demands strong encryption, authentication, and robust access control measures \cite{alcaraz2022security}. Future directions involve embedding security directly within digital twin designs, exploring privacy-preserving methods like federated learning and blockchain-based data provenance, and developing specialized intrusion detection frameworks tailored specifically for digital twins.

\subsection{Resource Consumption and Energy Efficiency}
Operating an end-to-end network digital twin is highly resource-intensive, demanding significant computational power, storage, and energy. The twin must continuously process data from numerous sources and update complex predictive models. Techniques like adaptive fidelity, model compression, resource scheduling, and specialized hardware acceleration are essential to manage these resources efficiently \cite{liu2024twotimescale}. Research into sustainable digital twin solutions aligned with the green objectives of 6G is critical for their widespread adoption \cite{alkhateeb2023realtime}. Exploring lean modeling, specialized hardware acceleration (GPUs, FPGAs), adaptive fidelity adjustments, and energy-aware orchestration strategies will significantly enhance the efficiency and sustainability of digital twin implementations.


\section{Conclusion}\label{sec:Summary}
In this study, we present a comprehensive survey of NDT for 6G\plus from both academic and industrial perspectives. In particular, we provide an in-depth guide to NDT and its supporting underlying technologies, including RAN/ORAN, transport networks, 5GCORE\plus{}, cloud and edge computing, applications, NTNs, and quantum networking. We also demonstrate how ORAN and 5GCORE\plus{}are applied in NTN from the perspective of practical deployment in an industrial perspective. Finally, we discuss current challenges, open issues and introduce some potential technologies as future research directions for NDT, including real-time synchronization and data fusion, high-fidelity modeling of dynamic environment, explainable AI, security and privacy, resource consumption and energy efficiency.

\bibliographystyle{IEEEtran}

\bibliography{Bibtex/RAN,Bibtex/Optical,Bibtex/5GCORE,Bibtex/Computing,
Bibtex/AIML,Bibtex/Introduction,Bibtex/NTN,Bibtex/Quantum,Bibtex/section9}

\begin{thebibliography}{100}
\providecommand{\url}[1]{#1}
\csname url@samestyle\endcsname
\providecommand{\newblock}{\relax}
\providecommand{\bibinfo}[2]{#2}
\providecommand{\BIBentrySTDinterwordspacing}{\spaceskip=0pt\relax}
\providecommand{\BIBentryALTinterwordstretchfactor}{4}
\providecommand{\BIBentryALTinterwordspacing}{\spaceskip=\fontdimen2\font plus
\BIBentryALTinterwordstretchfactor\fontdimen3\font minus
  \fontdimen4\font\relax}
\providecommand{\BIBforeignlanguage}[2]{{%
\expandafter\ifx\csname l@#1\endcsname\relax
\typeout{** WARNING: IEEEtran.bst: No hyphenation pattern has been}%
\typeout{** loaded for the language `#1'. Using the pattern for}%
\typeout{** the default language instead.}%
\else
\language=\csname l@#1\endcsname
\fi
#2}}
\providecommand{\BIBdecl}{\relax}
\BIBdecl

\bibitem{jiang2021road}
W.~Jiang, B.~Han, M.~A. Habibi, and H.~D. Schotten, ``The road towards 6g: A
  comprehensive survey,'' \emph{IEEE Open Journal of the Communications
  Society}, vol.~2, pp. 334--366, 2021.

\bibitem{nguyen20216g}
D.~C. Nguyen, M.~Ding, P.~N. Pathirana, A.~Seneviratne, J.~Li, D.~Niyato,
  O.~Dobre, and H.~V. Poor, ``6g internet of things: A comprehensive survey,''
  \emph{IEEE Internet of Things Journal}, vol.~9, no.~1, pp. 359--383, 2021.

\bibitem{khan2022digital}
L.~U. Khan, W.~Saad, D.~Niyato, Z.~Han, and C.~S. Hong, ``Digital-twin-enabled
  6g: Vision, architectural trends, and future directions,'' \emph{IEEE
  Communications Magazine}, vol.~60, no.~1, pp. 74--80, 2022.

\bibitem{tao2024wireless}
Z.~Tao, W.~Xu, Y.~Huang, X.~Wang, and X.~You, ``{Wireless Network Digital Twin
  for 6G: Generative AI as a Key Enabler},'' \emph{IEEE Wireless
  Communications}, vol.~31, no.~4, pp. 24--31, 2024.

\bibitem{grieves2017digital}
M.~Grieves and J.~Vickers, ``Digital twin: Mitigating unpredictable,
  undesirable emergent behavior in complex systems,'' \emph{Transdisciplinary
  perspectives on complex systems: New findings and approaches}, pp. 85--113,
  2017.

\bibitem{mihai2022digital}
S.~Mihai, M.~Yaqoob, D.~V. Hung, W.~Davis, P.~Towakel, M.~Raza, M.~Karamanoglu,
  B.~Barn, D.~Shetve, R.~V. Prasad \emph{et~al.}, ``Digital twins: A survey on
  enabling technologies, challenges, trends and future prospects,'' \emph{IEEE
  Communications Surveys \& Tutorials}, vol.~24, no.~4, pp. 2255--2291, 2022.

\bibitem{IBM_2025}
\BIBentryALTinterwordspacing
IBM, ``What is a digital twin?'' Jan 2025. [Online]. Available:
  \url{https://www.ibm.com/think/topics/what-is-a-digital-twin}
\BIBentrySTDinterwordspacing

\bibitem{10742580}
H.~Al-Hraishawi, M.~Alsenwi, J.~Ur~Rehman, E.~Lagunas, and S.~Chatzinotas,
  ``Digital twin for enhanced resource allocation in 6g non-terrestrial
  networks,'' \emph{IEEE Communications Magazine}, vol.~63, no.~3, pp. 47--53,
  2025.

\bibitem{wang2023survey}
Y.~Wang, Z.~Su, S.~Guo, M.~Dai, T.~H. Luan, and Y.~Liu, ``A survey on digital
  twins: Architecture, enabling technologies, security and privacy, and future
  prospects,'' \emph{IEEE Internet of Things Journal}, vol.~10, no.~17, pp.
  14\,965--14\,987, 2023.

\bibitem{tang2022survey}
F.~Tang, X.~Chen, T.~K. Rodrigues, M.~Zhao, and N.~Kato, ``Survey on digital
  twin edge networks (diten) toward 6g,'' \emph{IEEE Open Journal of the
  Communications Society}, vol.~3, pp. 1360--1381, 2022.

\bibitem{kuruvatti2022empowering}
N.~P. Kuruvatti, M.~A. Habibi, S.~Partani, B.~Han, A.~Fellan, and H.~D.
  Schotten, ``Empowering 6g communication systems with digital twin technology:
  A comprehensive survey,'' \emph{IEEE access}, vol.~10, pp.
  112\,158--112\,186, 2022.

\bibitem{lin20236g}
X.~Lin, L.~Kundu, C.~Dick, E.~Obiodu, T.~Mostak, and M.~Flaxman, ``{6G} digital
  twin networks: From theory to practice,'' \emph{IEEE Communications
  Magazine}, vol.~61, no.~11, pp. 72--78, 2023.

\bibitem{itur2022future}
{ITU-R}, ``Future technology trends of terrestrial {IMT} systems towards 2030
  and beyond,'' \emph{Report ITU-R M.2516-0}, 2022.

\bibitem{itut2022digital}
{ITU-T}, ``Digital twin network--requirements and architecture,''
  \emph{Recommendation ITU-T Y.3090}, 2022.

\bibitem{oran2024digital}
{O-RAN nGRG}, ``Digital twin {RAN}: Key enablers,'' \emph{Report ID:
  RR-2024-09}, 2024.

\bibitem{hoydis2024learning}
J.~Hoydis, F.~A{\"\i}t~Aoudia, S.~Cammerer, F.~Euchner, M.~Nimier-David,
  S.~Ten~Brink, and A.~Keller, ``Learning radio environments by differentiable
  ray tracing,'' \emph{IEEE Transactions on Machine Learning in Communications
  and Networking}, vol.~2, pp. 1527--1539, 2024.

\bibitem{ruah2024calibrating}
C.~Ruah, O.~Simeone, J.~Hoydis, and B.~Al-Hashimi, ``Calibrating wireless ray
  tracing for digital twinning using local phase error estimates,'' \emph{IEEE
  Transactions on Machine Learning in Communications and Networking}, vol.~2,
  pp. 1193--1215, 2024.

\bibitem{ren2023end}
Y.~Ren, S.~Guo, B.~Cao, and X.~Qiu, ``End-to-end network {SLA} quality
  assurance for {C-RAN}: A closed-loop management method based on digital twin
  network,'' \emph{IEEE Transactions on Mobile Computing}, vol.~23, no.~5, pp.
  4405--4422, 2023.

\bibitem{3gpp2025study}
{3GPP}, ``Study on management aspect of network digital twin,'' \emph{TR
  28.915}, 2025.

\bibitem{3gpp2025management}
------, ``Management and orchestration; management aspects of network digital
  twins,'' \emph{TS 28.561}, 2025.

\bibitem{akgun2024advancing}
B.~Akgun, A.~Jolly, B.~Sachdev, D.~Ravichandran, R.~Amiri, V.~Jain,
  M.~Jayabalan, Y.~Chen, H.~Pathak, V.~Chande \emph{et~al.}, ``Advancing next
  generation wireless networks with digital twin: Construction, validation, and
  real-world applications on an indoor over-the-air testbed,'' \emph{IEEE
  Access}, vol.~12, pp. 166\,298--166\,319, 2024.

\bibitem{thomas2024digital}
Y.~Thomas, S.~Toumpis, and N.~Smyrnioudis, ``Digital twin approach to
  estimating and utilizing the capacity region of wireless ad hoc networks,''
  \emph{Computer Networks}, vol. 241, p. 110213, 2024.

\bibitem{gao2023digital}
H.~Gao, P.~Ky{\"o}sti, X.~Zhang, and W.~Fan, ``Digital twin enabled {6G} radio
  testing: Concepts, challenges and solutions,'' \emph{IEEE Communications
  Magazine}, vol.~61, no.~11, pp. 88--94, 2023.

\bibitem{wiesmayr2024design}
R.~Wiesmayr, S.~Cammerer, F.~A. Aoudia, J.~Hoydis, J.~Zakrzewski, and
  A.~Keller, ``Design of a standard-compliant real-time neural receiver for {5G
  NR},'' in \emph{IEEE International Conference on Machine Learning for
  Communication and Networking}, 2025, pp. 1--6.

\bibitem{salehi2024multiverse}
B.~Salehi, U.~Demir, D.~Roy, S.~Pradhan, J.~Dy, S.~Ioannidis, and K.~Chowdhury,
  ``Multiverse at the edge: Interacting real world and digital twins for
  wireless beamforming,'' \emph{IEEE/ACM Transactions on Networking}, vol.~32,
  no.~4, pp. 3092--3110, 2024.

\bibitem{oran2024research}
{O-RAN nGRG}, ``Research report on digital twin {RAN} use cases,'' \emph{Report
  ID: RR-2024-07}, 2024.

\bibitem{groshev2021toward}
M.~Groshev, C.~Guimaraes, J.~Mart{\'\i}n-P{\'e}rez, and A.~de~la Oliva,
  ``Toward intelligent cyber-physical systems: Digital twin meets artificial
  intelligence,'' \emph{IEEE Communications Magazine}, vol.~59, no.~8, pp.
  14--20, 2021.

\bibitem{10437430}
A.~Kak, V.-Q. Pham, H.-T. Thieu, and N.~Choi, ``Ransight: Programmable
  telemetry for next-generation open radio access networks,'' in \emph{GLOBECOM
  2023 - 2023 IEEE Global Communications Conference}, 2023, pp. 5391--5396.

\bibitem{li2024digitalpart1}
L.~Li, H.~Sun, Z.~Wang, W.~Wang, and W.~Fan, ``Digital twins of electromagnetic
  propagation environments for live {5G} networks--{Part I}: Channel
  acquisition, {EM} simulation and verification,'' \emph{IEEE Transactions on
  Antennas and Propagation}, vol.~73, no.~4, pp. 2053--2064, 2025.

\bibitem{li2024digitalpart2}
L.~Li, S.~Lin, Y.~Zhong, Z.~Wang, and W.~Wang, ``Digital twins of
  electromagnetic propagation environments for live {5G} networks—{Part II}:
  High-fidelity emulation in the {MPAC} setup,'' \emph{IEEE Transactions on
  Antennas and Propagation}, vol.~73, no.~4, pp. 2065--2073, 2025.

\bibitem{pegurri2024toward}
R.~Pegurri, F.~Linsalata, E.~Moro, J.~Hoydis, and U.~Spagnolini, ``Toward
  digital network twins: Integrating {Sionna RT} in {NS3} for {6G} multi-{RAT}
  networks simulations,'' \emph{arXiv preprint arXiv:2501.00372}, 2024.

\bibitem{nguyen2021digital}
H.~X. Nguyen, R.~Trestian, D.~To, and M.~Tatipamula, ``Digital twin for {5G}
  and beyond,'' \emph{IEEE Communications Magazine}, vol.~59, no.~2, pp.
  10--15, 2021.

\bibitem{chen20215g}
Y.~Chen, X.~Lin, T.~Khan, M.~Afshang, and M.~Mozaffari, ``{5G} air-to-ground
  network design and optimization: A deep learning approach,'' in \emph{IEEE
  93rd Vehicular Technology Conference (VTC2021-Spring)}, 2021, pp. 1--6.

\bibitem{zhang2023gnn}
H.~Zhang, X.~Ma, X.~Liu, L.~Li, and K.~Sun, ``{GNN}-based power allocation and
  user association in digital twin network for the terahertz band,'' \emph{IEEE
  Journal on Selected Areas in Communications}, vol.~41, no.~10, pp.
  3111--3121, 2023.

\bibitem{liu2024coverage}
H.~Liu, T.~Li, F.~Jiang, W.~Su, and Z.~Wang, ``Coverage optimization for
  large-scale mobile networks with digital twin and multi-agent reinforcement
  learning,'' \emph{IEEE Transactions on Wireless Communications}, vol.~23,
  no.~12, pp. 18\,316--18\,330, 2024.

\bibitem{huang2024digital}
X.~Huang, H.~Yang, C.~Zhou, M.~He, X.~Shen, and W.~Zhuang, ``When digital twin
  meets generative {AI}: Intelligent closed-loop network management,''
  \emph{IEEE Network}, pp. 1--1, 2024.

\bibitem{villa2024colosseum}
D.~Villa, M.~Tehrani-Moayyed, C.~P. Robinson, L.~Bonati, P.~Johari, M.~Polese,
  and T.~Melodia, ``Colosseum as a digital twin: Bridging real-world
  experimentation and wireless network emulation,'' \emph{IEEE Transactions on
  Mobile Computing}, vol.~23, no.~10, pp. 9150--9166, 2024.

\bibitem{chen2024distributed}
Z.~Chen, W.~Yi, A.~Nallanathan, and J.~A. Chambers, ``{Distributed digital twin
  migration in multi-tier computing systems},'' \emph{IEEE Journal of Selected
  Topics in Signal Processing}, vol.~18, no.~1, pp. 109--123, 2024.

\bibitem{TranTWC22}
D.-H. Tran, V.-D. Nguyen, S.~Chatzinotas, T.~X. Vu, and B.~Ottersten, ``{UAV
  Relay-Assisted Emergency Communications in IoT Networks: Resource Allocation
  and Trajectory Optimization},'' \emph{IEEE Transactions on Wireless
  Communications}, vol.~21, no.~3, pp. 1621--1637, 2022.

\bibitem{PhuIoT21}
P.~X. Nguyen, D.-H. Tran, O.~Onireti, P.~T. Tin, S.~Q. Nguyen, S.~Chatzinotas,
  and H.~Vincent~Poor, ``{Backscatter-Assisted Data Offloading in OFDMA-Based
  Wireless-Powered Mobile Edge Computing for IoT Networks},'' \emph{IEEE
  Internet of Things Journal}, vol.~8, no.~11, pp. 9233--9243, 2021.

\bibitem{TranTVT20}
D.-H. Tran, T.~X. Vu, S.~Chatzinotas, S.~ShahbazPanahi, and B.~Ottersten,
  ``{Coarse Trajectory Design for Energy Minimization in UAV-Enabled},''
  \emph{IEEE Transactions on Vehicular Technology}, vol.~69, no.~9, pp.
  9483--9496, 2020.

\bibitem{alliance2023ORAN}
O.~Alliance, ``{O-RAN Empowering Vertical Industry: Scenarios, Solutions and
  Best Practice},'' \emph{White Paper, Dec.}, 2023.

\bibitem{ORANArchitecture24}
------, ``{O-RAN Architecture Description 11.0},'' \emph{White Paper, Feb.},
  2024.

\bibitem{ORANSlicingArchitecture}
------, ``{O-RAN.WG1.Slicing-Architecture-R003-v09.00},'' \emph{Specification,
  March}, 2023.

\bibitem{EricssionNDT22}
P.~Ohlen, C.~Johnston, H.~Olofsson, S.~Terrill, and F.~Chernogorov, ``{Network
  digital twins - outlook and opportunities},'' \emph{Ericsson Technology
  Review}, vol. 2022, no.~12, pp. 2--11, 2022.

\bibitem{almasanNDT22}
P.~Almasan, M.~Ferriol-Galmes, J.~Paillisse, J.~Suarez-Varela, D.~Perino,
  D.~Lopez, A.~A.~P. Perales, P.~Harvey, L.~Ciavaglia, L.~Wong, V.~Ram,
  S.~Xiao, X.~Shi, X.~Cheng, A.~Cabellos-Aparicio, and P.~Barlet-Ros,
  ``{Network Digital Twin: Context, Enabling Technologies, and
  Opportunities},'' \emph{IEEE Communications Magazine}, vol.~60, no.~11, pp.
  22--27, 2022.

\bibitem{NDT24}
{China Mobile, Telefonica, Huawei, Orange}, ``{Network Digital Twin: Concepts
  and Reference Architecture},'' \emph{Specification, March}, 2024.

\bibitem{mirzaei2023network}
J.~Mirzaei, I.~Abualhaol, and G.~Poitau, ``{Network Digital Twin for Open RAN:
  The Key Enablers, Standardization, and Use Cases},'' \emph{arXiv preprint
  arXiv:2308.02644}, 2023.

\bibitem{10.1145/3572864.3580329}
\BIBentryALTinterwordspacing
V.-Q. Pham, H.-T. Thieu, A.~Kak, and N.~Choi, ``Hexric: Building a better
  near-real time network controller for the open ran ecosystem,'' in
  \emph{Proceedings of the 24th International Workshop on Mobile Computing
  Systems and Applications}, ser. HotMobile '23.\hskip 1em plus 0.5em minus
  0.4em\relax New York, NY, USA: Association for Computing Machinery, 2023, p.
  15–21. [Online]. Available: \url{https://doi.org/10.1145/3572864.3580329}
\BIBentrySTDinterwordspacing

\bibitem{KumarNDTORAN23}
R.~Kumar, P.~Kumar, A.~Aljuhani, A.~Jolfaei, A.~N. Islam, and N.~Mohammad,
  ``{Secure Data Dissemination Scheme for Digital Twin Empowered Vehicular
  Networks in Open RAN},'' \emph{IEEE Transactions on Vehicular Technology},
  pp. 1--13, 2023.

\bibitem{AkramNDTORAN24}
J.~Akram, A.~Anaissi, R.~S. Rathore, R.~H. Jhaveri, and A.~Akram, ``{Digital
  Twin-Driven Trust Management in Open RAN-Based Spatial Crowdsourcing Drone
  Services},'' \emph{IEEE Transactions on Green Communications and Networking},
  pp. 1--1, 2024.

\bibitem{rumeshfederated}
Y.~Rumesh, D.~Attanayaka, P.~Porambage, J.~Pinola, J.~Groen, and K.~Chowdhury,
  ``{Federated Learning for Anomaly Detection in Open RAN: Security
  Architecture Within a Digital Twin},'' 2024.

\bibitem{DemoNDTORAN}
P.~Li, A.~Aijaz, T.~Farnham, S.~Gufran, and S.~Chintalapati, ``{Demo: A Digital
  Twin of the 5G Radio Access Network for Anomaly Detection Functionality},''
  in \emph{2023 IEEE 31st International Conference on Network Protocols
  (ICNP)}, 2023, pp. 1--2.

\bibitem{colosseum}
M.~Polese, L.~Bonati, S.~D'Oro, P.~Johari, D.~Villa, S.~Velumani, R.~Gangula,
  M.~Tsampazi, C.~P. Robinson, G.~Gemmi \emph{et~al.}, ``{Colosseum: The Open
  RAN Digital Twin},'' \emph{arXiv preprint arXiv:2404.17317}, 2024.

\bibitem{DTORAN_Energy}
A.~Ndikumana, K.~K. Nguyen, and M.~Cheriet, ``{Digital Twin Assisted
  Closed-Loops for Energy-Efficient Open RAN-Based Fixed Wireless Access
  Provisioning in Rural Areas},'' in \emph{GLOBECOM 2023 - 2023 IEEE Global
  Communications Conference}, 2023, pp. 6285--6290.

\bibitem{TinIoT2024}
P.~T. Tin, M.-S.~V. Nguyen, D.-H. Tran, C.~T. Nguyen, S.~Chatzinotas, Z.~Ding,
  and M.~Voznak, ``{Performance Analysis of User Pairing for Active RIS-Enabled
  Cooperative NOMA in 6G Cognitive Radio Networks},'' \emph{IEEE Internet of
  Things Journal}, vol.~11, no.~23, pp. 37\,675--37\,692, 2024.

\bibitem{long2025deep}
N.~Q. Long, V.-H. Dang, T.~D. Ho, H.~Tran, S.~Chatzinotas, D.-H. Tran,
  S.~Sanguanpong, C.~So-In \emph{et~al.}, ``{Deep Learning-Driven Throughput
  Maximization in Covert Communication for UAV-RIS Cognitive Systems},''
  \emph{IEEE Open Journal of the Communications Society}, 2025.

\bibitem{Oliveri2022Towards}
G.~Oliveri, M.~Salucci, and A.~Massa, ``Towards efficient reflectarray digital
  twins - an em-driven machine learning perspective,'' \emph{IEEE Transactions
  on Antennas and Propagation}, vol.~70, no.~7, pp. 5078--5093, 2022.

\bibitem{Cui2023Digital}
Y.~Cui, T.~Lv, W.~Ni, and A.~Jamalipour, ``Digital twin-aided learning for
  managing reconfigurable intelligent surface-assisted, uplink, user-centric
  cell-free systems,'' \emph{IEEE Journal on Selected Areas in Communications},
  vol.~41, no.~10, pp. 3175--3190, 2023.

\bibitem{Wang2021optical}
D.~Wang, Z.~Zhang, M.~Zhang, M.~Fu, J.~Li, S.~Cai, C.~Zhang, and X.~Chen, ``The
  role of digital twin in optical communication: Fault management, hardware
  configuration, and transmission simulation,'' \emph{IEEE Communications
  Magazine}, vol.~59, no.~1, pp. 133--139, 2021.

\bibitem{Sequeira2023IQ}
D.~Sequeira, M.~Ruiz, N.~Costa, A.~Napoli, J.~Pedro, and L.~Velasco, ``Ocata: a
  deep-learning-based digital twin for the optical time domain,'' \emph{Journal
  of Optical Communications and Networking}, vol.~15, no.~2, pp. 87--97, 2023.

\bibitem{Devigili2024IQ}
M.~Devigili, M.~Ruiz, N.~Costa, C.~Castro, A.~Napoli, J.~Pedro, and L.~Velasco,
  ``Applications of the ocata time domain digital twin: from qot estimation to
  failure management,'' \emph{Journal of Optical Communications and
  Networking}, vol.~16, no.~2, pp. 221--232, 2024.

\bibitem{Vilalta2023validDT}
R.~Vilalta, L.~Gifre, R.~Casellas, R.~Muñoz, R.~Martínez, A.~Mozo, A.~Pastor,
  D.~López, and J.~P. Fernández-Palacios, ``Applying digital twins to optical
  networks with cloud-native sdn controllers,'' \emph{IEEE Communications
  Magazine}, vol.~61, no.~12, pp. 128--134, 2023.

\bibitem{Zhang2023large}
Y.~Zhang, M.~Zhang, Y.~Song, Y.~Shi, C.~Zhang, C.~Ju, B.~Guo, S.~Huang, and
  D.~Wang, ``Building a digital twin for large-scale and dynamic c+l-band
  optical networks,'' \emph{Journal of Optical Communications and Networking},
  vol.~15, no.~12, pp. 985--998, 2023.

\bibitem{Yichen2023EDFA1}
Y.~Liu, X.~Liu, Y.~Zhang, M.~Cai, M.~Fu, X.~Zhong, L.~Yi, W.~Hu, and Q.~Zhuge,
  ``Building a digital twin of an edfa for optical networks: a gray-box
  modeling approach,'' \emph{Journal of Optical Communications and Networking},
  vol.~15, no.~11, pp. 830--838, 2023.

\bibitem{Yichen2023EDFA2}
\BIBentryALTinterwordspacing
X.~Liu, Y.~Zhang, Y.~Chen, Y.~Liu, M.~Cai, Q.~Qiu, M.~Fu, L.~Yi, W.~Hu, and
  Q.~Zhuge, ``{Digital twin modeling and controlling of optical power evolution
  enabling autonomous-driving optical networks: a Bayesian approach},''
  \emph{Advanced Photonics}, vol.~6, no.~2, p. 026006, 2024. [Online].
  Available: \url{https://doi.org/10.1117/1.AP.6.2.026006}
\BIBentrySTDinterwordspacing

\bibitem{Wang2024Rodam}
R.~Wang, J.~Zhang, Z.~Gu, M.~Ibrahimi, B.~Zhang, F.~Musumeci, M.~Tornatore, and
  Y.~Ji, ``Digital-twin-assisted meta learning for soft-failure localization in
  roadm-based optical networks,'' \emph{Journal of Optical Communications and
  Networking}, vol.~16, no.~7, pp. C11--C19, 2024.

\bibitem{Qu2022FedTwin}
Y.~Qu, L.~Gao, Y.~Xiang, S.~Shen, and S.~Yu, ``Fedtwin: Blockchain-enabled
  adaptive asynchronous federated learning for digital twin networks,''
  \emph{IEEE Network}, vol.~36, no.~6, pp. 183--190, 2022.

\bibitem{Aloqaily2023Reinforcing}
M.~Aloqaily, I.~A. Ridhawi, and S.~Kanhere, ``Reinforcing industry 4.0 with
  digital twins and blockchain-assisted federated learning,'' \emph{IEEE
  Journal on Selected Areas in Communications}, vol.~41, no.~11, pp.
  3504--3516, 2023.

\bibitem{Lv2023Blockchain}
Z.~Lv, C.~Cheng, and H.~Lv, ``Blockchain-based decentralized learning for
  security in digital twins,'' \emph{IEEE Internet of Things Journal}, vol.~10,
  no.~24, pp. 21\,479--21\,488, 2023.

\bibitem{Prathiba2024Fortifying}
S.~B. Prathiba, Y.~Govindarajan, V.~P.~A. Ganesan, A.~Ramachandran, A.~K.
  Selvaraj, A.~K. Bashir, and T.~R. Gadekallu, ``Fortifying federated learning
  in iiot: Leveraging blockchain and digital twin innovations for enhanced
  security and resilience,'' \emph{IEEE Access}, pp. 1--1, 2024.

\bibitem{Mu2023Digital}
J.~Mu, W.~Ouyang, T.~Hong, W.~Yuan, Y.~Cui, and Z.~Jing, ``Digital
  twins-enabled federated learning in mobile networks: From the perspective of
  communication-assisted sensing,'' \emph{IEEE Journal on Selected Areas in
  Communications}, vol.~41, no.~10, pp. 3230--3241, 2023.

\bibitem{kaada2024multi}
S.~Kaada, D.-H. Tran, N.~Van~Huynh, M.-L.~A. Morel, S.~Jelassi, and G.~Rubino,
  ``{Multi-Agent Deep Reinforcement Learning for Resilience Optimization in 5G
  RAN},'' \emph{arXiv preprint arXiv:2407.18066}, 2024.

\bibitem{Nokia6GSystem}
Nokia, ``{5G System Principles},'' \emph{IEEE Courses}, 2019.

\bibitem{Dell6GCore}
G.~Gaurav and G.~Kevin, ``{The 5G Core Network Demystified},''
  \emph{https://infohub.delltechnologies.com/en-us/p/the-5g-core-network-demystified/},
  2023.

\bibitem{ETSI6GCore}
ETSI, ``{System Architecture for the 5G System (3GPP TS 23.501 version 15.3.0
  Release 15)},''
  \emph{https://www.etsi.org/deliver/etsi{\_}ts/123500{\_}123599/123501/15.03.00
  {\_}60/ts{\_}123501v150300p.pdf}, 2023.

\bibitem{zheng2022research}
Y.~Zheng, C.~Lin, S.~Huang, W.~Li, K.~Wang, and J.~Lou, ``{Research and
  Application of Configuration System in Core Network Based on Digital Twin},''
  in \emph{2022 IEEE 5th Advanced Information Management, Communicates,
  Electronic and Automation Control Conference (IMCEC)}, vol.~5.\hskip 1em plus
  0.5em minus 0.4em\relax IEEE, 2022, pp. 293--297.

\bibitem{ZerotouchNDT23}
S.~Rani, H.~Babbar, M.~Krichen, K.~Yu, and F.~H. Memon, ``{Network Slicing for
  Zero-Touch Networks: A Top-Notch Technology},'' \emph{IEEE Network}, vol.~37,
  no.~5, pp. 16--24, 2023.

\bibitem{SanzCore23}
M.~Sanz~Rodrigo, D.~Rivera, J.~I. Moreno, M.~Alvarez-Campana, and D.~R. Lopez,
  ``{Digital Twins for 5G Networks: A Modeling and Deployment Methodology},''
  \emph{IEEE Access}, vol.~11, pp. 38\,112--38\,126, 2023.

\bibitem{TaoCore24}
Z.~Tao, Y.~Guo, G.~He, Y.~Huang, and X.~You, ``{Deep Learning-Based Modeling of
  5G Core Control Plane for 5G Network Digital Twin},'' \emph{IEEE Transactions
  on Cognitive Communications and Networking}, vol.~10, no.~1, pp. 238--251,
  2024.

\bibitem{wang2024digital}
J.~Wang, J.~Li, and J.~Liu, ``{Digital Twin-Assisted Flexible Slice Admission
  Control for 5G Core Network: A Deep Reinforcement Learning Approach},''
  \emph{Future Generation Computer Systems}, vol. 153, pp. 467--476, 2024.

\bibitem{GeiBlerCORE23}
S.~GeiBler, F.~Wamser, W.~Bauer, S.~Gebert, S.~Kounev, and T.~HoBfeld,
  ``{MVNOCoreSim: A Digital Twin for Virtualized IoT-Centric Mobile Core
  Networks},'' \emph{IEEE Internet of Things Journal}, vol.~10, no.~15, pp.
  13\,974--13\,987, 2023.

\bibitem{Bilen5GCORE22}
T.~Bilen, E.~Ak, B.~Bal, and B.~Canberk, ``{A Proof of Concept on Digital
  Twin-Controlled WiFi Core Network Selection for In-Flight Connectivity},''
  \emph{IEEE Communications Standards Magazine}, vol.~6, no.~3, pp. 60--68,
  2022.

\bibitem{YigitDDOSCORE22}
Y.~Yigit, B.~Bal, A.~Karameseoglu, T.~Q. Duong, and B.~Canberk, ``{Digital
  Twin-Enabled Intelligent DDoS Detection Mechanism for Autonomous Core
  Networks},'' \emph{IEEE Communications Standards Magazine}, vol.~6, no.~3,
  pp. 38--44, 2022.

\bibitem{mohamed2023leveraging}
N.~Mohamed, J.~Al-Jaroodi, I.~Jawhar, and N.~Kesserwan, ``Leveraging digital
  twins for healthcare systems engineering,'' \emph{Ieee Access}, 2023.

\bibitem{liu2019novel}
Y.~Liu, L.~Zhang, Y.~Yang, L.~Zhou, L.~Ren, F.~Wang, R.~Liu, Z.~Pang, and M.~J.
  Deen, ``A novel cloud-based framework for the elderly healthcare services
  using digital twin,'' \emph{IEEE access}, vol.~7, pp. 49\,088--49\,101, 2019.

\bibitem{zheng2021towards}
Y.~Zheng, R.~Lu, Y.~Guan, S.~Zhang, and J.~Shao, ``Towards private similarity
  query based healthcare monitoring over digital twin cloud platform,'' in
  \emph{2021 IEEE/ACM 29th International Symposium on Quality of Service
  (IWQOS)}.\hskip 1em plus 0.5em minus 0.4em\relax IEEE, 2021, pp. 1--10.

\bibitem{han2022cloud}
J.~Han, Q.~Hong, M.~H. Syed, M.~A.~U. Khan, G.~Yang, G.~Burt, and C.~Booth,
  ``Cloud-edge hosted digital twins for coordinated control of distributed
  energy resources,'' \emph{IEEE Transactions on Cloud Computing}, vol.~11,
  no.~2, pp. 1242--1256, 2022.

\bibitem{zhang2022practical}
S.~Zhang, A.~Pandey, X.~Luo, M.~Powell, R.~Banerji, L.~Fan, A.~Parchure, and
  E.~Luzcando, ``Practical adoption of cloud computing in power
  systems—drivers, challenges, guidance, and real-world use cases,''
  \emph{IEEE Transactions on Smart Grid}, vol.~13, no.~3, pp. 2390--2411, 2022.

\bibitem{stergiou2022digital}
C.~L. Stergiou and K.~E. Psannis, ``Digital twin intelligent system for
  industrial iot-based big data management and analysis in cloud,''
  \emph{Virtual Reality \& Intelligent Hardware}, vol.~4, no.~4, pp. 279--291,
  2022.

\bibitem{mikhailov2022new}
A.~Mikhailov, S.~Tretyakov, and Y.~Andreev, ``A new approach to build
  industrial internet of things (iiot) systems based on digital twin's
  technologies,'' in \emph{2022 International conference on industrial
  engineering, applications and manufacturing (ICIEAM)}.\hskip 1em plus 0.5em
  minus 0.4em\relax IEEE, 2022, pp. 1091--1095.

\bibitem{huang2021digital}
H.~Huang, L.~Yang, Y.~Wang, X.~Xu, and Y.~Lu, ``Digital twin-driven online
  anomaly detection for an automation system based on edge intelligence,''
  \emph{Journal of Manufacturing Systems}, vol.~59, pp. 138--150, 2021.

\bibitem{pan2021digital}
Y.~Pan, T.~Qu, N.~Wu, M.~Khalgui, and G.~Huang, ``Digital twin based real-time
  production logistics synchronization system in a multi-level computing
  architecture,'' \emph{Journal of Manufacturing Systems}, vol.~58, pp.
  246--260, 2021.

\bibitem{dang2021cloud}
H.~V. Dang, M.~Tatipamula, and H.~X. Nguyen, ``Cloud-based digital twinning for
  structural health monitoring using deep learning,'' \emph{IEEE Transactions
  on Industrial Informatics}, vol.~18, no.~6, pp. 3820--3830, 2021.

\bibitem{dai2022adaptive}
Y.~Dai and Y.~Zhang, ``Adaptive digital twin for vehicular edge computing and
  networks,'' \emph{Journal of Communications and Information Networks},
  vol.~7, no.~1, pp. 48--59, 2022.

\bibitem{9830363}
T.~Q. Duong, D.~Van~Huynh, Y.~Li, E.~Garcia-Palacios, and K.~Sun, ``Digital
  twin-enabled 6g aerial edge computing with ultra-reliable and low-latency
  communications : (invited paper),'' in \emph{2022 1st International
  Conference on 6G Networking (6GNet)}, 2022, pp. 1--5.

\bibitem{karobi2024ecoedgetwin}
S.~H. Karobi, S.~Ahmed, S.~R. Sabuj, and A.~Khokhar, ``Ecoedgetwin: Enhanced 6g
  network via mobile edge computing and digital twin integration,'' \emph{arXiv
  preprint arXiv:2405.06507}, 2024.

\bibitem{lu2021adaptive}
Y.~Lu, S.~Maharjan, and Y.~Zhang, ``Adaptive edge association for wireless
  digital twin networks in 6g,'' \emph{IEEE Internet of Things Journal},
  vol.~8, no.~22, pp. 16\,219--16\,230, 2021.

\bibitem{yang2023edge}
P.~Yang, J.~Hou, L.~Yu, W.~Chen, and Y.~Wu, ``Edge-coordinated energy-efficient
  video analytics for digital twin in 6g,'' \emph{China Communications},
  vol.~20, no.~2, pp. 14--25, 2023.

\bibitem{wang2024human}
C.~Wang, Z.~Cai, and Y.~Li, ``Human activity recognition in mobile edge
  computing: A low-cost and high-fidelity digital twin approach with deep
  reinforcement learning,'' \emph{IEEE Transactions on Consumer Electronics},
  2024.

\bibitem{xiang2023realizing}
H.~Xiang, C.~Yi, K.~Wu, J.~Chen, J.~Cai, D.~Niyato, and X.~Shen, ``Realizing
  immersive communications in human digital twin by edge computing empowered
  tactile internet: Visions and case study,'' \emph{IEEE Communications
  Magazine}, 2023.

\bibitem{martinez2019cardio}
R.~Martinez-Velazquez, R.~Gamez, and A.~El~Saddik, ``Cardio twin: A digital
  twin of the human heart running on the edge,'' in \emph{2019 IEEE
  international symposium on medical measurements and applications
  (MeMeA)}.\hskip 1em plus 0.5em minus 0.4em\relax IEEE, 2019, pp. 1--6.

\bibitem{ouahabi2021distributed}
N.~Ouahabi, A.~Chebak, M.~Zegrari, O.~Kamach, and M.~Berquedich, ``A
  distributed digital twin architecture for shop floor monitoring based on
  edge-cloud collaboration,'' in \emph{2021 Third International Conference on
  Transportation and Smart Technologies (TST)}.\hskip 1em plus 0.5em minus
  0.4em\relax IEEE, 2021, pp. 72--78.

\bibitem{glatt2021edge}
M.~Glatt, P.~K{\"o}lsch, C.~Siedler, P.~Langlotz, S.~Ehmsen, and J.~C. Aurich,
  ``Edge-based digital twin to trace and ensure sustainability in cross-company
  production networks,'' \emph{Procedia CIRP}, vol.~98, pp. 276--281, 2021.

\bibitem{girletti2020intelligent}
L.~Girletti, M.~Groshev, C.~Guimar{\~a}es, C.~J. Bernardos, and A.~de~la Oliva,
  ``An intelligent edge-based digital twin for robotics,'' in \emph{2020 IEEE
  Globecom Workshops (GC Wkshps}.\hskip 1em plus 0.5em minus 0.4em\relax IEEE,
  2020, pp. 1--6.

\bibitem{van2022edge}
D.~Van~Huynh, S.~R. Khosravirad, A.~Masaracchia, O.~A. Dobre, and T.~Q. Duong,
  ``Edge intelligence-based ultra-reliable and low-latency communications for
  digital twin-enabled metaverse,'' \emph{IEEE Wireless Communications
  Letters}, vol.~11, no.~8, pp. 1733--1737, 2022.

\bibitem{Dong2019Deep}
R.~Dong, C.~She, W.~Hardjawana, Y.~Li, and B.~Vucetic, ``{Deep Learning for
  Hybrid 5G Services in Mobile Edge Computing Systems: Learn From a Digital
  Twin},'' \emph{IEEE Transactions on Wireless Communications}, vol.~18,
  no.~10, pp. 4692--4707, 2019.

\bibitem{Lv2022Edge}
Z.~Lv and R.~Lou, ``Edge-fog-cloud secure storage with deep-learning-assisted
  digital twins,'' \emph{IEEE Internet of Things Magazine}, vol.~5, no.~2, pp.
  36--40, 2022.

\bibitem{Sanchez2024Building}
L.~Roda-Sanchez, F.~Cirillo, G.~Solmaz, T.~Jacobs, C.~Garrido-Hidalgo,
  T.~Olivares, and E.~Kovacs, ``{Building a Smart Campus Digital Twin: System,
  Analytics, and Lessons Learned From a Real-World Project},'' \emph{IEEE
  Internet of Things Journal}, vol.~11, no.~3, pp. 4614--4627, 2024.

\bibitem{Lu2021Comm}
Y.~Lu, X.~Huang, K.~Zhang, S.~Maharjan, and Y.~Zhang,
  ``{Communication-Efficient Federated Learning for Digital Twin Edge Networks
  in Industrial IoT},'' \emph{IEEE Transactions on Industrial Informatics},
  vol.~17, no.~8, pp. 5709--5718, 2021.

\bibitem{Yang2023Hyper}
J.~Yang, W.~Jiang, and L.~Nie, ``{Hypernetworks-Based Hierarchical Federated
  Learning on Hybrid Non-IID Datasets for Digital Twin in Industrial IoT},''
  \emph{IEEE Transactions on Network Science and Engineering}, vol.~11, no.~2,
  pp. 1413--1423, 2024.

\bibitem{Zhang2024Digital}
R.~Zhang, Z.~Xie, D.~Yu, W.~Liang, and X.~Cheng, ``{Digital Twin-Assisted
  Federated Learning Service Provisioning Over Mobile Edge Networks},''
  \emph{IEEE Transactions on Computers}, vol.~73, no.~2, pp. 586--598, 2024.

\bibitem{tran2024deep}
D.-H. Tran, N.~Van~Huynh, S.~Kaada, V.~N. Vo, E.~Lagunas, and S.~Chatzinotas,
  ``{Deep Reinforcement Learning for Network Energy Saving in 6G and Beyond
  Networks},'' \emph{arXiv preprint arXiv:2408.10974}, 2024.

\bibitem{Elayan2021Digital}
H.~Elayan, M.~Aloqaily, and M.~Guizani, ``Digital twin for intelligent
  context-aware iot healthcare systems,'' \emph{IEEE Internet of Things
  Journal}, vol.~8, no.~23, pp. 16\,749--16\,757, 2021.

\bibitem{Yu2023FMCPNN}
Z.~Yu, K.~Wang, Z.~Wan, S.~Xie, and Z.~Lv, ``Fmcpnn in digital twins smart
  healthcare,'' \emph{IEEE Consumer Electronics Magazine}, vol.~12, no.~4, pp.
  66--73, 2023.

\bibitem{Tai2022Digital}
Y.~Tai, L.~Zhang, Q.~Li, C.~Zhu, V.~Chang, J.~J. P.~C. Rodrigues, and
  M.~Guizani, ``Digital-twin-enabled iomt system for surgical simulation using
  rac-gan,'' \emph{IEEE Internet of Things Journal}, vol.~9, no.~21, pp.
  20\,918--20\,931, 2022.

\bibitem{Khan2023ANovel}
S.~Khan, A.~Alzaabi, Z.~Iqbal, T.~Ratnarajah, and T.~Arslan, ``A novel digital
  twin (dt) model based on wifi csi, signal processing and machine learning for
  patient respiration monitoring and decision-support,'' \emph{IEEE Access},
  vol.~11, pp. 103\,554--103\,568, 2023.

\bibitem{Dang2022Cloud}
H.~V. Dang, M.~Tatipamula, and H.~X. Nguyen, ``Cloud-based digital twinning for
  structural health monitoring using deep learning,'' \emph{IEEE Transactions
  on Industrial Informatics}, vol.~18, no.~6, pp. 3820--3830, 2022.

\bibitem{Mortlock2022Graph}
T.~Mortlock, D.~Muthirayan, S.-Y. Yu, P.~P. Khargonekar, and M.~Abdullah
  Al~Faruque, ``Graph learning for cognitive digital twins in manufacturing
  systems,'' \emph{IEEE Transactions on Emerging Topics in Computing}, vol.~10,
  no.~1, pp. 34--45, 2022.

\bibitem{Ren2022Machine}
Z.~Ren, J.~Wan, and P.~Deng, ``Machine-learning-driven digital twin for
  lifecycle management of complex equipment,'' \emph{IEEE Transactions on
  Emerging Topics in Computing}, vol.~10, no.~1, pp. 9--22, 2022.

\bibitem{Sampedro20233D}
G.~A.~R. Sampedro, M.~A.~P. Putra, and M.~Abisado, ``3d-amplifai: An ensemble
  machine learning approach to digital twin fault monitoring for additive
  manufacturing in smart factories,'' \emph{IEEE Access}, vol.~11, pp.
  64\,128--64\,140, 2023.

\bibitem{Jyeniskhan2023Integrating}
N.~Jyeniskhan, A.~Keutayeva, G.~Kazbek, M.~H. Ali, and E.~Shehab, ``Integrating
  machine learning model and digital twin system for additive manufacturing,''
  \emph{IEEE Access}, vol.~11, pp. 71\,113--71\,126, 2023.

\bibitem{Qiao2024Resources}
D.~Qiao, M.~Li, S.~Guo, J.~Zhao, and B.~Xiao, ``Resources-efficient adaptive
  federated learning for digital twin-enabled iiot,'' \emph{IEEE Transactions
  on Network Science and Engineering}, pp. 1--13, 2024.

\bibitem{Castellani2021Real}
A.~Castellani, S.~Schmitt, and S.~Squartini, ``Real-world anomaly detection by
  using digital twin systems and weakly supervised learning,'' \emph{IEEE
  Transactions on Industrial Informatics}, vol.~17, no.~7, pp. 4733--4742,
  2021.

\bibitem{Li2023ANovel}
Z.~Li, M.~Duan, B.~Xiao, and S.~Yang, ``A novel anomaly detection method for
  digital twin data using deconvolution operation with attention mechanism,''
  \emph{IEEE Transactions on Industrial Informatics}, vol.~19, no.~5, pp.
  7278--7286, 2023.

\bibitem{Wang2023DTITD}
Z.~Q. Wang and A.~El~Saddik, ``Dtitd: An intelligent insider threat detection
  framework based on digital twin and self-attention based deep learning
  models,'' \emph{IEEE Access}, vol.~11, pp. 114\,013--114\,030, 2023.

\bibitem{Boche2024On}
H.~Boche, R.~F. Schaefer, H.~V. Poor, and F.~H.~P. Fitzek, ``On the need of
  neuromorphic twins to detect denial-of-service attacks on communication
  networks,'' \emph{IEEE/ACM Transactions on Networking}, pp. 1--13, 2024.

\bibitem{Lv2024Secure}
Z.~Lv, D.~Chen, B.~Cao, H.~Song, and H.~Lv, ``Secure deep learning in defense
  in deep-learning-as-a-service computing systems in digital twins,''
  \emph{IEEE Transactions on Computers}, vol.~73, no.~3, pp. 656--668, 2024.

\bibitem{Kumar2023Blockchain}
P.~Kumar, R.~Kumar, A.~Kumar, A.~A. Franklin, S.~Garg, and S.~Singh,
  ``Blockchain and deep learning for secure communication in digital twin
  empowered industrial iot network,'' \emph{IEEE Transactions on Network
  Science and Engineering}, vol.~10, no.~5, pp. 2802--2813, 2023.

\bibitem{Belousov2024AMachine}
Y.~Belousov, G.~Quétant, B.~Pulfer, R.~Chaban, J.~Tutt, O.~Taran, T.~Holotyak,
  and S.~Voloshynovskiy, ``A machine learning-based digital twin for
  anti-counterfeiting applications with copy detection patterns,'' \emph{IEEE
  Transactions on Information Forensics and Security}, vol.~19, pp. 3395--3408,
  2024.

\bibitem{Zhao2023ELITE}
L.~Zhao, Z.~Bi, A.~Hawbani, K.~Yu, Y.~Zhang, and M.~Guizani, ``Elite: An
  intelligent digital twin-based hierarchical routing scheme for softwarized
  vehicular networks,'' \emph{IEEE Transactions on Mobile Computing}, vol.~22,
  no.~9, pp. 5231--5247, 2023.

\bibitem{Zhou2023Digital}
X.~Zhou, X.~Zheng, X.~Cui, J.~Shi, W.~Liang, Z.~Yan, L.~T. Yang, S.~Shimizu,
  and K.~I.-K. Wang, ``Digital twin enhanced federated reinforcement learning
  with lightweight knowledge distillation in mobile networks,'' \emph{IEEE
  Journal on Selected Areas in Communications}, vol.~41, no.~10, pp.
  3191--3211, 2023.

\bibitem{wigard2023ubiquitous}
J.~Wigard, E.~Juan, J.~Stanczak, M.~Lauridsen, A.~Marcone, S.~Hoppe,
  A.~Ahmadzadeh, A.~Masri, and D.-H. Tran, ``{Ubiquitous 6G Service Through
  Non-Terrestrial Networks},'' \emph{IEEE Wireless Communications}, vol.~30,
  no.~6, pp. 12--18, 2023.

\bibitem{nguyen2024emerging}
C.~T. Nguyen, Y.~M. Saputra, N.~Van~Huynh, T.~N. Nguyen, D.~T. Hoang, D.~N.
  Nguyen, V.-Q. Pham, M.~Voznak, S.~Chatzinotas, and D.-H. Tran, ``{Emerging
  technologies for 6G non-terrestrial-networks: From academia to industrial
  applications},'' \emph{IEEE Open Journal of the Communications Society},
  2024.

\bibitem{nguyen2023security}
T.~N. Nguyen, T.~Van~Chien, D.-H. Tran, V.-D. Phan, M.~Voznak, S.~Chatzinotas,
  Z.~Ding, and H.~V. Poor, ``{Security-Reliability Tradeoffs for
  Satellite--Terrestrial Relay Networks With a Friendly Jammer and Imperfect
  CSI},'' \emph{IEEE Transactions on Aerospace and Electronic Systems},
  vol.~59, no.~5, pp. 7004--7019, 2023.

\bibitem{nguyen2022security}
T.~N. Nguyen, D.-H. Tran, T.~Van~Chien, V.-D. Phan, M.~Voznak, and
  S.~Chatzinotas, ``{Security and reliability analysis of satellite-terrestrial
  multirelay networks with imperfect CSI},'' \emph{IEEE Systems Journal}, 2022.

\bibitem{nguyen2022outage}
T.~N. Nguyen, L.-T. Tu, D.-H. Tran, V.-D. Phan, M.~Voznak, S.~Chatzinotas, and
  Z.~Ding, ``{Outage performance of satellite terrestrial full-duplex relaying
  networks with co-channel interference},'' \emph{IEEE Wireless Communications
  Letters}, vol.~11, no.~7, pp. 1478--1482, 2022.

\bibitem{Zhou2023HDTSatnet}
Y.~Zhou, R.~Zhang, J.~Liu, T.~Huang, Q.~Tang, and F.~R. Yu, ``A hierarchical
  digital twin network for satellite communication networks,'' \emph{IEEE
  Communications Magazine}, vol.~61, no.~11, pp. 104--110, 2023.

\bibitem{Zhao2022interlink}
L.~Zhao, C.~Wang, K.~Zhao, D.~Tarchi, S.~Wan, and N.~Kumar, ``Interlink: A
  digital twin-assisted storage strategy for satellite-terrestrial networks,''
  \emph{IEEE Transactions on Aerospace and Electronic Systems}, vol.~58, no.~5,
  pp. 3746--3759, 2022.

\bibitem{Fan2023contact}
H.~Fan, J.~Long, L.~Liu, and Z.~Yang, ``Dynamic digital twin and online
  scheduling for contact window resources in satellite network,'' \emph{IEEE
  Transactions on Industrial Informatics}, vol.~19, no.~5, pp. 7217--7227,
  2023.

\bibitem{Xu2023MlSatIot}
X.~Xu, H.~Wen, H.~Song, and Y.~Zhao, ``A dt machine learning-based satellite
  orbit prediction for iot applications,'' \emph{IEEE Internet of Things
  Magazine}, vol.~6, no.~2, pp. 96--100, 2023.

\bibitem{tran2022throughput}
D.-H. Tran, S.~Chatzinotas, and B.~Ottersten, ``{Throughput maximization for
  backscatter-and cache-assisted wireless powered UAV technology},'' \emph{IEEE
  Transactions on Vehicular Technology}, vol.~71, no.~5, pp. 5187--5202, 2022.

\bibitem{tran2022satellite}
------, ``{Satellite-and cache-assisted UAV: A joint cache placement, resource
  allocation, and trajectory optimization for 6G aerial networks},'' \emph{IEEE
  Open Journal of Vehicular Technology}, vol.~3, pp. 40--54, 2022.

\bibitem{Lei2021Toward}
L.~Lei, G.~Shen, L.~Zhang, and Z.~Li, ``Toward intelligent cooperation of uav
  swarms: When machine learning meets digital twin,'' \emph{IEEE Network},
  vol.~35, no.~1, pp. 386--392, 2021.

\bibitem{Sun2022Dynamic}
W.~Sun, N.~Xu, L.~Wang, H.~Zhang, and Y.~Zhang, ``{Dynamic Digital Twin and
  Federated Learning With Incentives for Air-Ground Networks},'' \emph{IEEE
  Transactions on Network Science and Engineering}, vol.~9, no.~1, pp.
  321--333, 2022.

\bibitem{Sun2023Lightweight}
W.~Sun, S.~Lian, H.~Zhang, and Y.~Zhang, ``{Lightweight Digital Twin and
  Federated Learning With Distributed Incentive in Air-Ground 6G Networks},''
  \emph{IEEE Transactions on Network Science and Engineering}, vol.~10, no.~3,
  pp. 1214--1227, 2023.

\bibitem{Tang2023Digital}
X.~Tang, X.~Li, R.~Yu, Y.~Wu, J.~Ye, F.~Tang, and Q.~Chen,
  ``{Digital-Twin-Assisted Task Assignment in Multi-UAV Systems: A Deep
  Reinforcement Learning Approach},'' \emph{IEEE Internet of Things Journal},
  vol.~10, no.~17, pp. 15\,362--15\,375, 2023.

\bibitem{Hazarika2024Hybrid}
B.~Hazarika, K.~Singh, A.~Paul, and T.~Q. Duong, ``{Hybrid Machine Learning
  Approach for Resource Allocation of Digital Twin in UAV-Aided
  Internet-of-Vehicles Networks},'' \emph{IEEE Transactions on Intelligent
  Vehicles}, vol.~9, no.~1, pp. 2923--2939, 2024.

\bibitem{Zhao2024traject}
L.~Zhao, S.~Li, Y.~Guan, S.~Wan, A.~Hawbani, Y.~Bi, and M.~Guizani, ``Adaptive
  multi-uav trajectory planning leveraging digital twin technology for urban
  iiot applications,'' \emph{IEEE Transactions on Network Science and
  Engineering}, vol.~11, no.~6, pp. 5349--5363, 2024.

\bibitem{Shen2022flock}
G.~Shen, L.~Lei, Z.~Li, S.~Cai, L.~Zhang, P.~Cao, and X.~Liu, ``Deep
  reinforcement learning for flocking motion of multi-uav systems: Learn from a
  digital twin,'' \emph{IEEE Internet of Things Journal}, vol.~9, no.~13, pp.
  11\,141--11\,153, 2022.

\bibitem{Tang2023task}
X.~Tang, X.~Li, R.~Yu, Y.~Wu, J.~Ye, F.~Tang, and Q.~Chen,
  ``Digital-twin-assisted task assignment in multi-uav systems: A deep
  reinforcement learning approach,'' \emph{IEEE Internet of Things Journal},
  vol.~10, no.~17, pp. 15\,362--15\,375, 2023.

\bibitem{Deng2023task2}
M.~Deng, Z.~Yao, X.~Li, H.~Wang, A.~Nallanathan, and Z.~Zhang, ``Dynamic
  multi-objective awpso in dt-assisted uav cooperative task assignment,''
  \emph{IEEE Journal on Selected Areas in Communications}, vol.~41, no.~11, pp.
  3444--3460, 2023.

\bibitem{Zhou2024FdDtUav}
L.~Zhou, S.~Leng, and Q.~Wang, ``A federated digital twin framework for
  uavs-based mobile scenarios,'' \emph{IEEE Transactions on Mobile Computing},
  vol.~23, no.~6, pp. 7377--7393, 2024.

\bibitem{Bilen2022Flight}
T.~Bilen, E.~Ak, B.~Bal, and B.~Canberk, ``A proof of concept on digital
  twin-controlled wifi core network selection for in-flight connectivity,''
  \emph{IEEE Communications Standards Magazine}, vol.~6, no.~3, pp. 60--68,
  2022.

\bibitem{darius}
M.~Ahmadian, M.~Ruiz, J.~Comellas, and L.~Velasco, ``Darius: A digital twin to
  improve the performance of quantum key distribution,'' \emph{Journal of
  Lightwave Technology}, vol.~42, no.~5, pp. 1356--1367, 2024.

\bibitem{Chau2023}
H.~K. Chau, C.~Bridges, and P.~Nisbet-Jones, ``{Portable Frequency Stabilized
  Lasers for Quantum Technologies Using Digital Techniques},'' \emph{IEEE
  Transactions on Instrumentation and Measurement}, vol.~72, pp. 1--6, 2023.

\bibitem{Amir2022}
M.~Amir, C.~Bauckhage, A.~Chircu, C.~Czarnecki, C.~Knopf, N.~Piatkowski, and
  E.~Sultanow, ``{What can we expect from Quantum (Digital) Twins?}''
  \emph{17th International Conference on Wirtschaftsinformatik, WI 2022}, 2022.

\bibitem{Khan2022}
A.~Khan, F.~Shahid, C.~Maple, A.~Ahmad, and G.~Jeon, ``{Toward Smart
  Manufacturing Using Spiral Digital Twin Framework and Twinchain},''
  \emph{IEEE Transactions on Industrial Informatics}, vol.~18, no.~2, pp.
  1359--1366, 2022.

\bibitem{van2024dynamic}
N.~Van~Huynh, B.~Zhang, D.-H. Tran, D.~T. Hoang, D.~N. Nguyen, G.~Zheng,
  D.~Niyato, and Q.-V. Pham, ``{Dynamic Spectrum Access for Ambient Backscatter
  Communication-assisted D2D Systems with Quantum Reinforcement Learning},''
  \emph{arXiv preprint arXiv:2410.17971}, 2024.

\bibitem{jaschke2022ab}
D.~Jaschke, A.~Pagano, S.~Weber, and S.~Montangero, ``Ab-initio
  tree-tensor-network digital twin for quantum computer benchmarking in 2d,''
  \emph{arXiv preprint arXiv:2210.03763}, 2022.

\bibitem{ahmadian2022designing}
M.~Ahmadian, M.~Ruiz, M.~On, S.~K. Singh, J.~Comellas, R.~Proietti, S.~B. Yoo,
  and L.~Velasco, ``Designing a digital twin for quantum key distribution,'' in
  \emph{2022 European Conference on Optical Communication (ECOC)}.\hskip 1em
  plus 0.5em minus 0.4em\relax IEEE, 2022, pp. 1--4.

\bibitem{qu2023dtqfl}
Z.~Qu, Y.~Li, B.~Liu, D.~Gupta, and P.~Tiwari, ``Dtqfl: A digital twin-assisted
  quantum federated learning algorithm for intelligent diagnosis in 5g mobile
  network,'' \emph{IEEE journal of biomedical and health informatics}, 2023.

\bibitem{wang2023efficient}
Y.~Wang, Y.~He, F.~R. Yu, B.~Song, and V.~C. Leung, ``Efficient resource
  allocation for building the metaverse with uavs: A quantum collective
  reinforcement learning approach,'' \emph{IEEE Wireless Communications},
  vol.~30, no.~5, pp. 152--159, 2023.

\bibitem{tuli2024integration}
E.~A. Tuli, J.-M. Lee, and D.-S. Kim, ``Integration of quantum technologies
  into metaverse: Applications, potentials, and challenges,'' \emph{IEEE
  Access}, vol.~12, pp. 29\,995--30\,019, 2024.

\bibitem{nouma2023post}
S.~E. Nouma and A.~A. Yavuz, ``Post-quantum hybrid digital signatures with
  hardware-support for digital twins,'' \emph{arXiv preprint arXiv:2305.12298},
  2023.

\bibitem{otgonbaatar2023quantum}
S.~Otgonbaatar, O.~Nurmi, M.~Johansson, J.~M{\"a}kel{\"a}, P.~Gawron,
  Z.~Pucha{\l}a, J.~Mielzcarek, A.~Miroszewski, C.~O. Dumitru \emph{et~al.},
  ``Quantum computing for climate change detection, climate modeling, and
  climate digital twin,'' \emph{Authorea Preprints}, 2023.

\bibitem{Khalid2023}
U.~Khalid, M.~S. Ulum, A.~Farooq, T.~Q. Duong, O.~A. Dobre, and H.~Shin,
  ``{Quantum Semantic Communications for Metaverse: Principles and
  Challenges},'' \emph{IEEE Wireless Communications}, vol.~30, no.~4, pp.
  26--36, 2023.

\bibitem{martin2024service}
R.~Martin, B.~Lopez, I.~Vidal, F.~Valera, and B.~Nogales, ``Service for
  deploying digital twins of qkd networks,'' \emph{Applied Sciences}, vol.~14,
  no.~3, p. 1018, 2024.

\bibitem{cao2024softwarized}
H.~Cao, S.~Garg, S.~Mumtaz, M.~Alrashoud, L.~Yang, and G.~Kaddoum,
  ``Softwarized resource allocation in digital twins-empowered networks for
  future quantum-enabled consumer applications,'' \emph{IEEE Transactions on
  Consumer Electronics}, 2024.

\bibitem{vaezi2023delay}
M.~Vaezi, K.~Noroozi, T.~D. Todd, D.~Zhao, and G.~Karakostas, ``Digital twin
  placement for minimum application request delay with data age targets,''
  \emph{IEEE Internet of Things Journal}, vol.~10, no.~13, pp.
  11\,547--11\,557, Jul. 2023.

\bibitem{zheng2023vehicular}
J.~Zheng, T.~H. Luan, Y.~Zhang, R.~Li, Y.~Hui, L.~Gao, and M.~Dong, ``Data
  synchronization in vehicular digital twin network: A game theoretic
  approach,'' \emph{IEEE Transactions on Wireless Communications}, vol.~22,
  no.~11, pp. 7635--7647, Nov. 2023.

\bibitem{alkhateeb2023realtime}
A.~Alkhateeb, S.~Jiang, and G.~Charan, ``Real-time digital twins: Vision and
  research directions for 6g and beyond,'' \emph{arXiv preprint
  arXiv:2301.11283}, 2023.

\bibitem{lin2023from}
X.~Lin, L.~Kundu, C.~Dick, E.~Obiodu, T.~Mostak, and M.~Flaxman, ``6g digital
  twin networks: From theory to practice,'' \emph{IEEE Communications
  Magazine}, vol.~61, no.~11, pp. 72--78, 2023.

\bibitem{wang2024XAI}
S.~Wang, M.~A. Qureshi, L.~Miralles-Pechuán, T.~Huynh-The, and M.~Liyanage,
  ``Explainable ai for 6g use cases: Technical aspects and research
  challenges,'' \emph{IEEE Open Journal of the Communications Society}, vol.~5,
  pp. 1--18, 2024.

\bibitem{alcaraz2022security}
C.~Alcaraz and J.~Lopez, ``Digital twin: A comprehensive survey of security
  threats,'' \emph{IEEE Communications Surveys \& Tutorials}, vol.~24, no.~3,
  pp. 1475--1503, 2022.

\bibitem{liu2024twotimescale}
W.~Liu, Y.~Fu, Y.~Guo, F.~Wang, W.~Sun, and Y.~Zhang, ``Two-timescale
  synchronization and migration for digital twin networks: A multi-agent deep
  reinforcement learning approach,'' \emph{IEEE Transactions on Wireless
  Communications}, 2024, early Access, doi:10.1109/TWC.2024.3452689.

\end{thebibliography}

\end{document}